\documentclass[aps,prd,onecolumn,nofootinbib,superscriptaddress,floatfix]{revtex4-2}
\pdfoutput=1
\usepackage{amsmath,amssymb}
\usepackage{graphicx}
\usepackage[dvipsnames,table]{xcolor}
\usepackage[T1]{fontenc}
\usepackage[colorlinks=true, linkcolor=blue, urlcolor=blue, citecolor=blue, anchorcolor=blue]{hyperref}
\usepackage{multirow,makecell}
\usepackage{comment}

\newcommand{\bec}{\begin{center}}
\newcommand{\eec}{\end{center}}
\newcommand{\beq}{\begin{equation}}
\newcommand{\eeq}{\end{equation}}
\newcommand{\bea}{\begin{eqnarray}}
\newcommand{\eea}{\end{eqnarray}}

\usepackage{orcidlink}

\begin{document}

\title{Diagnosing Complex Langevin using Configurational Temperature in the Three-Dimensional XY Model at Finite Chemical Potential}

\author{Anosh Joseph~\orcidlink{0000-0003-4288-8207}}
\email{anosh.joseph@wits.ac.za}
\affiliation{National Institute for Theoretical and Computational Sciences, School of Physics, and Mandelstam Institute for Theoretical Physics,
University of the Witwatersrand, Johannesburg, Wits 2050, South Africa}

\author{Arpith Kumar~\orcidlink{0000-0002-5887-3803}}
\email{arpith.kumar@uni-bielefeld.de}
\affiliation{Fakult\"at f\"ur Physik, Universit\"at Bielefeld, D-33615 Bielefeld, Germany}
\affiliation{Key Laboratory of Quark and Lepton Physics (MOE) and Institute of Particle Physics, Central China Normal University, Wuhan 430079, China}

\begin{abstract}
We study complex Langevin (CL) dynamics for the three-dimensional XY model at finite chemical potential, where the sign-problem-free world-line formulation provides reference values with only statistical errors.
We use this benchmark to test the configurational temperature estimator, which is a sampling-consistency check whose target value follows from a Schwinger--Dyson identity that also holds for the complex weight.
We compare the estimator with the Nagata--Nishimura--Shimasaki drift-decay criterion and with the world-line action density on a single $8^3$ lattice, for $0.1 \leq \beta \leq 1.0$ and $-0.2 \leq \mu^2 \leq 0.2$.
We find that in the ordered phase, away from the transition, CL agrees with the world-line results and the estimator agrees with its target value.
In the disordered phase, CL fails whether the sign problem is mild or severe, in line with earlier studies, and the estimator lies $1$--$12\%$ below its target.
Near the transition, where the CL action density differs by a few percent, the estimator fails to resolve the failure. At the same time, the drift distributions show extended tails over the sampled range.
An exploratory map of the estimator in the $(\beta, \mu^2)$ plane up to $\mu^2 = 14$ indicates that, on this lattice, the region of large deviations moves with the order--disorder boundary.
Dynamical stabilization restores agreement of the action density with the world-line value at $\beta = 0.1$ and preserves it at $\beta = 0.8$, but leaves a $1$--$4\%$ deviation at $\beta = 0.4$, where the sign problem is severe.
We discuss the limitations of the estimator, including the effect of its poles on its statistical uncertainty and its insensitivity to ergodicity problems.
\end{abstract}

\maketitle 

\section{Introduction} 
\label{sec:intro} 

Quantum field theories with complex actions are notoriously hard to study non-perturbatively. 
The partition function involves a weight 
\begin{equation} 
e^{- S} = | e^{- S} | e^{i \varphi}, 
\end{equation} 
which is not real. 
It, therefore, cannot be interpreted as a probability. 
As a result, standard importance-sampling techniques typically fail. 
This difficulty is known as the {\it sign problem}. 
This problem is particularly pressing in QCD at nonzero baryon chemical potential. 
There, a reliable determination of the phase diagram in the temperature-chemical potential plane is still missing \cite{deForcrand:2009zkb}. 
For a recent review of the sign problem and approaches to it, see Ref.~\cite{Aarts:2026uiu}. 

Many methods have been developed to probe restricted regions of the QCD phase diagram \cite{Fodor:2001au, Fodor:2002km, Allton:2002zi, Gavai:2003mf, deForcrand:2002hgr, DElia:2002tig, DElia:2009bzj, Kratochvila:2005mk, Alexandru:2005ix, Ejiri:2008xt, Fodor:2007vv}.
The sign problem itself has been studied extensively, which has led to new formulations \cite{Anagnostopoulos:2001yb, Ambjorn:2002pz} and to a better understanding of how phase fluctuations affect observables \cite{Akemann:2004dr, Splittorff:2006fu, Han:2008xj, Bloch:2008cf, Lombardo:2009aw, Danzer:2009dk, Hands:2010zp}.
In some theories, the path integral can be rewritten with a manifestly real and positive weight, which removes the sign problem altogether \cite{Chandrasekharan:1999cm, Endres:2006xu, Chandrasekharan:2008gp, Banerjee:2010kc}; the sign problem is thus a property of the formulation rather than of the theory.
No such reformulation is known for QCD. 

The complex Langevin method (CLM) offers an alternative and more general approach \cite{Parisi:1984cs, Klauder:1983nn}. 
In this method, the fields $\phi$ evolve in a fictitious Langevin time $\theta$ according to a stochastic equation. 
The key modification is that the fields are complexified when the action is complex. 
Since stochastic evolution replaces importance sampling, CLM can bypass the sign problem. 
For real actions, proofs based on the associated Fokker--Planck equation guarantee correct convergence \cite{Damgaard:1987rr}. 
Unfortunately, these proofs do not hold for complex actions. 
One can use formal arguments based on the holomorphicity and the Cauchy--Riemann equations to suggest the conditions under which CLM can still converge correctly. 
(See Ref. \cite{Joseph:2025tfw} for a recent review.) 

Several developments concern the convergence of CLM.
The choice of kernel and integration cycle can strongly affect it, as studied early on \cite{Okamoto:1988ru} and, more recently, with machine-learning kernels \cite{Alvestad:2022abf, Alvestad:2023jgl} and deformed integration cycles \cite{Mandl:2024zvz, Hansen:2024kjm}.
{\it Dynamical stabilization} adds a non-holomorphic force that keeps the complexified fields close to the real manifold \cite{Attanasio:2018rtq, Aarts:2016qhx, Attanasio:2020spv, Hansen:2024lkn}, and has also been applied to the XY model \cite{Attanasio:2017rxk, Cai:2021hjq}.
Diffusion models have been used to learn the distributions that complex Langevin processes sample \cite{Habibi:2024fbn, Aarts:2025lpi}. 

CLM has been applied to a wide range of models \cite{Berges:2005yt, Berges:2006xc, Berges:2007nr, Bloch:2017sex, Aarts:2008rr, Pehlevan:2007eq, Aarts:2008wh, Aarts:2009hn, Aarts:2010gr, Aarts:2011zn}, including the relativistic Bose gas at finite chemical potential, low-dimensional QCD-like systems and spin models at nonzero density, as well as supersymmetric matrix models \cite{Ito:2016efb, Ito:2016hlj, Anagnostopoulos:2017gos, Joseph:2019sof, Joseph:2020gdh, Kumar:2022fas, Kumar:2022giw, Kumar:2023nya}, large-$N$ unitary matrix models with Gross--Witten--Wadia transitions~\cite{Basu:2018dtm, Gross:1980he, Wadia:2012fr, Wadia:1980cp}, and real-time gauge theories, including $3+1$-dimensional Yang--Mills dynamics \cite{Boguslavski:2022dee, Boguslavski:2023unu, Boguslavski:2026ker}.
For QCD at finite density \cite{Sexty:2013ica}, CLM stabilized by gauge cooling has been compared with reweighting \cite{Fodor:2015doa}, and used to study the thermodynamics and equation of state of hot and dense QCD \cite{Sexty:2019vqx, Kogut:2019qmi, Attanasio:2022mjd, Mandl:2026ngb, Mandl:2026eos} and the transition temperature \cite{Scherzer:2020kiu}.
Its region of validity has been examined at low temperature and high density \cite{Nagata:2018mkb, Ito:2020mys} and near the deconfinement transition \cite{Tsutsui:2025jez}. 

Despite this progress, the method still has two main shortcomings: non-convergence and the notorious wrong convergence. 
It remains an open challenge to control convergence robustly. 
This has motivated the development of {\it reliability criteria} that distinguish correct from incorrect convergence. 
The underlying problems are usually traced to the probability distribution in the complexified configuration space. 
If it does not decay fast enough at infinity or at the poles of the drift, boundary terms appear. 
These boundary terms undermine the formal proof of correctness \cite{Aarts:2009uq, Aarts:2011ax}. 
Conditions that relate incorrect convergence to these boundary terms were formulated in Ref. \cite{Seiler:2023kes}; see also Refs. \cite{Scherzer:2018hid, Seiler:2020mkh} and, for the XY model, see Refs. \cite{Scherzer:2019lrh, Sexty:2021nov}. 
The criterion of Nagata, Nishimura, and Shimasaki (NNS) links the presence of boundary terms to the growth of the drift term. 
It requires the probability distribution of the drift to decay exponentially or faster \cite{Nagata:2016vkn}. 
Recent analyses established necessary and sufficient conditions for correctness \cite{Mandl:2025mav} and compared several criteria systematically \cite{Mandl:2026vdc}. 
More recently, the present authors proposed an alternative diagnostic based on the configurational temperature estimator \cite{Joseph:2025fcd}. 
This diagnostic has since been extended to matrix models and random matrix ensembles \cite{Joseph:2026rwh}. 
The estimator is part of a broader probability-geometric framework, which relates it to Fisher information and Schwinger--Dyson identities \cite{Joseph:2026hom}. 

In this work, we return to the three-dimensional XY model at finite chemical potential, a well-studied testbed for CLM.
Aarts and James showed that CLM works in the ordered phase of this model and fails in the disordered phase, and that the failure is not due to the sign problem \cite{Aarts:2010aq}.
The boundary-term analysis of Ref.~\cite{Scherzer:2019lrh} reached a consistent picture. 
Earlier investigations revealed severe numerical instabilities that require adaptive step-size control \cite{Aarts:2009dg}. 
The same instabilities are encountered in QCD in the heavy-dense limit \cite{Aarts:2008rr, Aarts:2009dg}. 
The XY model also exhibits Roberge--Weiss periodicity at imaginary chemical potential \cite{Roberge:1986mm}. 
It is closely related to the relativistic Bose gas, where CLM has proven successful at weak coupling \cite{Aarts:2008wh, Aarts:2009hn}. 
Furthermore, this model admits a sign-problem-free world-line formulation that can be solved with the worm algorithm \cite{Chandrasekharan:2008gp, Banerjee:2010kc, Prokofev:2001ddj}. 
This makes it a well-suited benchmark for assessing the reliability of the CLM. 

Since the failure region of CLM in this model is known and the world-line formulation provides a benchmark free of the sign problem, the model is well suited to test how a correctness criterion performs.
We compare the configurational temperature estimator $\beta_{\rm config}$ with the NNS drift criterion and with the world-line action density, from strong to weak coupling and from imaginary to real chemical potential.
In addition, we
(i) derive the target value of the estimator from a Schwinger--Dyson identity that holds for the complex weight, and relate it to the consistency conditions and boundary terms of Refs.~\cite{Aarts:2011ax, Scherzer:2018hid};
(ii) analyze the poles of the estimator and their consequences for its interpretation and its statistical uncertainty;
(iii) map the estimator over the $(\beta, \mu^2)$ plane up to $\mu^2 = 14$; and
(iv) test dynamical stabilization against the world-line results and examine what the estimator indicates for the stabilized process.
Throughout, we use the world-line results as a reference to quantify how well the estimator detects the known failure of CLM, and we identify where it does not: near the transition and in ergodicity problems.
All results are for a single $8^3$ lattice, on which the order--disorder transition is a crossover.
Robust, quantitative diagnostics of this kind are essential if CLM is to be used with confidence in more challenging settings, such as lattice QCD at finite baryon density. 

The remainder of this paper is organized as follows. 
In Sec.~\ref{sec:Complex_Langevin_Dynamics}, we give a brief overview of complex Langevin dynamics and of the stochastic evolution equations relevant for theories with complex actions. 
Sec.~\ref{sec:3D_XY_Model} introduces the three-dimensional XY model at finite chemical potential. 
It discusses the Roberge--Weiss periodicity and the strong-coupling behavior of the model, and quantifies the severity of the sign problem using the phase-quenched ensemble. 
It also presents the sign-problem-free world-line benchmark and the details of the complex Langevin simulation. 
Sec.~\ref{sec:Criteria_for_Correctness} introduces and applies the configurational temperature estimator. 
We study how it depends on the severity of the sign problem and on the order--disorder phase boundary across the full $(\beta, \mu^2)$ plane. 
We then turn to the NNS drift-decay criterion and compare the two correctness criteria, with a summary in Table~\ref{tab:summary}. 
The section closes with methods for obtaining correct results in the disordered phase (Sec.~\ref{sec:ds}); we test one of them, dynamical stabilization, using the estimator and the world-line results. 
Sec.~\ref{sec:discussion} summarizes our findings and discusses implications for broader applications of complex Langevin dynamics.

\section{A Brief Overview of Complex Langevin Dynamics}
\label{sec:Complex_Langevin_Dynamics}

In stochastic quantization \cite{Damgaard:1987rr}, the fields $\phi_x \equiv \phi(x)$ are given an additional, auxiliary Langevin time $\theta$.
They evolve according to the stochastic differential equation
\begin{equation} 
\label{eq:eqphi} 
\frac{\partial \phi_{x}(\theta)}{\partial \theta} = - \frac{\delta S[\phi]}{\delta \phi_{x}(\theta)} + \eta_x(\theta), 
\end{equation} 
where $\eta_x(\theta)$ is Gaussian noise.
Here and in the following, $x$ labels the sites of the Euclidean spacetime lattice.

When the action $S$ is complex, the fields must be complexified,
\begin{equation} 
\phi_x \to \phi_x^{\rm R} + i \phi_x^{\rm I}.  
\end{equation} 
This leads to coupled Langevin equations for the real and imaginary components,
\begin{subequations} 
\begin{align} 
\frac{\partial \phi_{x}^{\rm R}}{\partial \theta} &= K_x^{\rm R} + \eta^{\rm R}_{x}, 
& K_x^{\rm R} = - \left. {\rm Re} \frac{\delta S}{\delta \phi_x} \right|_{\phi_x \to \phi_x^{\rm R} + i \phi_x^{\rm I}}, \\ 
\frac{\partial \phi_x^{\rm I}}{\partial \theta} &= K_x^{\rm I} + \eta^{\rm I}_x, 
& K_x^{\rm I} = - \left. {\rm Im} \frac{\delta S}{\delta \phi_x} \right|_{\phi_x \to \phi_x^{\rm R} + i \phi_x^{\rm I}}. 
\end{align} 
\label{eq:eqphic} 
\end{subequations} 

The real and imaginary noise components obey
\begin{subequations} 
\begin{align} 
& \langle \eta_x^{\rm R}(\theta) \rangle = \langle \eta_x^{\rm I}(\theta) \rangle = 0, \\ 
& \langle \eta^{\rm R}_x(\theta) \eta^{\rm I}_y(\theta') \rangle = 0, \\ 
& \langle \eta^{\rm R}_x(\theta) \eta^{\rm R}_y(\theta') \rangle = 2 N^{\rm R} \delta(x - y) \delta(\theta - \theta'), \\ 
& \langle \eta^{\rm I}_x(\theta) \eta^{\rm I}_y(\theta') \rangle = 2 N^{\rm I} \delta(x - y) \delta(\theta - \theta'). 
\end{align} 
\end{subequations} 
In the above, the parameters $N^{\rm R}$ and $N^{\rm I}$ control the relative strength of the real and imaginary noise components.
They are subject to the constraints $N^{\rm R} - N^{\rm I} = 1$ and $N^{\rm I} \geq 0$.
In this work, we restrict ourselves to real noise, $N^{\rm R} = 1$ and $N^{\rm I} = 0$.

CLM has a key advantage.
The complex action enters only through the drift.
The process therefore samples a real, positive distribution over the complexified field space, rather than the complex weight $e^{-S}$.
This opens the possibility of evading the sign problem.

In the limit of infinite Langevin time, we can define the expectation values as averages over the probability density $P[\phi_x^{\rm R}, \phi_x^{\rm I}; \theta]$.
This density evolves according to the Fokker--Planck equation
\begin{equation} 
\frac{\partial P}{\partial \theta} = L^{T} P, 
\end{equation} 
with the Fokker--Planck operator
\begin{equation} 
L^T = \sum_x \left[ \frac{\delta}{\delta \phi_x^{\rm R}} \left( N^{\rm R} \frac{\delta}{\delta \phi_x^{\rm R}} - K_x^{\rm R} \right) + 
\frac{\delta}{\delta \phi_x^{\rm I}} 
\left( N^{\rm I} \frac{\delta}{\delta \phi_x^{\rm I}} - K_x^{\rm I} \right) \right]. 
\end{equation} 

Formally, expectation values are then
\begin{equation} 
\label{eq:eqP} 
\langle O \rangle_{P(\theta)} = \frac{\int D\phi^{\rm R} D\phi^{\rm I} P[\phi^{\rm R}, \phi^{\rm I}; \theta] O[\phi^{\rm R} + i \phi^{\rm I}]}
{\int D\phi^{\rm R} D\phi^{\rm I} P[\phi^{\rm R}, \phi^{\rm I}; \theta]}, 
\end{equation} 
with
\begin{equation} 
\int D\phi^{\rm R} D\phi^{\rm I} \equiv \int \prod_x d \phi_x^{\rm R} d \phi_x^{\rm I}, 
\end{equation} 
where the product over $x$ represents a functional integral over spacetime.

One may also define expectation values with respect to a complex weight $\rho[\phi; \theta]$.
This weight evolves under the {\it complex} Fokker--Planck equation,
\begin{equation} 
\frac{\partial \rho}{\partial \theta} = L_0^{T} \rho, \qquad L_0^T = \sum_x \frac{\delta}{\delta \phi_x} 
\left( \frac{\delta}{\delta \phi_x} + \frac{\delta S}{\delta \phi_x} \right).  
\end{equation} 
It admits the stationary solution
\begin{equation} 
\rho[\phi] \propto e^{- S[\phi]}. 
\end{equation} 

Under the assumptions of holomorphicity and the validity of partial integration, we find \cite{Aarts:2009uq}
\begin{equation} 
\langle O \rangle_{P(\theta)} = \langle O \rangle_{\rho(\theta)}.  
\end{equation} 
If convergence as $\theta \to \infty$ is achieved,
\begin{equation} 
\lim_{\theta \to \infty} \langle O \rangle_{\rho(\theta)} = \langle O \rangle_{\rho(\infty)}, \quad\quad \rho(\phi; \infty) \sim \exp(- S[\phi]), 
\end{equation} 
then CLM reproduces the correct quantum expectation values.

The equality \(\langle O \rangle_{P} = \langle O \rangle_{\rho}\) is established for observables $O$ that are holomorphic functions of the complexified fields.
The argument has two ingredients.
First, it uses the Cauchy--Riemann equations to trade derivatives with respect to $\phi^{\rm R}$ and $\phi^{\rm I}$ for complex derivatives.
Second, it integrates by parts in the complexified field space.
This second step requires that there are no boundary terms, neither at infinity nor at singularities of the drift or of the observable \cite{Aarts:2009uq, Aarts:2011ax, Aarts:2017vrv}.

However, explicit studies have revealed limitations.
For example, in simple models, CLM fails when complex noise is present ($N^{\rm I} > 0$), but converges correctly for real noise ($N^{\rm I} = 0$) \cite{Aarts:2009uq}.
More generally, the stochastic process can explore regions of the complexified field space that are not representative of the underlying theory.
It may then converge to the wrong limit.
This motivates the need for {\it diagnostic criteria} that identify when CLM results can be trusted.

\section{The Three-dimensional XY Model at Finite Chemical Potential}
\label{sec:3D_XY_Model}

The three-dimensional XY model is a well-suited testing ground for complex Langevin dynamics at finite chemical potential. 
On the one hand, it shares several features with finite-density QCD.
It has a genuine sign problem in its conventional formulation.
It shows Roberge--Weiss periodicity at imaginary chemical potential.
It also suffers from instabilities that require adaptive step-size control \cite{Aarts:2009dg}.
However, it is far simpler, since it has neither gauge fields nor a fermion determinant.
On the other hand, it admits a world-line (flux) representation with a real and positive weight.
This representation can be simulated efficiently with the worm algorithm \cite{Chandrasekharan:2008gp, Banerjee:2010kc, Prokofev:2001ddj}.
Together, these two properties make the model a cost-effective laboratory.
Complex Langevin results can be compared directly with sign-problem-free simulations, which is a stringent test of the correctness criteria.

The model is defined by the lattice action
\begin{equation} 
S = - \beta \sum_{x} \sum_{\nu = 0}^{2} \cos \big(\phi_{x} - \phi_{x + \hat{\nu}} - i \mu \delta_{\nu, 0} \big), 
\label{eq:action} 
\end{equation} 
where the angular variables satisfy $0 \leq \phi_x < 2\pi$.
The model lives on a three-dimensional cubic lattice $\Lambda$ of volume $\Omega = N_\tau N_s^2$ with periodic boundary conditions.
Here $x$ is a point on the lattice, $x \equiv (x_0, x_1, x_2)$.
The index $\nu = 1, 2$ labels the spatial directions, while $\nu = 0$ denotes the temporal direction.
The non-linear U(1) bosonic field of the theory takes the form $\exp(i \phi_x)$.
The coupling constant $\beta$ controls the strength of the interactions.
The chemical potential $\mu$ couples to the Noether charge of the global symmetry $\phi_x \to \phi_x + \alpha$, where $\alpha$ is any constant real number in $[0, 2 \pi)$ \cite{Hasenfratz:1983ba}.

At nonzero chemical potential, the action becomes complex. 
It satisfies the relation
\begin{equation} 
S^*(\mu) = S(-\mu^*). 
\end{equation} 
At $\mu = 0$, the theory has a continuous phase transition, at $\beta_c(0) \simeq 0.4542$ \cite{Campostrini:2000iw, Banerjee:2010kc}.
For $\beta > \beta_c(\mu)$, the theory is in a spontaneously broken (ordered) phase.
For $\beta < \beta_c(\mu)$, it is in the symmetric (disordered) phase.
On the finite lattice used here, the transition is rounded into a crossover.

In the complex Langevin formulation, the real and imaginary drift components read
\begin{subequations} 
\begin{align} 
K^{\rm R}_x = - \beta \sum_\nu & \Bigg[ \sin \left( \phi_x^{\rm R} - \phi_{x + \hat{\nu}}^{\rm R} \right) 
\cosh \left( \phi_x^{\rm I} - \phi_{x + \hat{\nu}}^{\rm I} - \mu \delta_{\nu, 0} \right) 
+ \sin \left( \phi_x^{\rm R} - \phi_{x - \hat{\nu}}^{\rm R} \right) \cosh \left( \phi_x^{\rm I} - \phi_{x - \hat{\nu}}^{\rm I} + \mu \delta_{\nu, 0} \right) \Bigg], \\ 
K^{\rm I}_x = - \beta \sum_\nu & \Bigg[ \cos \left( \phi_x^{\rm R} - \phi_{x + \hat{\nu}}^{\rm R} \right) 
\sinh \left( \phi_x^{\rm I} - \phi_{x + \hat{\nu}}^{\rm I} - \mu \delta_{\nu, 0} \right) 
+ \cos \left( \phi_x^{\rm R} - \phi_{x - \hat{\nu}}^{\rm R} \right) \sinh \left( \phi_x^{\rm I} - \phi_{x - \hat{\nu}}^{\rm I} + \mu \delta_{\nu, 0} \right) \Bigg]. 
\end{align} 
\end{subequations} 
These drift terms follow from the lattice action in Eq. \eqref{eq:action} after complexification, $\phi_x \to \phi_x^{\rm R} + i \phi_x^{\rm I}$.

We evolve the fields with the Euler-discretized Langevin equations.
The Langevin time is discretized as $\theta_n = \sum_{m < n} \epsilon_m$ with adaptive step size $\epsilon_n$. 
The discretized field evolution equations take the form
\begin{subequations} 
\begin{align} 
\phi_x^{\rm R}(n + 1) {} &= \phi_x^{\rm R}(n) + \epsilon_n K_x^{\rm R}(n) + \sqrt{\epsilon_n} \eta_x(n), \\ 
\phi_x^{\rm I}(n + 1) {} &= \phi_x^{\rm I}(n) + \epsilon_n K_x^{\rm I}(n), 
\end{align} 
\label{eq:eqphic_disc} 
\end{subequations} 
where we use real noise,
\begin{equation} 
\langle \eta_x(n) \eta_{x'}(n') \rangle = 2 \delta_{xx'} \delta_{nn'}. 
\end{equation} 

The drift terms become unbounded when $\phi_x^{\rm I} \neq 0$.
The system is therefore prone to instabilities and runaway trajectories.
In practice, stable simulations require an adaptive step size \cite{Aarts:2009dg}.
It is determined at each update by
\beq 
\epsilon_n = \min \left\{ \bar{\epsilon}, \bar{\epsilon} \frac{\langle K^{\rm max} \rangle}{K^{\rm max}_n} \right\}, 
\eeq 
where
\beq 
K^{\rm max}_n = \max_x \left| K^{\rm R}_x(n) + i K^{\rm I}_x(n) \right|, 
\eeq 
and $\bar{\epsilon}$ is the target step size. 
We estimate $\langle K^{\rm max} \rangle$ during thermalization.
The observables are then collected at equal intervals of Langevin time to ensure statistically consistent sampling.

Our primary observable is the action density, $\langle S \rangle / \Omega$.
After complexification the action decomposes as $S = S^{\rm R} + i S^{\rm I}$ with
\begin{subequations} 
\begin{align} 
S^{\rm R} & = - \beta \sum_{x, \nu} \cos \left( \phi_x^{\rm R} - \phi_{x + \hat{\nu}}^{\rm R} \right) 
\cosh \left( \phi_x^{\rm I} - \phi_{x + \hat{\nu}}^{\rm I} - \mu \delta_{\nu, 0} \right), \\ 
S^{\rm I} & = \beta \sum_{x, \nu} \sin \left( \phi_x^{\rm R} - \phi_{x + \hat{\nu}}^{\rm R} \right) 
\sinh \left( \phi_x^{\rm I} - \phi_{x + \hat{\nu}}^{\rm I} - \mu \delta_{\nu, 0} \right). 
\end{align} 
\end{subequations} 
For real $\mu$, the relation $S^*(\mu) = S(-\mu^*)$, together with the invariance of the action under $\phi_x \to -\phi_x$, $\mu \to -\mu$, implies that $\langle S \rangle$ is real and even in $\mu$. Accordingly, $\langle S^{\rm I} \rangle$ is consistent with zero in our simulations.

\subsection{Imaginary chemical potential and RW periodicity}

When the chemical potential is imaginary, $\mu = i \mu_{\rm I}$, the action \eqref{eq:action} becomes real.
In this case, it reduces to
\begin{equation} 
S_{\rm imag} = - \beta \sum_{x, \nu} \cos\left( \phi_x - \phi_{x + \hat{\nu}} + \mu_{\rm I} \delta_{\nu, 0} \right). 
\end{equation} 

The drift then has only a real component, which we denote as $K_x = K_x^{\rm R}$. 
We have 
\begin{equation} 
\label{eq:K_x} 
K_x = - \beta \sum_\nu \Bigg[ \sin \left( \phi_x - \phi_{x + \hat{\nu}} + \mu_{\rm I} \delta_{\nu, 0} \right) 
+ \sin \left( \phi_x - \phi_{x - \hat{\nu}} - \mu_{\rm I} \delta_{\nu, 0} \right) \Bigg]. 
\end{equation} 
In this case we can apply the standard Monte Carlo methods without a sign problem.
In this work, we use real Langevin dynamics.
This gives a controlled testing ground for the continuation from $\mu^2 < 0$ (imaginary chemical potential) to $\mu^2 > 0$ (real chemical potential).

Here we write $x = ({\bf x}, \tau)$, where $\tau = 0, \cdots, N_\tau - 1$ denotes the Euclidean time coordinate.
In the continuum and in the final time slice (FTS) formulation, the theory is periodic under shifts $\mu_{\rm I} \to \mu_{\rm I} + 2 \pi / N_\tau$.
This leads to Roberge--Weiss (RW) transitions at $\mu_{\rm I} = \pi / N_\tau$, as in QCD \cite{Roberge:1986mm}.
In QCD, these transitions are closely tied to confinement-deconfinement dynamics.
Reproducing them correctly in the XY model is therefore a useful analog test.
However, the standard lattice discretization of the action does not make this periodicity manifest.
The reason is that the chemical potential is spread uniformly over all temporal links, rather than implemented as a twist at the temporal boundary.
As a result, simulations with local updates may miss the RW transition, as we discuss below. 
The periodicity can be restored by moving the entire chemical potential dependence to the final time slice, through the field redefinition
\beq 
\phi_{{\bf x}, \tau} \to \phi'_{{\bf x}, \tau} = \phi_{{\bf x}, \tau} - \mu_{\rm I} \tau. 
\eeq 
This yields the final-time-slice formulation, which is valid for arbitrary complex $\mu$. 
The action is
\begin{equation} 
S_{\rm FTS} = - \beta \sum_{x, \nu} \cos \left( \phi_x - \phi_{x + \hat{\nu}} - i N_\tau \mu \delta_{\tau, N_\tau - 1} \delta_{\nu, 0} \right). 
\end{equation} 
The redefinition is a constant shift of each angular variable.
The two formulations therefore have identical partition functions and expectation values for any $\mu$, and differ only in how a local algorithm explores the configuration space.
In the standard formulation, reaching the correct branch beyond the RW point requires a global change of the winding of the fields along the temporal direction, which local updates rarely achieve in the ordered phase.
Ref.~\cite{Aarts:2010aq} showed that the FTS formulation captures the RW transition even at large $\beta$ (e.g., $\beta = 0.7$), where simulations with the standard discretization miss it.

\subsection{Action density at strong coupling}
\label{sec:strong_coupling}

We use expansions at small chemical potential and at strong coupling to assess the applicability of complex Langevin dynamics.
In this regime, observables at real and imaginary values of the chemical potential can be compared via analytic continuation in $\mu^2$.
Figs.~\ref{fig:act_den_plots_a} and~\ref{fig:act_den_plots_b} show the real part of the action density as a function of $\mu^2$ for several values of $\beta$.
These values span the ordered phase at large $\beta$ and the disordered phase at small $\beta$.
At large $\beta$, the observable is smooth across $\mu^2 = 0$.
This strongly suggests that complex Langevin may be reliable in this region.

At small $\beta$, the results can be cross-checked against the strong-coupling expansion.
Writing the partition function in terms of the free energy density,
\beq 
Z = \exp(- \Omega f), 
\eeq 
one finds, for $N_\tau > 4$,
\begin{equation} 
f(\beta) = - \tfrac{3}{4} \beta^2 - \tfrac{21}{64} \beta^4 + \mathcal{O}(\beta^6), 
\end{equation} 
so that the action density becomes
\begin{equation} 
\label{eq:act_density} 
\frac{\langle S \rangle}{\Omega} = \beta f'(\beta) = - \tfrac{3}{2}\beta^2 - \tfrac{21}{16}\beta^4 + \mathcal{O}(\beta^6), 
\end{equation} 
where the prime denotes differentiation with respect to $\beta$.
The chemical potential enters only through world lines that wind around the temporal direction, whose contribution is of order $\beta^{N_\tau} = \beta^8$ here.
To the order shown, $f$, and hence the action density, is therefore independent of $\mu$, in line with the absence of a response to $\mu$ in the gapped disordered phase.

Using Eq.~\eqref{eq:act_density}, the strong-coupling expansion yields
\begin{equation} 
\begin{array}{c|cccc} 
\beta & 0.1 & 0.2 & 0.3 & 0.4 \\ \hline 
\langle S \rangle / \Omega & -0.01513 & -0.0621 & -0.1456 & -0.2736 
\end{array} 
\end{equation} 
up to corrections of order $\beta^6$.
The difference between the simulation results (Fig.~\ref{fig:act_den_plots_a}) and the truncated series, which indicates the size of the neglected terms, is negligible at $\beta = 0.1$, but reaches $\approx 2.6 \times 10^{-2}$ (about $9\%$) at $\beta = 0.4$.
The above predictions agree well with the simulation results up to $\beta=0.3$.
Beyond this value, the agreement begins to deteriorate.

\subsection{Severity of the sign problem}
\label{sec:sign_problem_severity}

We can quantify the severity of the sign problem through the expectation value of the complex phase factor
\begin{equation} 
\langle e^{i\varphi} \rangle = \left \langle \frac{ e^{- S} }{ | e^{- S} | } \right\rangle_{\rm pq}, ~~~ \text{where} ~~~ S = S^{\rm R} + i S^{\rm I}, 
\end{equation} 
so that $e^{i \varphi} = e^{- i S^{\rm I}}$.
The expectation value is evaluated in the {\it phase-quenched theory}, where only the real part of the action contributes to the Boltzmann weight \cite{deForcrand:2009zkb}.
For the XY model, the phase-quenched theory is an {\it anisotropic XY model} with action
\begin{equation} 
S_{\rm pq} = - \sum_{x, \nu} \beta_\nu \cos\left( \phi_x - \phi_{x + \hat\nu} \right), 
\end{equation} 
where $\beta_0 = \beta \cosh \mu$ for real $\mu$ and $\beta_{1, 2} = \beta$.

In the phase-quenched theory, the corresponding free energy density is
\begin{equation} 
f_{\rm pq} = - \tfrac{1}{4} \beta^2 \left( 2 + \cosh^2 \mu \right) - \tfrac{1}{64} \beta^4 \left(6 + 16 \cosh^2 \mu - \cosh^4 \mu \right) + \mathcal{O}(\beta^6). 
\end{equation} 
Note that the strong-coupling expansion depends on $\mu$ only through even powers.
This is consistent with analyticity in $\mu^2$.

The severity of the sign problem can then be quantified by the average phase factor,
\begin{equation} 
\langle e^{i \varphi} \rangle_{\rm pq} = \frac{Z}{Z_{\rm pq}} = \exp \left(- \Omega \Delta f \right), ~~~ ~~~ \Delta f = f - f_{\rm pq}. 
\end{equation} 
At strong coupling, this difference is
\begin{equation} 
\Delta f = \tfrac{1}{4} \beta^2 \left(\cosh^2 \mu - 1 \right) + \tfrac{1}{64} \beta^4 \left( \cosh^2 \mu - 1 \right) \left( 15 - \cosh^2 \mu \right) + \mathcal{O}(\beta^6). 
\label{eq:delta_f_sc} 
\end{equation} 
In Sec.~\ref{sec:sign_problem_dependence} (Fig.~\ref{fig:pq_vs_musq}), we will see that our simulation results for $\langle e^{i \varphi} \rangle_{\rm pq}$ at $\beta \leq 0.4$ agree with the above expression, within errors.

On a finite lattice and for small $\mu$, the sign problem is therefore mild, since the suppression factor scales as
\begin{equation} 
\exp \left(- \Omega \Delta f\right) \approx \exp \left( - \tfrac{1}{4} \Omega \beta^2 \mu^2 \right). 
\end{equation} 

If the controlling parameter satisfies
\begin{equation} 
\tfrac{1}{4}\Omega \beta^2 \mu^2 \ll 1, 
\end{equation} 
then
\begin{equation} 
\langle e^{i \varphi} \rangle_{\rm pq} \approx 1 - \tfrac{1}{4} \Omega \beta^2 \mu^2. 
\end{equation} 
Moreover, there are no catastrophic cancellations.

For example, take $\mu^2 = 0.1$ and $\beta = 0.2$.
For the lattice volume used in our simulations, Eq.~\eqref{eq:delta_f_sc} gives $\Omega \Delta f \approx 0.55$ and $\langle e^{i \varphi} \rangle_{\rm pq}\approx 0.58$. 
This is not small, and there is no exponential catastrophe or severe sign problem.

We also quantify the severity of the sign problem systematically across the full parameter space.
To this end, we simulate the phase-quenched ensemble, i.e., the anisotropic XY model with $\beta_0 = \beta \cosh\mu$, $\beta_{1,2} = \beta$, using standard Metropolis dynamics.
The run parameters are given together with those of the other simulations in Sec. ~\ref{sec:sim_details_results}.
The average phase factor is measured directly as
\begin{equation} 
\langle e^{i \varphi} \rangle_{\rm pq} = \langle \cos(S^{\rm I}) \rangle_{\rm pq}, \qquad S^{\rm I} = - \beta \sinh\mu \sum_{x} \sin(\phi_x - \phi_{x + \hat 0}), 
\label{eq:pq_phase} 
\end{equation} 
where the imaginary part vanishes by the $\phi_x \to -\phi_x$ symmetry of $S_{\rm pq}$.

The dependence of the phase factor on $\beta$ can be understood from a cumulant expansion of Eq.~\eqref{eq:pq_phase}. For small $\mu$,
\begin{equation}
\ln \langle e^{i \varphi} \rangle_{\rm pq} \approx - \tfrac{1}{2} \langle (S^{\rm I})^2 \rangle_{\rm pq} = - \tfrac{1}{2} \beta^2 \sinh^2\mu \, \langle J^2 \rangle_{\rm pq}, \qquad J \equiv \sum_x \sin(\phi_x - \phi_{x + \hat 0}).
\label{eq:pq_cumulant}
\end{equation}
Deep in the disordered phase, the terms in $J$ are nearly uncorrelated, $\langle J^2 \rangle \approx \Omega/2$, which reproduces the leading term of Eq.~\eqref{eq:delta_f_sc}.
In the ordered phase, $\sin(\phi_x - \phi_{x + \hat 0}) \approx \phi_x - \phi_{x + \hat 0}$, and the sum along each temporal line is fixed by its winding number, which vanishes in typical configurations.
The temporal current $J$, and with it the sign problem, is then strongly suppressed.
This explains the non-monotonic dependence of the sign problem on $\beta$ found in Sec.~\ref{sec:sign_problem_dependence}.
The results across the full $(\beta, \mu^2)$ grid are discussed in Sec.~\ref{sec:Criteria_for_Correctness}, together with the CLM correctness diagnostics.

\subsection{World-line formulation and worm algorithm}
\label{sec:worldline_worm}

The world-line (flux) representation is an exact reformulation of the partition function.
At real $\mu$, it is free of the sign problem, and we use it as a benchmark for the complex Langevin results.
Starting from the action~\eqref{eq:action}, the Boltzmann weight on each link is expanded using the identity
\begin{equation} 
e^{\beta \cos(\phi_x - \phi_{x + \hat\nu} - i \mu\delta_{\nu, 0})} = \sum_{k = -\infty}^{\infty} I_k(\beta) \, e^{\mu k \, \delta_{\nu,0}} \, e^{i k (\phi_x - \phi_{x + \hat\nu})}.  
\label{eq:bessel_identity} 
\end{equation} 
In the above, $I_k(\beta)$ is the modified Bessel function of the first kind.
Integrating over the angular variables $\phi_x$ at every site enforces a local divergence-free constraint, $\sum_\nu (k_{x, \nu} - k_{x - \hat\nu, \nu}) = 0$.
This results in a dual partition function in terms of integer bond variables $k_{x, \nu} \in \mathbb{Z}$:
\begin{equation} 
Z = \sum_{\{k\}, \, \mathrm{div\text{-}free}} \prod_{x, \nu} I_{k_{x, \nu}}(\beta) \, e^{\mu k_{x, 0}}. 
\label{eq:Z_flux} 
\end{equation} 
Here $I_k(\beta) > 0$ for all $k$ and $\beta > 0$, and $e^{\mu k_{x, 0}}$ is strictly positive for real $\mu$.
Every term in the sum is therefore positive: the representation is free of the sign problem~\cite{Banerjee:2010kc}.
The reformulation is exact; the Monte Carlo estimates obtained from it carry statistical errors, which we show in the figures, but no sign-problem or step-size errors.

The action density is obtained from
\beq 
\langle S \rangle = - \beta \, \frac{\partial \ln Z}{\partial \beta}, 
\eeq 
using
\beq 
\partial_\beta \ln I_k(\beta) = \frac{I_{k -1}(\beta) + I_{k + 1}(\beta)}{2 I_k(\beta)}, 
\eeq
we have
\begin{equation} 
\frac{\langle S \rangle}{\Omega} = - \frac{\beta}{\Omega} \sum_{x, \nu} \left\langle \frac{I_{k_{x, \nu} - 1}(\beta) + I_{k_{x, \nu} + 1}(\beta)}{2 \, I_{k_{x, \nu}}(\beta)} \right\rangle. 
\label{eq:act_wv} 
\end{equation} 

We sample Eq.~\eqref{eq:Z_flux} with the Prokofev--Svistunov worm algorithm~\cite{Prokofev:2001ddj, Banerjee:2010kc}.
The algorithm moves a pair of defects (head and tail) through the lattice.
At the two defect sites, the divergence-free constraint is temporarily violated.
At each step, the head proposes moving to a random neighbor in the direction $\hat\nu$ with sign $s = \pm 1$.
This changes the bond $k_{x, \nu}$ by $\pm 1$.
The move is accepted with Metropolis probability $\min(1, r)$, where
\begin{equation} 
r = \frac{I_{k \pm 1}(\beta)}{I_k(\beta)} \times \begin{cases} e^{\pm\mu} & (\text{temporal bond}), \\ 1 & (\text{spatial bond}). \end{cases} 
\end{equation} 
When the head returns to the tail site, the worm closes, and observables are measured on the resulting valid configuration.
The modified Bessel functions $I_k(\beta)$ are tabulated via their series expansion for $|k| \leq 25$. The typical temporal flux is of order $\beta e^{\mu}/2$, which is $\mathcal{O}(1)$ both for the benchmark points ($\beta \leq 1$, $\mu^2 \leq 0.2$) and along the phase boundary $\beta_c(\mu^2)$ of Sec.~\ref{sec:phase_boundary_dependence}, so the truncation has no effect.
The worm algorithm results are compared and validated in Sec.~\ref{sec:sim_details_results}, where the complex Langevin simulation details are also given.

\subsection{Simulation details and action density results}
\label{sec:sim_details_results}

\begin{figure*}[htbp] 
\centering 
\includegraphics[width=0.485\columnwidth]{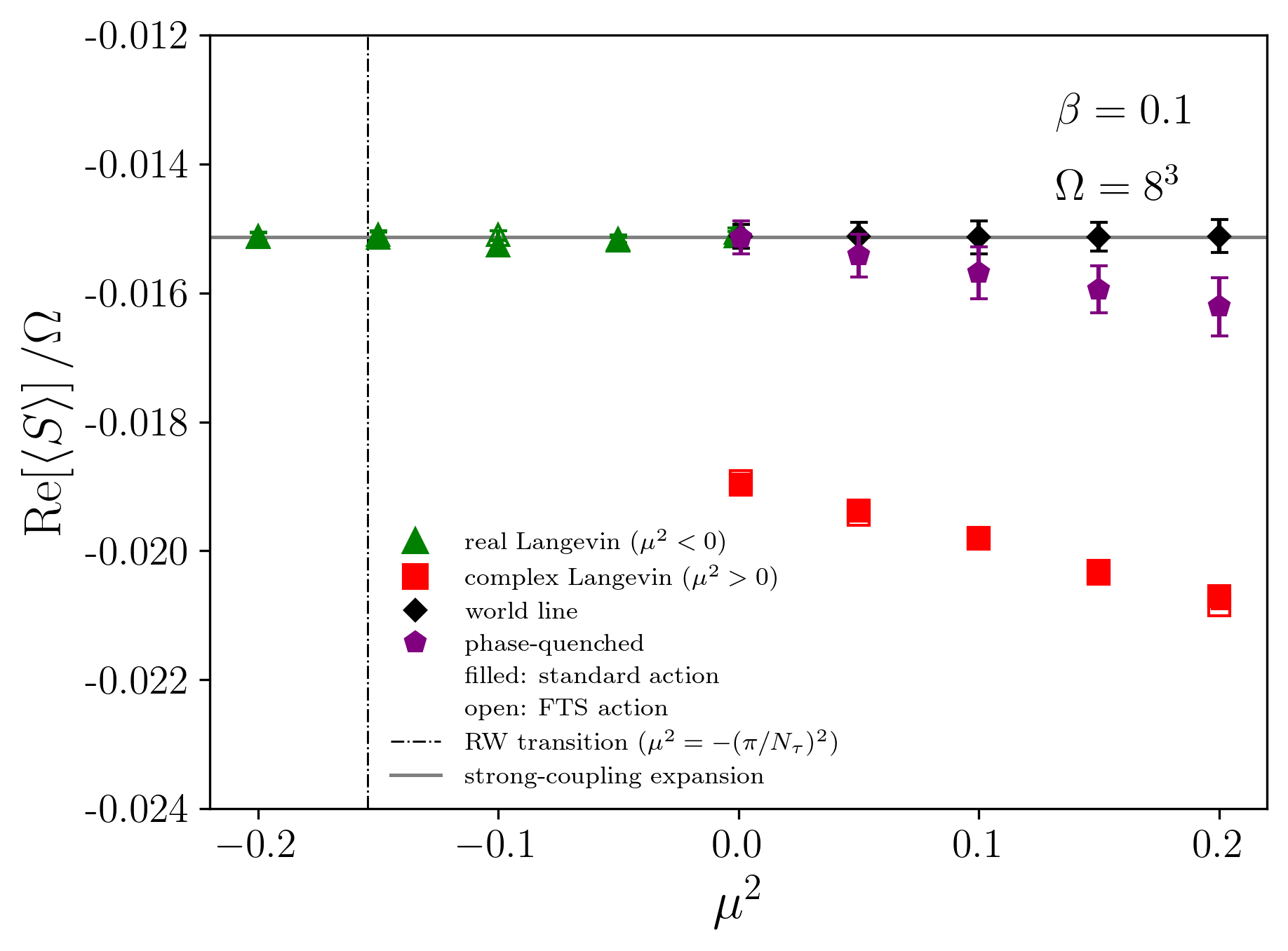} 
\includegraphics[width=0.485\columnwidth]{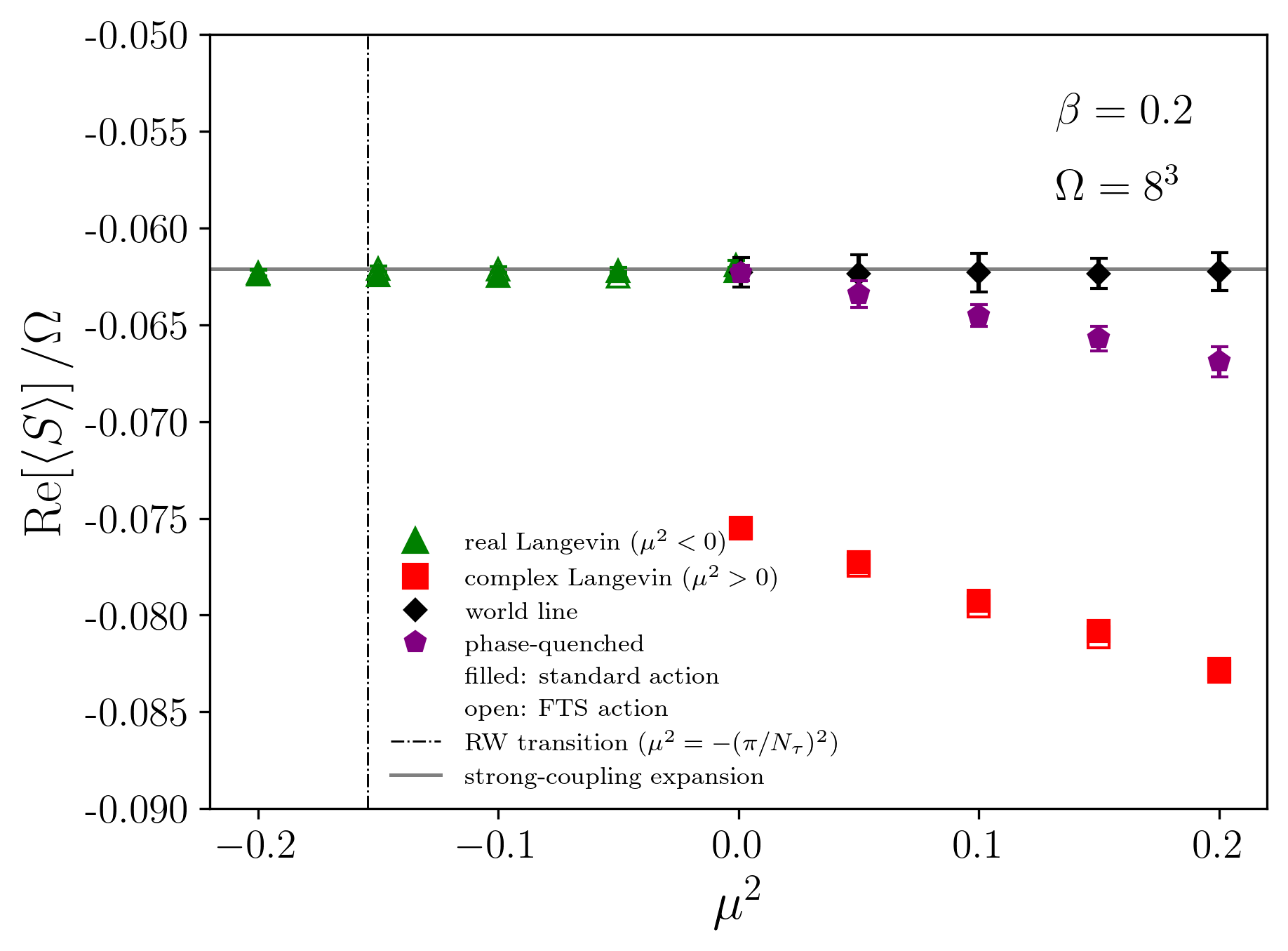} 

\includegraphics[width=0.485\columnwidth]{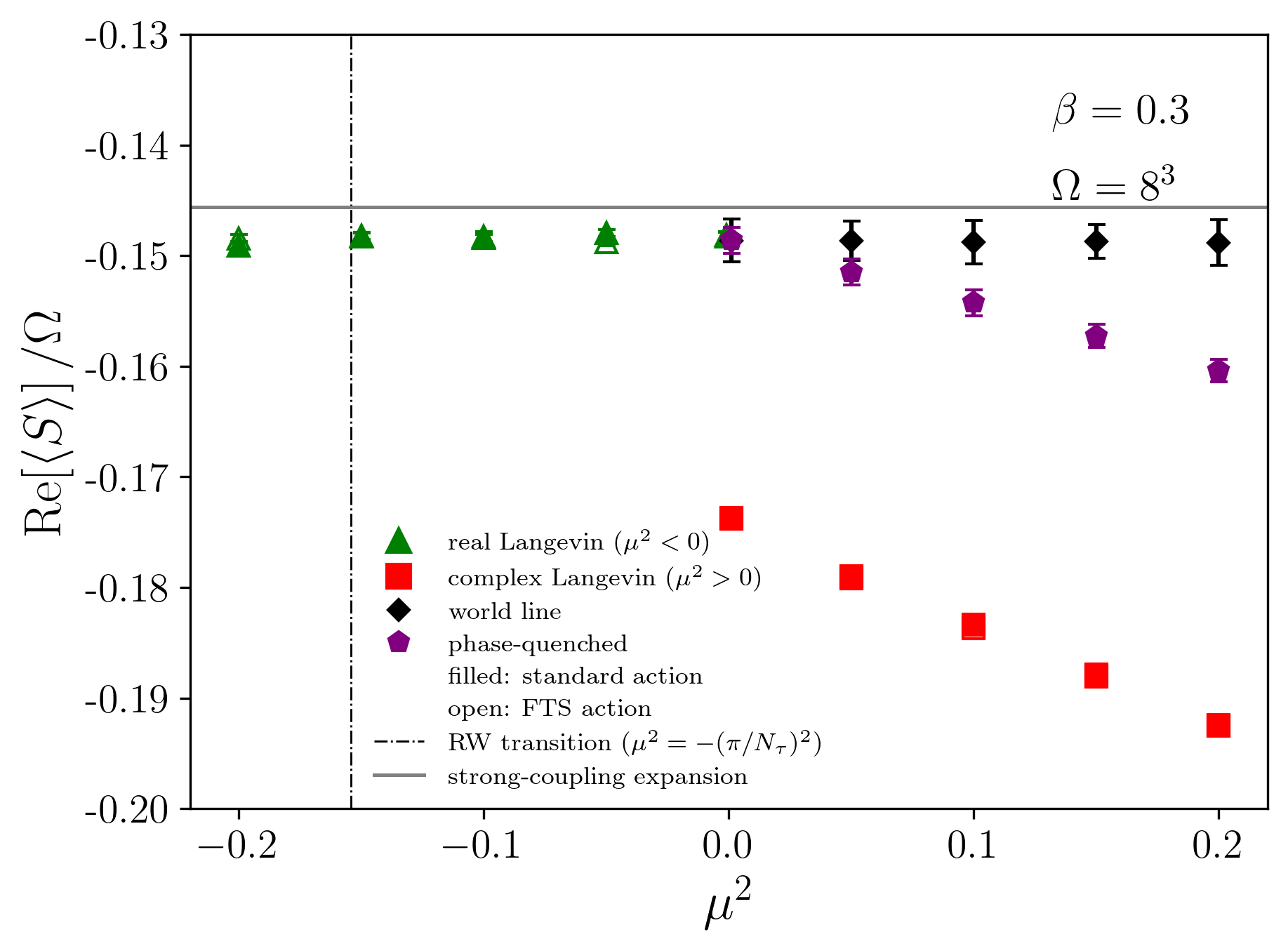} 
\includegraphics[width=0.485\columnwidth]{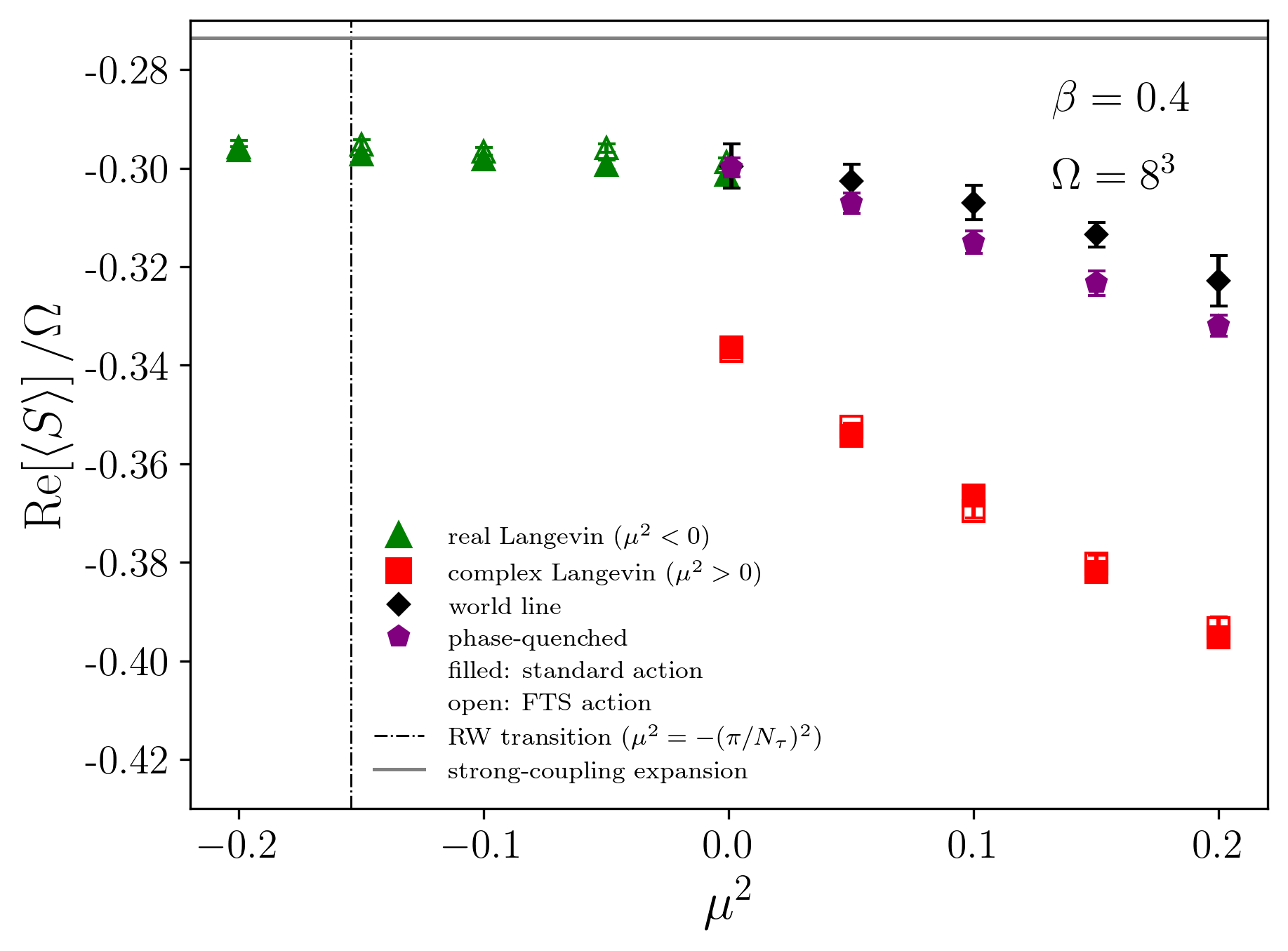} 

\includegraphics[width=0.485\columnwidth]{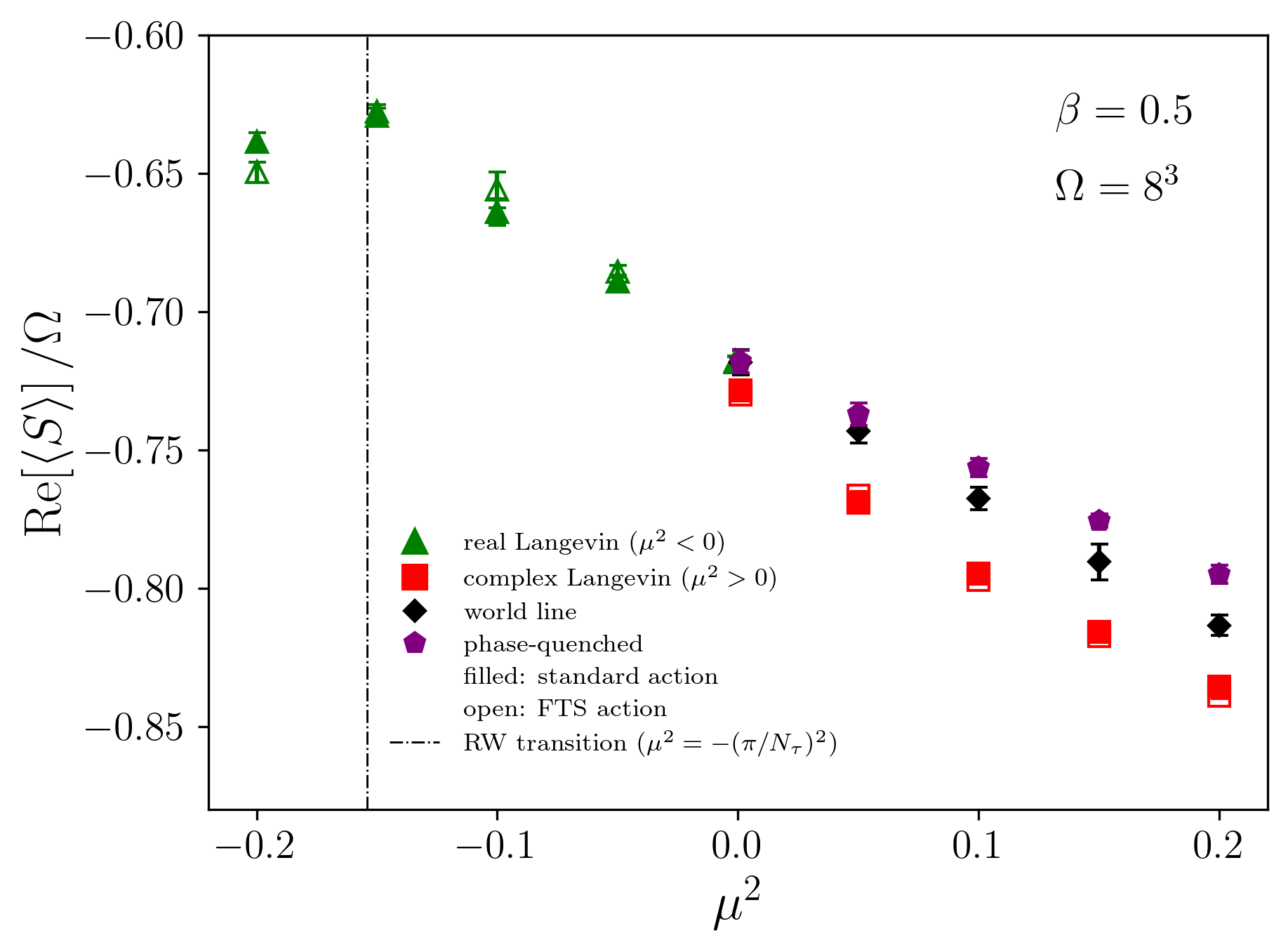} 
\includegraphics[width=0.485\columnwidth]{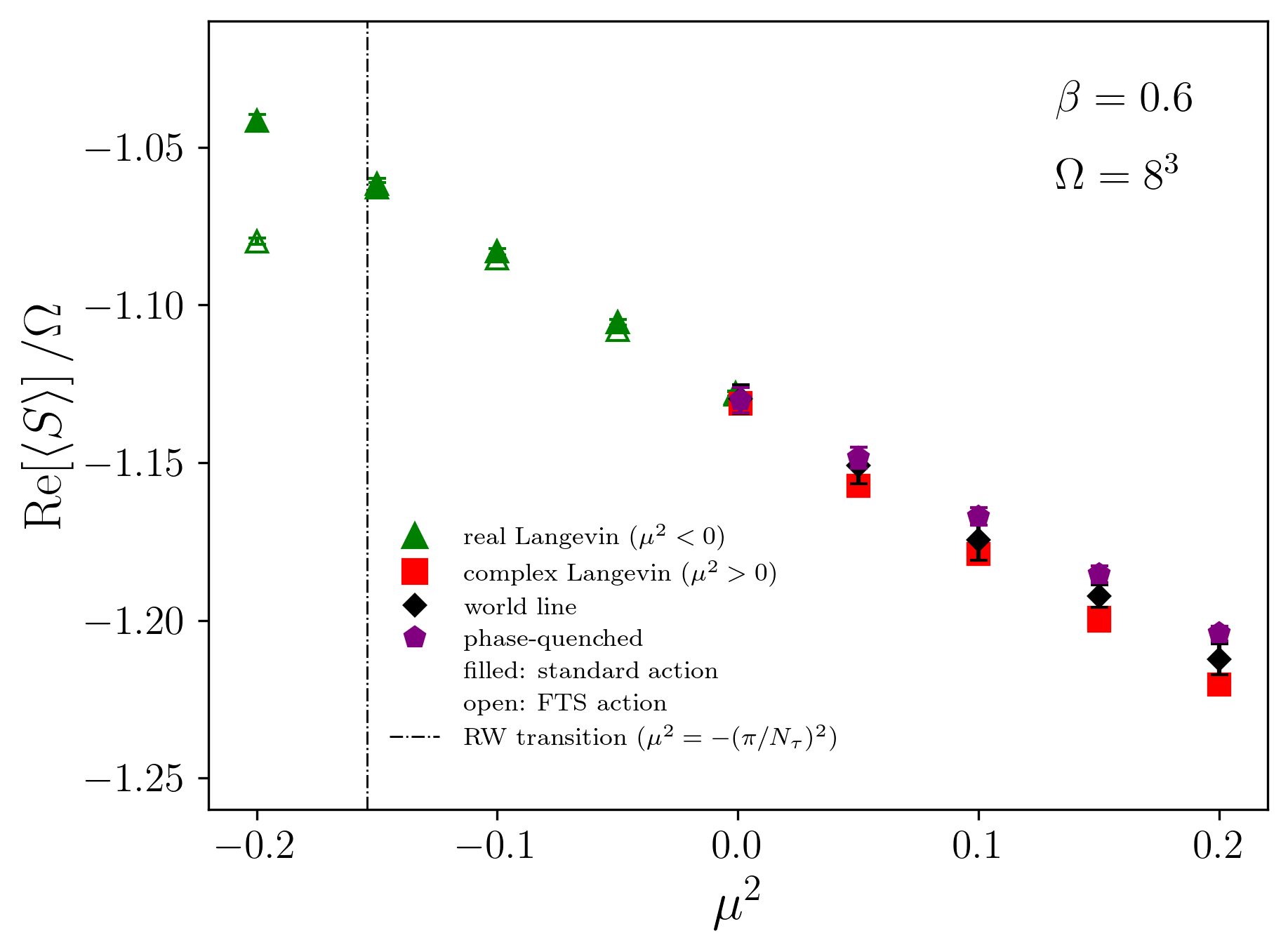} 

\caption {Real part of the action density ${\rm Re}[\langle S \rangle] / \Omega$ as a function of $\mu^2$, for $\beta = 0.1$ -- $0.6$ (disordered and near-transition regimes), on an $8^3$ lattice. 
Green triangles: real Langevin dynamics at imaginary chemical potential ($\mu^2 < 0$); red squares: complex Langevin dynamics at real chemical potential ($\mu^2 > 0$); filled and open symbols denote the standard and final-time-slice (FTS) action formulations, respectively. 
Filled black diamonds ($\mu^2 > 0$ only): world-line (worm-algorithm) benchmark. 
Filled purple pentagons ($\mu^2 > 0$ only): phase-quenched action density. 
The solid horizontal line represents the strong-coupling expansion predictions for $\beta \leq 0.4$, and the dashed vertical line marks the Roberge--Weiss transition of the FTS action, $\mu^2 = -(\pi/N_\tau)^2$.} 
\label{fig:act_den_plots_a} 
\end{figure*} 

\begin{figure*}[htbp] 
\centering 

\includegraphics[width=0.49\columnwidth]{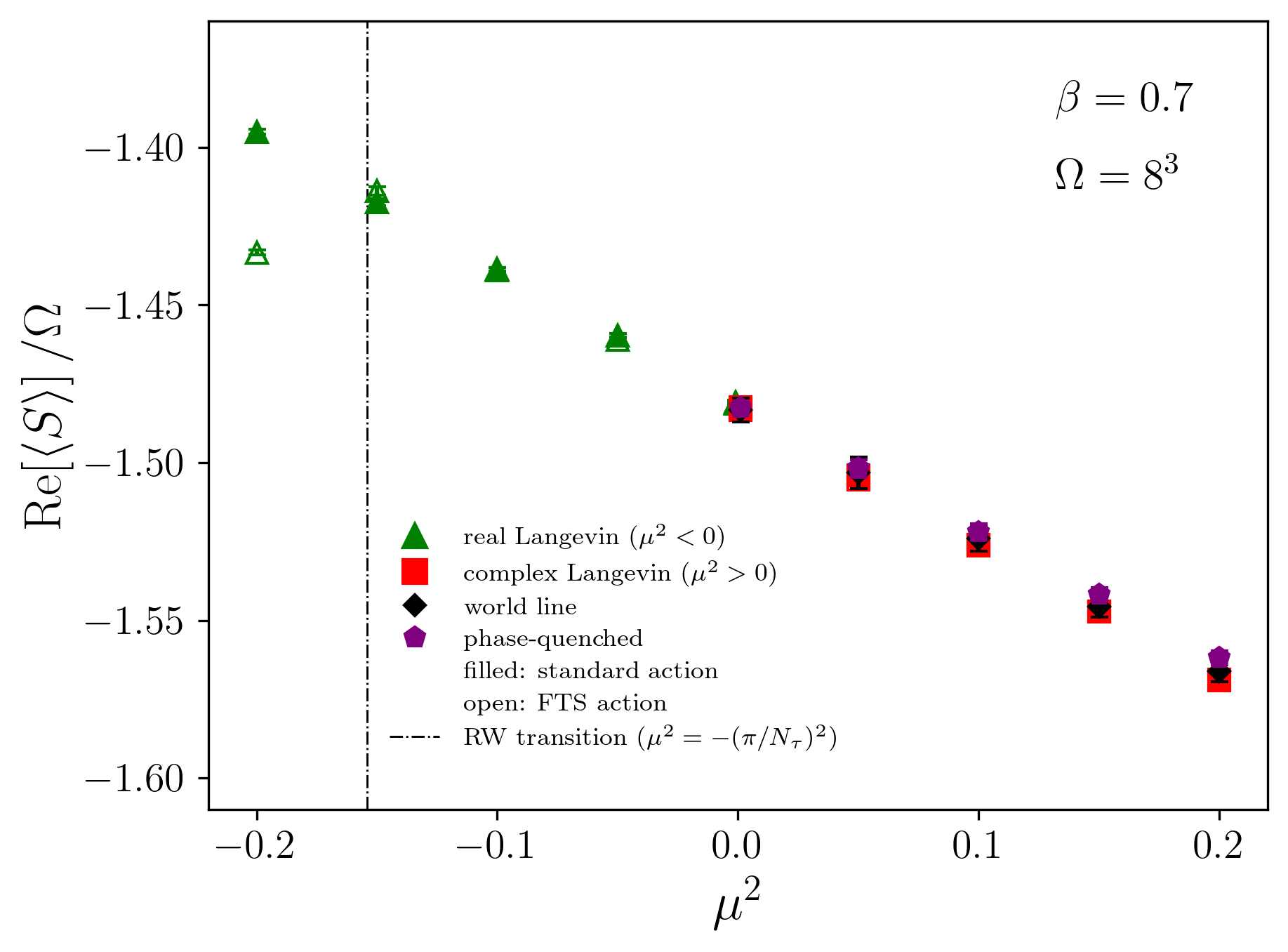} 
\includegraphics[width=0.49\columnwidth]{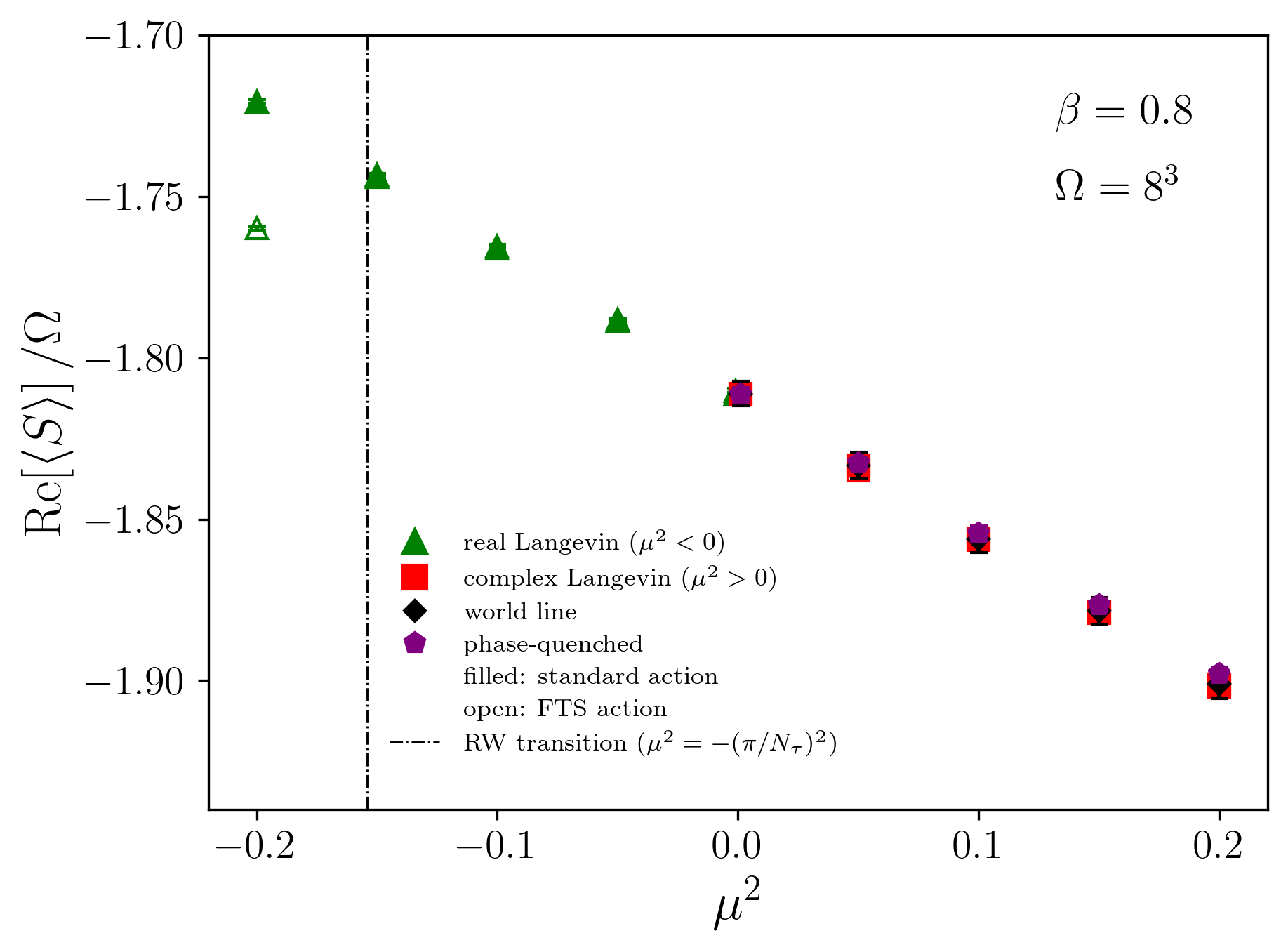} 

\includegraphics[width=0.49\columnwidth]{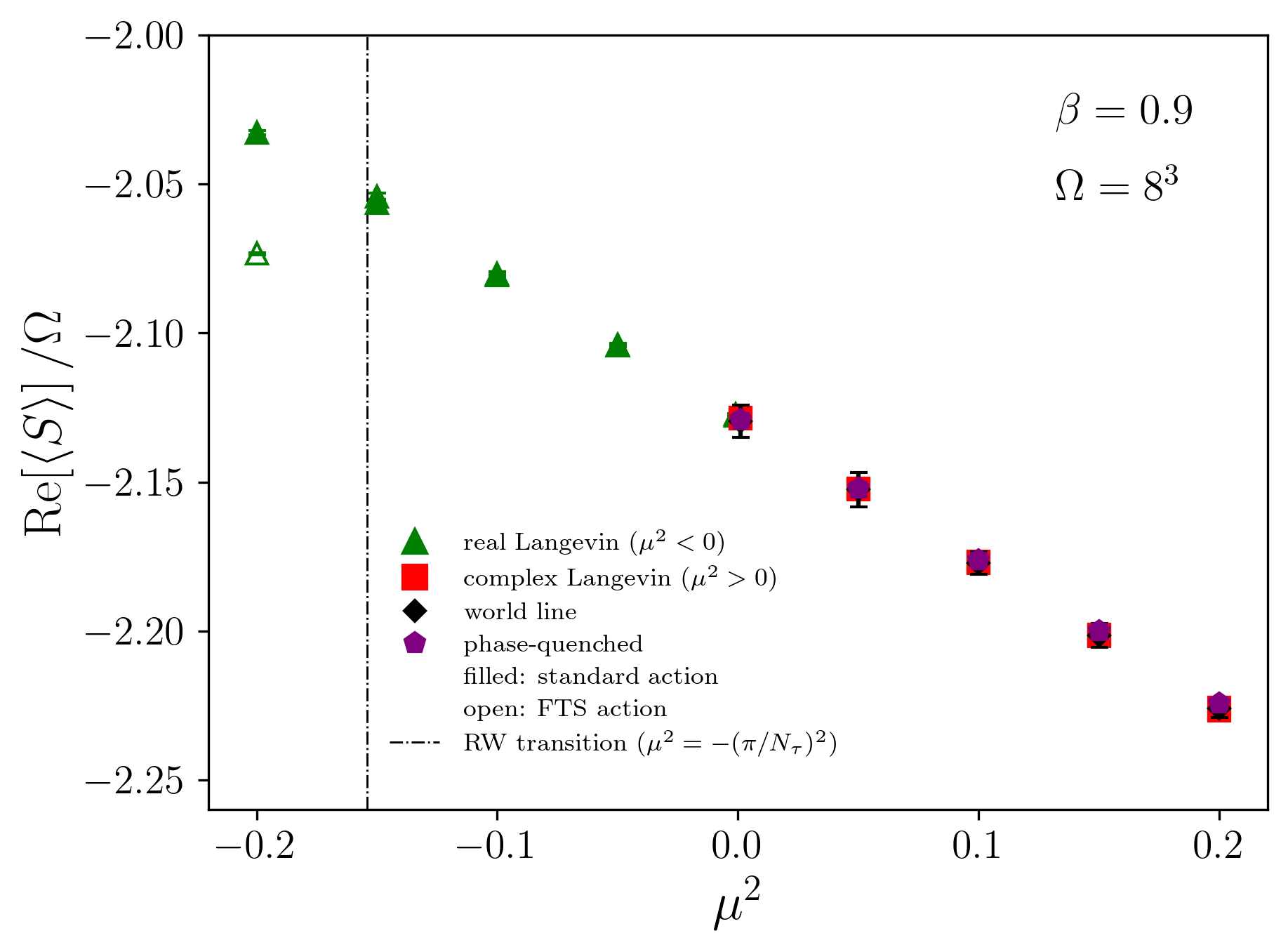} 
\includegraphics[width=0.49\columnwidth]{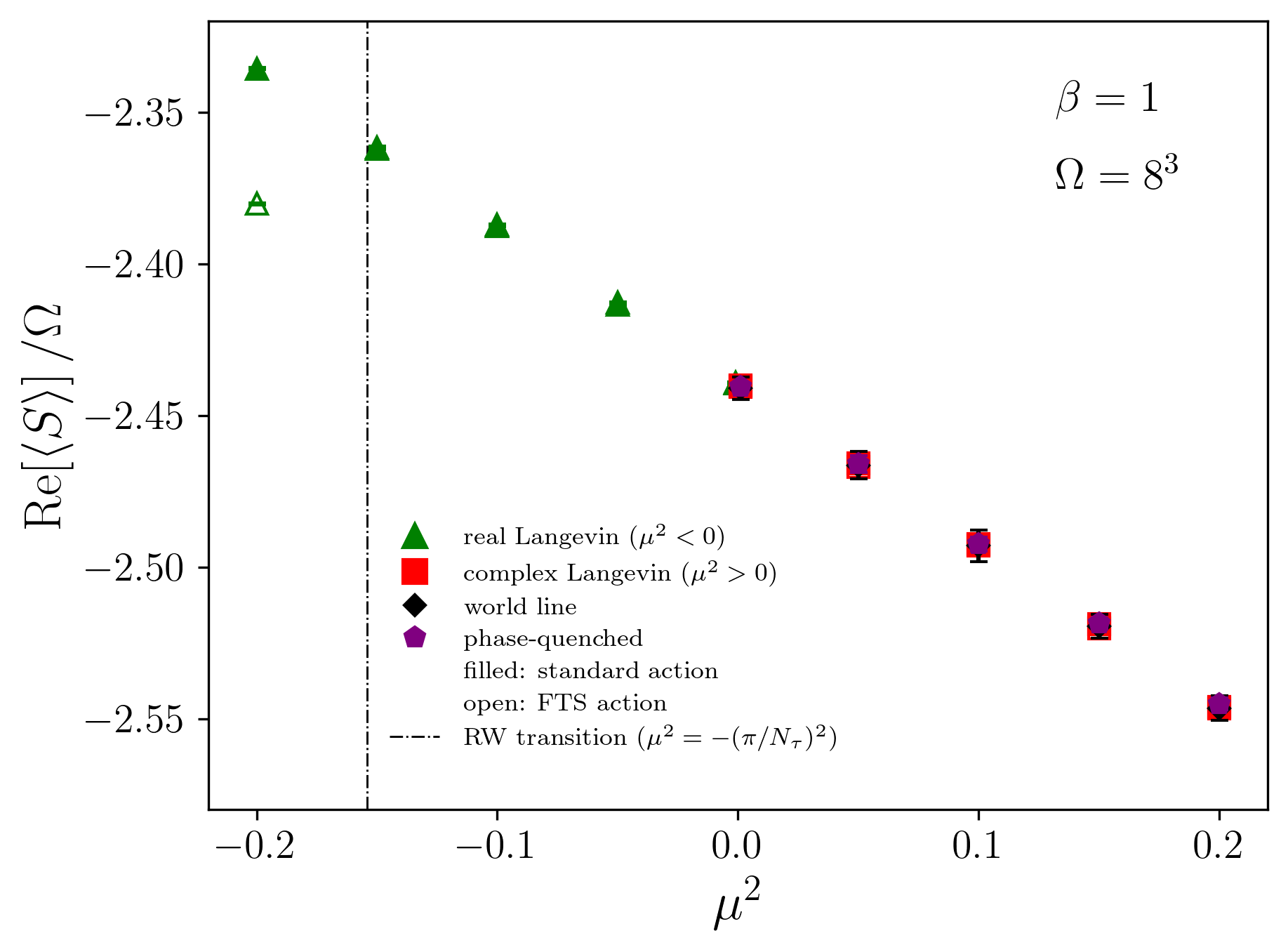} 

\caption {Same as Fig.~\ref{fig:act_den_plots_a}, but for $\beta = 0.7$ -- $1.0$ (ordered phase).} 
\label{fig:act_den_plots_b} 
\end{figure*} 

Simulations were performed on a cubic lattice\footnote{An earlier study of complex Langevin dynamics for the three-dimensional XY model reports that the region of failure does not depend on the lattice volume \cite{Aarts:2010aq}.} with $N_\tau = 8$ and $\Omega = 8^3$.
We integrate the discretized complex Langevin equations, Eqs. \eqref{eq:eqphic_disc}, from cold starts, with several random seeds per $(\beta, \mu^2)$ point.
We take the target step size of the adaptive scheme to be $\bar{\epsilon} \leq 5 \times 10^{-4}$. 
The runs extend to Langevin times up to $2500$, giving us about $10^4$ to $10^5$ measured configurations per parameter point.
The results are then averaged over the runs.
Errors are computed through jackknife binning. 
Jackknife errors assume a finite variance. Where the complex Langevin distribution has extended tails, in particular in the disordered phase (Sec.~\ref{sec:lattice_estimator}), statements in units of the statistical error are therefore indicative.

Euler integration introduces a bias of $\mathcal{O}(\bar\epsilon)$ in the stationary distribution, and we have not performed an extrapolation to $\bar\epsilon \to 0$.
Two observations constrain the bias.
First, the real-Langevin runs at $\mu^2 < 0$ use the same step sizes and reproduce the target value of the configurational estimator to within about $1\%$ (Sec.~\ref{sec:estimator_results}).
This bounds the step-size bias of the estimator at that level, but not that of the action density at the sub-percent level.
Second, for $\beta \geq 0.7$ the complex Langevin action density agrees with the world-line results within $0.3\%$, so its step-size bias there is below this level, unless it is compensated by a bias of complex Langevin itself.
Where we quote agreement or deviations at the sub-percent or few-percent level, in particular at $\beta = 0.5$, a residual step-size effect cannot be excluded. 
The run parameters and the data are available in Ref. \cite{Joseph:2026data}.
For the phase-quenched ensemble of Sec. \ref{sec:sign_problem_severity}, we use standard Metropolis with $2 \times 10^6$ measurement sweeps after $2 \times 10^5$ thermalization sweeps.
For the world-line benchmark of Sec. \ref{sec:worldline_worm}, we use $10^7$ worm sweeps per run with $10^6$ thermalization sweeps, and record observables every 200th worm closure.
Our results for the world-line simulations are consistent with those in Refs.~\cite{Aarts:2010aq, Banerjee:2010kc}.

Figs. \ref{fig:act_den_plots_a} and \ref{fig:act_den_plots_b} show ${\rm Re}[\langle S \rangle] / \Omega$ as a function of $\mu^2$ on an $\Omega = 8^3$ lattice.
Fig. \ref{fig:act_den_plots_a} covers $\beta = 0.1$ -- $0.6$, which spans the disordered and near-transition regimes.
Fig. \ref{fig:act_den_plots_b} covers $\beta = 0.7$ -- $1.0$, the ordered phase.
The standard (filled symbols) and final-time-slice (open symbols) action formulations agree well with each other for $\mu^2 \geq -(\pi/N_\tau)^2$.
Beyond the Roberge--Weiss line, at $\mu^2 = -0.2$, they differ visibly for $\beta \geq 0.5$.
Since the two formulations are exactly equivalent (Sec.~\ref{sec:3D_XY_Model}), this difference is a sampling effect.
The FTS results respect the RW periodicity, while the standard formulation remains on the metastable continuation of the branch with $\mu_{\rm I} < \pi / N_\tau$.
At real $\mu$, the field redefinition is an imaginary shift, and the complex Langevin processes for the two formulations differ only in their initial conditions; their agreement at $\mu^2 > 0$ shows that our results are the same for these two cold starts, which differ by an imaginary shift of the fields. It does not test the dependence on initial conditions in general.
The distinction between the two matters most at $\mu^2 \leq 0$, where only the FTS formulation manifestly captures the Roberge--Weiss periodicity.
For the disordered-phase couplings $\beta \leq 0.4$, the strong-coupling expansion prediction \eqref{eq:act_density} is shown as a solid horizontal line.
At imaginary chemical potential, the real Langevin results agree very well with these predictions up to $\beta = 0.3$, with deviations of at most $2.5\%$.

For $\mu^2 \leq 0$ (imaginary chemical potential, green triangles), the action density varies smoothly with $\mu^2$, as expected from analyticity.
The FTS formulation correctly captures the Roberge--Weiss transition at $\mu^2 = - (\pi / N_\tau)^2$ (dashed vertical line in each panel).
Near $\mu^2 = 0$, each panel shows two data points.
One is obtained from real Langevin dynamics (at $\mu^2 = - 0.001$) and the other from complex Langevin dynamics (at $\mu^2 = 0.001$).
Together they provide a direct continuity test.
For $\beta \geq 0.7$, deep in the ordered phase ($\beta_c \approx 0.454$), the two points are consistent with each other.
This confirms a smooth, analytic crossing. 
At $\beta = 0.6$ and $0.5$, just above $\beta_c$, the complex Langevin action density at $\mu^2 = 0.001$
differs from the real Langevin action density at $\mu^2 = - 0.001$ by $\sim 0.3\%$ and $\sim 1.5\%$, respectively. 
This is significantly larger than the change of the action density between these two values of $\mu^2$, which, estimated from the world-line data at small $\mu^2 > 0$ using that $\langle S \rangle$ is analytic in $\mu^2$, is only $\sim 0.08\%$ at $\beta = 0.6$ and $\sim 0.14\%$ at $\beta = 0.5$. 
Complex Langevin is therefore already slightly off at a negligible chemical potential.
At smaller $\beta$, $\beta < \beta_c$ (in the disordered phase), the two points again disagree visibly.
This signals that complex Langevin fails to reproduce the correct analytic continuation. 
This was also noted in Ref. \cite{Aarts:2010aq}.

For $\mu^2 > 0$, the worm algorithm \cite{Banerjee:2010kc} (filled black diamonds) provides sign-problem-free reference values, exact up to statistical errors.
In the ordered phase ($\beta \geq 0.5$), the complex Langevin results are close to the world-line data across the full range of $\mu^2$, with deviations $\sim 1\%$ at $\beta = 0.6$ and $\sim 4\%$ at $\beta = 0.5$.
For $\beta \geq 0.7$, the complex Langevin action density agrees with the world-line values within statistical errors at all $\mu^2$, with deviations of at most $0.3\%$.
In the disordered phase, $\beta < \beta_c$, the agreement deteriorates sharply.
Notably, the world-line data show a nearly flat, $\mu^2$-independent action density at $\beta = 0.1$ -- $0.3$.
In the disordered phase, the system does not respond to a chemical potential below the threshold, as expected on general grounds for a gapped phase and as found in the strong-coupling expansion (Sec.~\ref{sec:strong_coupling}).
Complex Langevin, by contrast, produces a spurious $\mu^2$ dependence and a substantially more negative action density in this regime.
At $\beta = 0.3$, the deviation at $\mu^2 = 0.2$ is about $30\%$ relative to the world-line benchmark, for both the standard and the FTS actions.
At $\beta = 0.5$, which lies near the phase boundary, the discrepancy is reduced to $\sim 3$ -- $4\%$.
Also shown for $\mu^2 > 0$ is the phase-quenched action density (filled purple pentagons).
It is computed in the same phase-quenched ensemble that we use to quantify the severity of the sign problem in Sec. \ref{sec:sign_problem_severity}.
The phase-quenched weight discards the re-weighting phase $e^{i\varphi}$, which the full theory retains.
The phase-quenched result therefore deviates more and more from the world-line result as $\mu^2$ increases.
In the next section, we quantitatively assess the reliability of the complex Langevin results.
We use the configurational temperature estimator and the drift-decay criterion as complementary diagnostics of correctness.

\section{Criteria for Correctness}
\label{sec:Criteria_for_Correctness}

We assess the reliability of CLM for the three-dimensional XY model with two complementary criteria for correctness.
The first is based on the configurational temperature estimator \cite{Joseph:2025fcd}.
The second relies on the behavior of the magnitude of the drift terms.

Sec.~\ref{sec:thermometer} introduces the configurational temperature estimator and the identity that fixes its target value.
Sec.~\ref{sec:estimator_results} presents its results across the $(\beta, \mu^2)$ plane and relates them to the sign problem and to the phase boundary.
Secs.~\ref{sec:Decay_Drift} and~\ref{sec:compare_correctness} study the decay of the drift terms and compare the two criteria.
In Sec.~\ref{sec:ds} we explore correctness with dynamical stabilization.

\subsection{Configurational temperature estimator}
\label{sec:thermometer}

The configurational temperature estimator was first introduced in classical statistical mechanics, where it has been studied extensively \cite{PhysRevLett.78.772, 1998JPhA...31.7761R, PhysRevE.62.4757, 10.1063/1.480995, PhysRevE.94.062113, 10.1063/1.1348024}. 
In the microcanonical ensemble, the inverse temperature has a geometric expression.
It is the phase-space average of the divergence of the normalized potential gradient \cite{PhysRevLett.78.772, 1998JPhA...31.7761R}.
Using the equivalence of ensembles in the thermodynamic limit, this construction can be carried over to the canonical ensemble.
The result holds up to $\mathcal{O}(1/N)$ averaging corrections, where $N$ is the number of degrees of freedom \cite{10.1063/1.477301, PhysRevE.62.4757}.
The configurational temperature has the form
\begin{equation} 
\langle \beta_{\rm config} \rangle = \left \langle \frac{1}{k_B T_{\rm config}} \right \rangle = \left\langle \nabla \cdot \frac{\nabla \Phi}{\nabla \Phi \cdot \nabla \Phi} \right\rangle + \mathcal{O}(1/N). 
\label{eq:rugh_microcanonical} 
\end{equation} 

Using a vector identity, we can write this as
\begin{equation}
\langle \beta_{\rm config} \rangle = \left\langle \frac{{\rm Tr} \, H}{\vec{g}^{\,T} \vec{g}} - 2 \, \frac{\vec{g}^{\,T} H \, \vec{g}}{(\vec{g}^{\,T} \vec{g})^2} \right\rangle + \mathcal{O}(1/N). 
\label{eq:beta_Hessian_Phi} 
\end{equation} 
Here $\Phi$ denotes the classical potential energy, $\vec{g} \equiv \nabla\Phi$ is the gradient vector, $H \equiv \nabla\nabla^{T}\Phi$ is the Hessian matrix, and $\langle \cdot \rangle$ denotes the ensemble average. 
The trace of the Hessian is the dominant term and captures the isotropic curvature.
The directional term $\vec{g}^{\,T} H\, \vec{g}$ is sub-leading and captures the curvature along the gradient \cite{PhysRevLett.78.772}.
An analogous estimator has been studied for classical spin systems \cite{PhysRevE.94.062113}.

It is possible to extend the configurational temperature estimator to quantum field theories formulated via Euclidean path integrals \cite{Dhindsa:2025xfv, Joseph:2025fcd}.
The extension is based on the formal analogy between classical statistical mechanics and Euclidean quantum field theory.
In the path integral, the weight $e^{- S[\phi]}$ has the same mathematical structure as a canonical ensemble at unit inverse temperature, with the action $S[\phi]$ in the role of the potential $\Phi$.
This motivates introducing by hand an overall factor $\beta_{\rm config}$ in front of the action, that is, $P[\phi] \propto e^{- \beta_{\rm config} S[\phi]}$.
This factor is distinct from the coupling $\beta$ inside $S$.
For correct sampling, its target value is, by analogy, $\beta_{\rm config} = 1$.

\subsubsection{Target value from a Schwinger--Dyson identity}
\label{sec:target_derivations}

The microcanonical construction above gives the configurational temperature in the canonical ensemble only up to $\mathcal{O}(1/N)$ corrections.
Its generalization to complex actions is non-trivial.
For a lattice field theory with action $S[\phi]$, an exact statement follows instead from a Schwinger--Dyson identity~\cite{Joseph:2026hom}. Reference~\cite{Joseph:2026hom} formulates this for positive weights; we give the derivation here and extend it to the complex weight in Sec.~\ref{sec:lattice_estimator}.
Consider any vector field $F_x[\phi]$ on the space of field configurations.
The invariance of the partition function under
\beq 
\phi_x \to \phi_x + \epsilon F_x[\phi], 
\eeq 
or equivalently, performing integration by parts, gives
\beq
\begin{aligned} 
\left\langle \sum_x \frac{\partial F_x}{\partial \phi_x} \right\rangle &= \frac{1}{Z} \int d^N\phi \, \sum_x \frac{\partial F_x}{\partial \phi_x} \, e^{-S} \\
&= \frac{1}{Z} \int d^N\phi \, \sum_x \left[ \frac{\partial}{\partial \phi_x} \big( F_x \, e^{-S} \big) + F_x \, \frac{\partial S}{\partial \phi_x} \, e^{- S} \right] \\ 
&= \left\langle \sum_x F_x \, \frac{\partial S}{\partial \phi_x} \right\rangle.  
\end{aligned} 
\label{eq:sd_master} 
\eeq
In the above, the total derivative integrates to zero if the boundary terms vanish.
For the XY model, the fields are angles, so a periodic, non-singular $F_x$ produces no boundary terms.

Two choices of $F$ are relevant here.
\begin{itemize}
\item[\textit{(i)}] Consider
\beq 
F_x = K_x \equiv \frac{\partial S}{\partial\phi_x}.  
\eeq 
(Here and in the rest of this section $K_x$ denotes the gradient $\partial S / \partial \phi_x$, i.e., minus the Langevin drift of Secs.~\ref{sec:Complex_Langevin_Dynamics} and \ref{sec:3D_XY_Model}. The estimator is invariant under $K_x \to -K_x$.) 
One obtains the Schwinger--Dyson equation
\begin{equation} 
\left\langle \sum_x H_{xx} \right\rangle = \left\langle \sum_x K_x^2 \right\rangle, \qquad H_{xy} \equiv \frac{\partial^2 S}{\partial\phi_x \partial\phi_y}. 
\label{eq:ds_trH} 
\end{equation} 
Each link term of the XY action depends on two sites.
It contributes $\beta\cos(\cdot)$ to the diagonal Hessian entry of each of them.
Hence $\sum_x H_{xx} = - 2 S$, and Eq.~\eqref{eq:ds_trH} reads $\langle \sum_x K_x^2 \rangle = - 2 \langle S \rangle$.
Since $\langle S \rangle$ is known from the world-line formulation without a sign problem, this also fixes $\langle \sum_x K_x^2 \rangle$, which provides a pole-free check against the benchmark; being quadratic in the drift, it is, however, sensitive to large drifts.
In the language of Ref.~\cite{Aarts:2011ax}, Eq.~\eqref{eq:ds_trH} is the consistency condition $\langle L_c S \rangle = 0$, with $L_c O = \sum_x \big( \partial^2 O / \partial \phi_x^2 - (\partial S / \partial \phi_x) (\partial O / \partial \phi_x) \big)$.
Expectation values of $L_c O$ in the Langevin ensemble are closely related to the boundary terms studied in Refs.~\cite{Scherzer:2018hid, Scherzer:2019lrh}.

\item[\textit{(ii)}] Consider
\beq 
F_x = \frac{K_x}{\sum_y K_y^2}. 
\eeq 
This choice defines the configurational temperature. 
One has
\beq 
\sum_x F_x K_x = 1, 
\eeq 
and Eq.~\eqref{eq:sd_master} becomes
\begin{align}
\left\langle \nabla \cdot \frac{\nabla S}{\nabla S \cdot \nabla S} \right\rangle &= \left\langle \sum_x \frac{\partial}{\partial \phi_x} \frac{K_x}{\sum_y K_y^2} \right\rangle \nonumber \\ 
&= \left\langle \sum_x \left[ \frac{1}{\sum_y K_y^2} \frac{\partial K_x}{\partial \phi_x} - \frac{K_x}{\big(\sum_y K_y^2\big)^2} \frac{\partial}{\partial \phi_x} \sum_y K_y^2 \right] \right\rangle. 
\end{align}

That is,
\begin{align}
\left\langle \beta_{\rm config} \right\rangle &= \left\langle \frac{\sum_x H_{xx}}{\sum_x K_x^2} - \frac{2 \sum_{x,y} K_x H_{xy} K_y}{\left(\sum_x K_x^2\right)^2} \right\rangle = 1, 
\label{eq:beta_full_identity} 
\end{align} 
Eq.~\eqref{eq:beta_full_identity} is a particular member of the Schwinger--Dyson hierarchy \cite{Joseph:2026hom}.
Unlike Eq.~\eqref{eq:ds_trH}, it is not of the form $\langle L_c O \rangle = 0$, since $F_x$ is not the gradient of a function.
The normalization by $\sum_y K_y^2$ makes the estimator dimensionless, with a universal target value, and suppresses the contribution of configurations with large drift.
The price is a pole where $\sum_x K_x^2$ vanishes, which we discuss in Sec.~\ref{sec:lattice_estimator}.
In contrast to the microcanonical route, it holds at any lattice size.
$\mathcal{O}(1/N)$ corrections appear only if the second, directional term is dropped (Sec.~\ref{sec:lattice_estimator}).
A deviation of $\langle \beta_{\rm config} \rangle$ from unity therefore shows that the sampled distribution is not the correct one \cite{Dhindsa:2025xfv, Joseph:2025fcd}.
\end{itemize}

In Euclidean quantum field theory, $\beta_{\rm config}$ is a fictitious inverse temperature that multiplies the entire action.
It does not correspond to a physical thermodynamic temperature in the usual sense.
The configurational temperature serves as a {\it sampling consistency diagnostic}; it tests whether the configurations generated by a given algorithm satisfy an identity involving the weight $e^{-S[\phi]}$.

\subsubsection{Lattice estimator and complex actions}
\label{sec:lattice_estimator}

For the three-dimensional XY model action with complexified fields $\phi_x = \phi_x^{\rm R} + i \phi_x^{\rm I}$, the holomorphic Hessian has the form 
\begin{align} 
H_{xx} &= \beta \sum_\nu \Bigl[ 
\cos\bigl(\phi_x - \phi_{x + \hat\nu} - i \mu \, \delta_{\nu, 0} \bigr) + \cos\bigl(\phi_x - \phi_{x - \hat\nu} + i \mu \, \delta_{\nu, 0}\bigr) 
\Bigr], \nonumber \\ 
H_{x, \, x \pm \hat\nu} &= - \beta \, \cos\bigl(\phi_x - \phi_{x \pm \hat\nu} \mp i \mu \, \delta_{\nu, 0} \bigr). 
\label{eq:XY_hessian} 
\end{align} 
In our simulations we evaluate, for each configuration $i$, only the diagonal part of Eq. \eqref{eq:beta_full_identity}. 
We define it as
\begin{equation} 
\beta_{\rm config}(i) \equiv \frac{\sum_x H_{xx}}{\sum_x K_x^2}, 
\label{eq:beta_config_M_def} 
\end{equation} 
and quote its Langevin-time average $\langle \beta_{\rm config} \rangle$.
The diagonal estimator has a simple pole where $\sum_x K_x^2$ vanishes, while the directional term, with $(\sum_x K_x^2)^2$ in the denominator, has a double pole.
For complex Langevin, the full estimator therefore has much heavier tails than the diagonal one (see below).
Dropping the directional term shifts the target value by $\mathcal{O}(1/N)$\footnote{In matrix models, the finite-$N$ corrections attributed to the directional (anisotropic) term were recently found to decrease as a power of the matrix size, roughly as $N^{-0.6}$ \cite{Joseph:2026rwh}.}.
The shift equals the average of the directional term.
On an $8^3$ lattice, Metropolis simulations at $\mu = 0$ give about $1.003$ for the diagonal estimator, i.e., the directional term contributes about $0.3\%$.
With real Langevin at $\mu^2 < 0$, its average stays below $\sim 0.3\%$ at all $\beta$ and depends weakly on $\beta$, as the real-Langevin points in Fig.~\ref{fig:beta_beta_m_plots_b} show.
For complex Langevin, the average of the directional term is meaningful only where the process stays away from the poles of the estimator (see below).
In the ordered phase it is again below $\sim 0.3\%$; in the disordered phase it is dominated by near-pole configurations, and we do not use it.
We therefore show both unity and $1.003$ as reference lines, and we do not interpret deviations of the diagonal estimator smaller than $\sim 0.3\%$.
The imaginary part of $\langle \beta_{\rm config} \rangle$ must vanish, and it is consistent with zero at all parameters studied.

For real $\mu$, the weight $e^{-S}$ of the XY model is complex.
The derivation of Eqs. \eqref{eq:sd_master} -- \eqref{eq:beta_full_identity} is linear in $e^{-S}$ and never uses its positivity.
The identities therefore hold unchanged, with $\langle \cdot \rangle$ now denoting averages with the complex weight $\rho \propto e^{-S}$ (assuming $Z \neq 0$).

Let us note two points before we proceed further. 
First, the denominator of the configurational-temperature identity, Eq. \eqref{eq:beta_full_identity}, is the holomorphic square $\sum_x K_x^2$, which is complex.
The step $\sum_x F_x K_x = 1$ fails if this denominator is replaced by $\sum_x |K_x|^2$.

Second, $F_x = K_x / \sum_y K_y^2$ is singular where $\sum_x K_x^2$ vanishes.
For a real action, this happens only at stationary points of the action, a set of measure zero that does not contribute.
For the complex weight on the real manifold, $\sum_x K_x^2 = 0$ requires $\sum_x ({\rm Re} K_x)^2 = \sum_x ({\rm Im} K_x)^2$ and $\sum_x {\rm Re} K_x \, {\rm Im} K_x = 0$.
This set has real codimension two, and Eq.~\eqref{eq:sd_master} acquires a contribution localized on it.
For real $\mu$, the imaginary part of the drift comes only from the temporal links and carries a factor $\sinh \mu$, ${\rm Im} K_x = \beta \sinh\mu \, [\cos(\phi_x - \phi_{x - \hat 0}) - \cos(\phi_x - \phi_{x + \hat 0})]$.
For $|\mu^2| \leq 0.2$, its typical norm is therefore well below that of the real part.
Since $\sum_x ({\rm Re} K_x)^2$ is a sum over $\Omega = 512$ sites with small relative fluctuations, configurations on the singular set lie far in the tail of the distribution, and we neglect the localized term for $|\mu^2| \leq 0.2$.
At large $\mu^2$, where $\tanh\mu \to 1$, this argument no longer applies, and we regard the map of the estimator at large $\mu^2$ (Sec.~\ref{sec:phase_boundary_dependence}) as qualitative.

Two distinct pole effects should be distinguished.
The first concerns the target value.
It is a property of the complex weight on the real manifold, and is affected only by the localized term just discussed.
The second concerns the complex Langevin estimate.
In the complexified field space, the process can approach the poles, which affects both whether complex Langevin reproduces the identity and the statistical properties of the measured estimator.
That the process approaches the poles in the disordered phase therefore does not change the target value, but it does affect the interpretation and the uncertainty of the measured deviation, as we now discuss.

The quantities $\sum_x H_{xx}$ and $\sum_x K_x^2$ are entire functions of the fields, and the estimator in Eq.~\eqref{eq:beta_full_identity} is meromorphic.
A correctly converging complex Langevin process therefore reproduces the two identities, Eqs.~\eqref{eq:ds_trH} and \eqref{eq:beta_full_identity}.
This holds provided the Langevin distribution carries negligible weight near the poles of the estimator, where $\sum_x K_x^2 = 0$ \cite{Aarts:2017vrv, Seiler:2020mkh}.
In complex Langevin, the fields are complexified, and the process can come close to these poles; we find that this happens in the disordered phase (Sec.~\ref{sec:ds}).
Near the poles, the diagonal estimator behaves as $1/w$, where $w = \sum_x K_x^2$ serves as a local complex coordinate transverse to the singular set.
If the Langevin density is smooth there, the probability of a value of $|\beta_{\rm config}|$ larger than $b$ falls off as $b^{-2}$.
The mean then exists, but the variance diverges logarithmically, so the jackknife errors in the disordered phase should be regarded as indicative.
We therefore quote errors on the estimator in the disordered phase for orientation only, and do not attach a statistical significance to them.
Our conclusions there rest on the systematic pattern of deviations of up to $12\%$, which lie below the target at every point, and on the world-line comparison.
A direct characterization of the distribution of $w$ near zero in the complex Langevin ensemble, and of the sensitivity of the estimator to excluding a neighborhood of the poles, would sharpen this analysis; we have not carried it out.
The directional term behaves as $1/w^2$, and its mean is not defined under the same assumption, which is a further reason to use the diagonal estimator.

Two consequences follow.
First, a deviation of the estimator from its target shows that the Langevin distribution is not correct for this observable.
Either it does not reproduce the complex weight, or it carries non-negligible weight near the poles of the estimator, so that boundary terms arise there \cite{Aarts:2017vrv, Seiler:2020mkh}.
Both signal an unreliable simulation, but in the second case the deviation of the estimator need not match the bias of holomorphic observables such as the action density.
Other observables may also be biased, and the estimator does not tell which ones, since boundary terms depend on the observable \cite{Scherzer:2019lrh}.
In the present model, the world-line benchmark tests the action density directly.
Second, agreement with the target is a necessary but not a sufficient condition for correctness \cite{Mandl:2026vdc}.
In particular, the estimator tests a local identity, and it cannot detect a failure of ergodicity, such as trapping in a metastable sector (Sec.~\ref{sec:sim_details_results}).

\subsection{Results for the configurational temperature}
\label{sec:estimator_results}

\begin{figure*}[htbp] 
\centering 

\includegraphics[width=0.49\columnwidth]{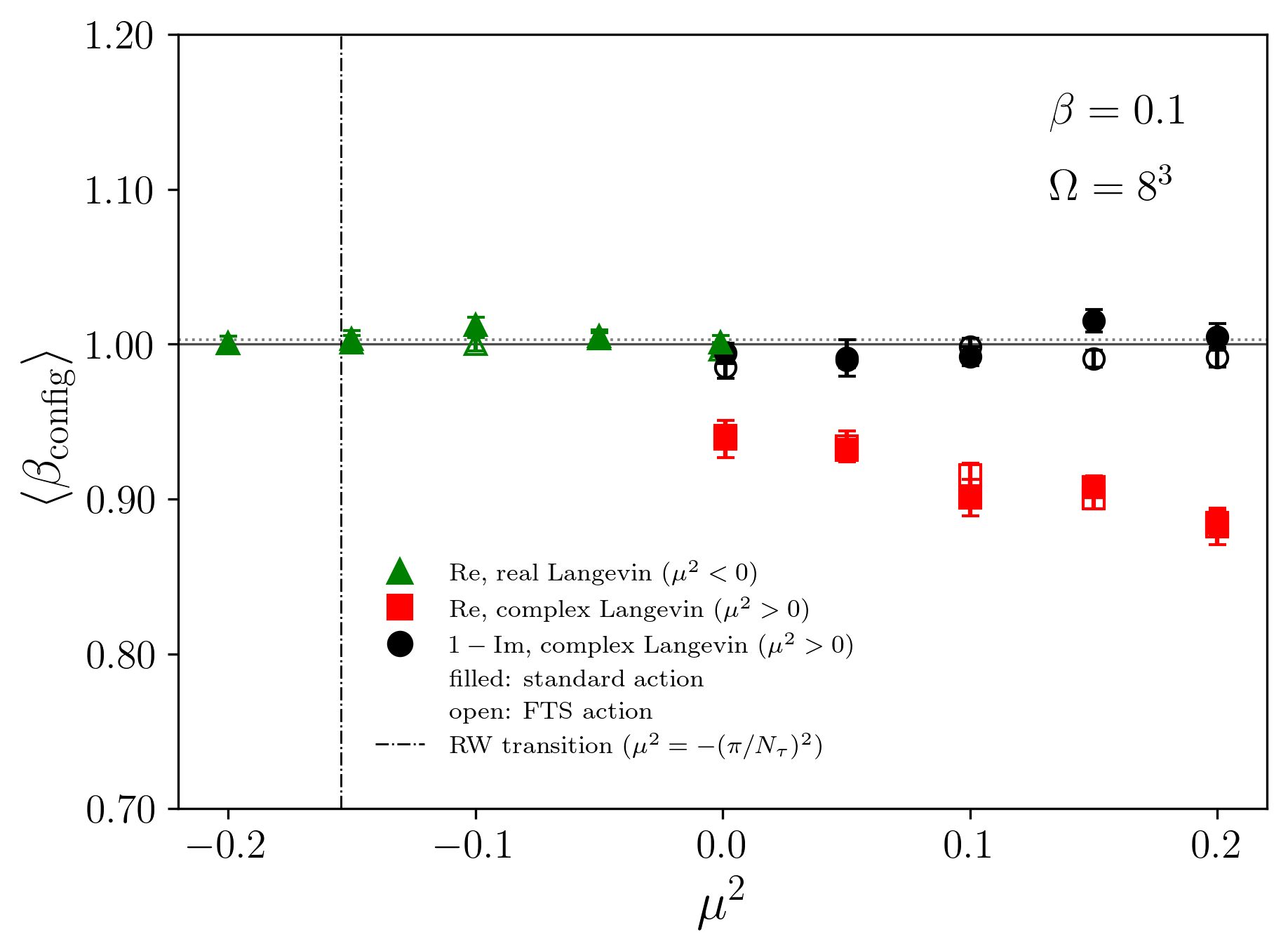} 
\includegraphics[width=0.49\columnwidth]{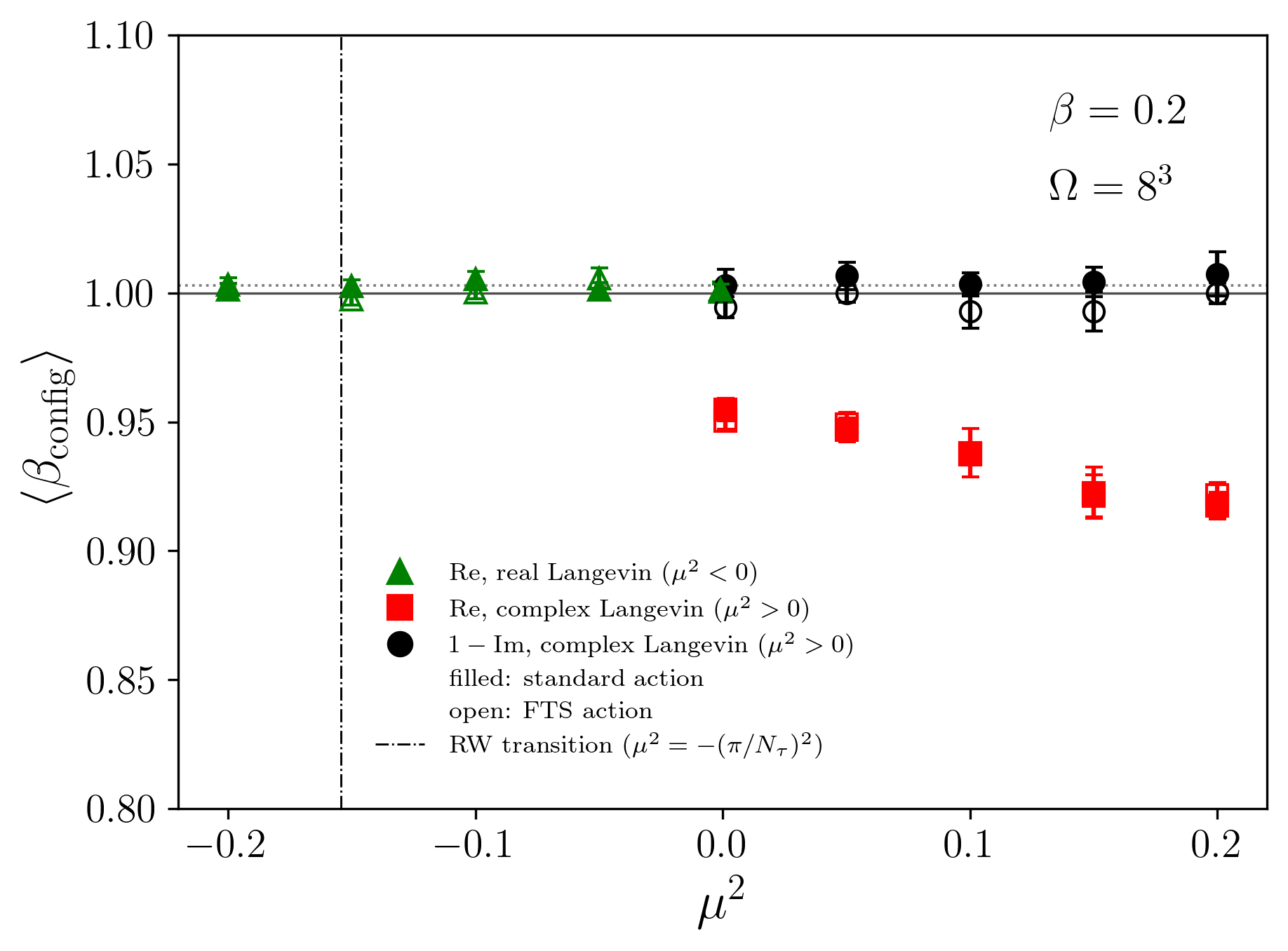} 

\includegraphics[width=0.49\columnwidth]{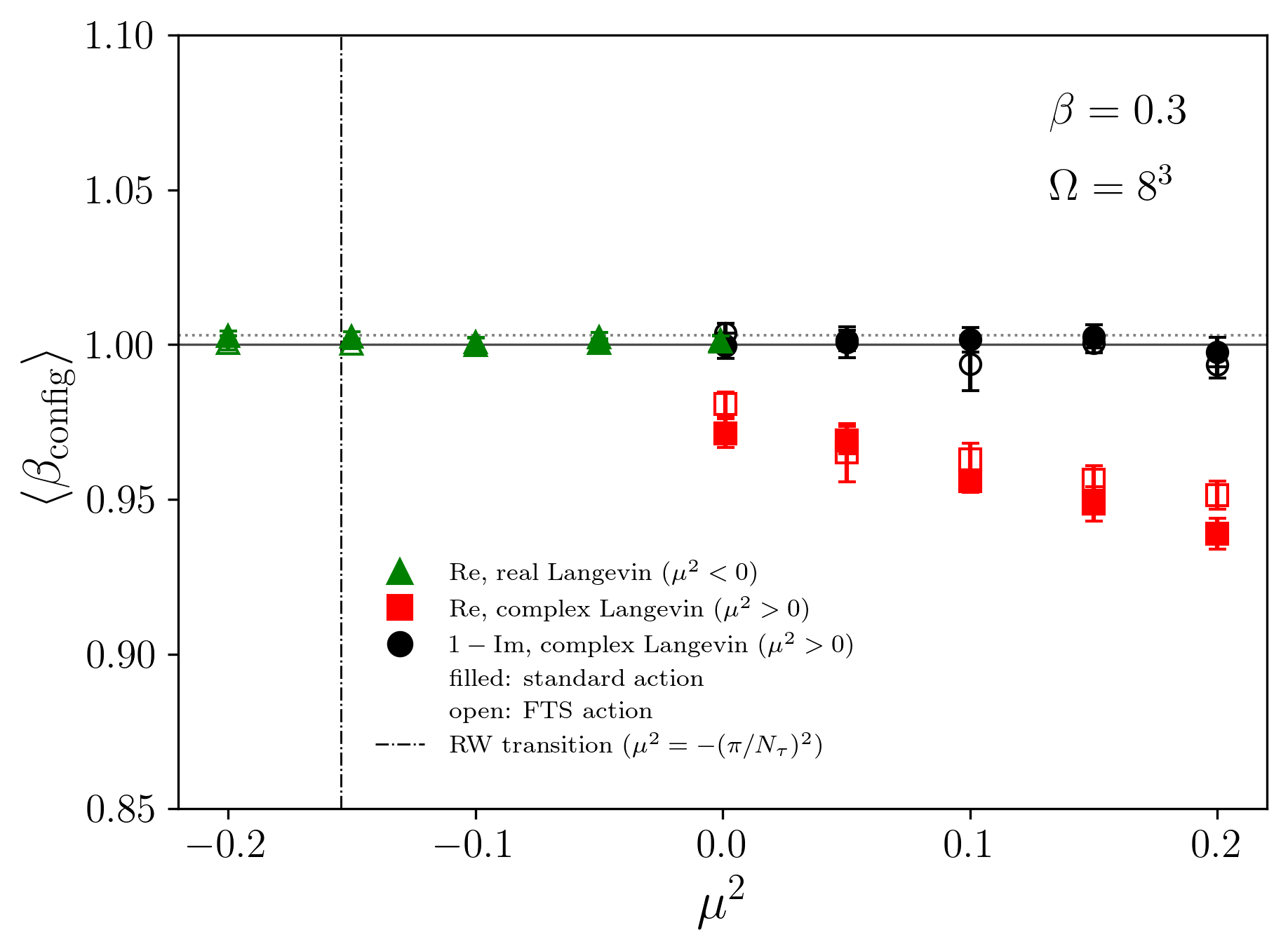} 
\includegraphics[width=0.49\columnwidth]{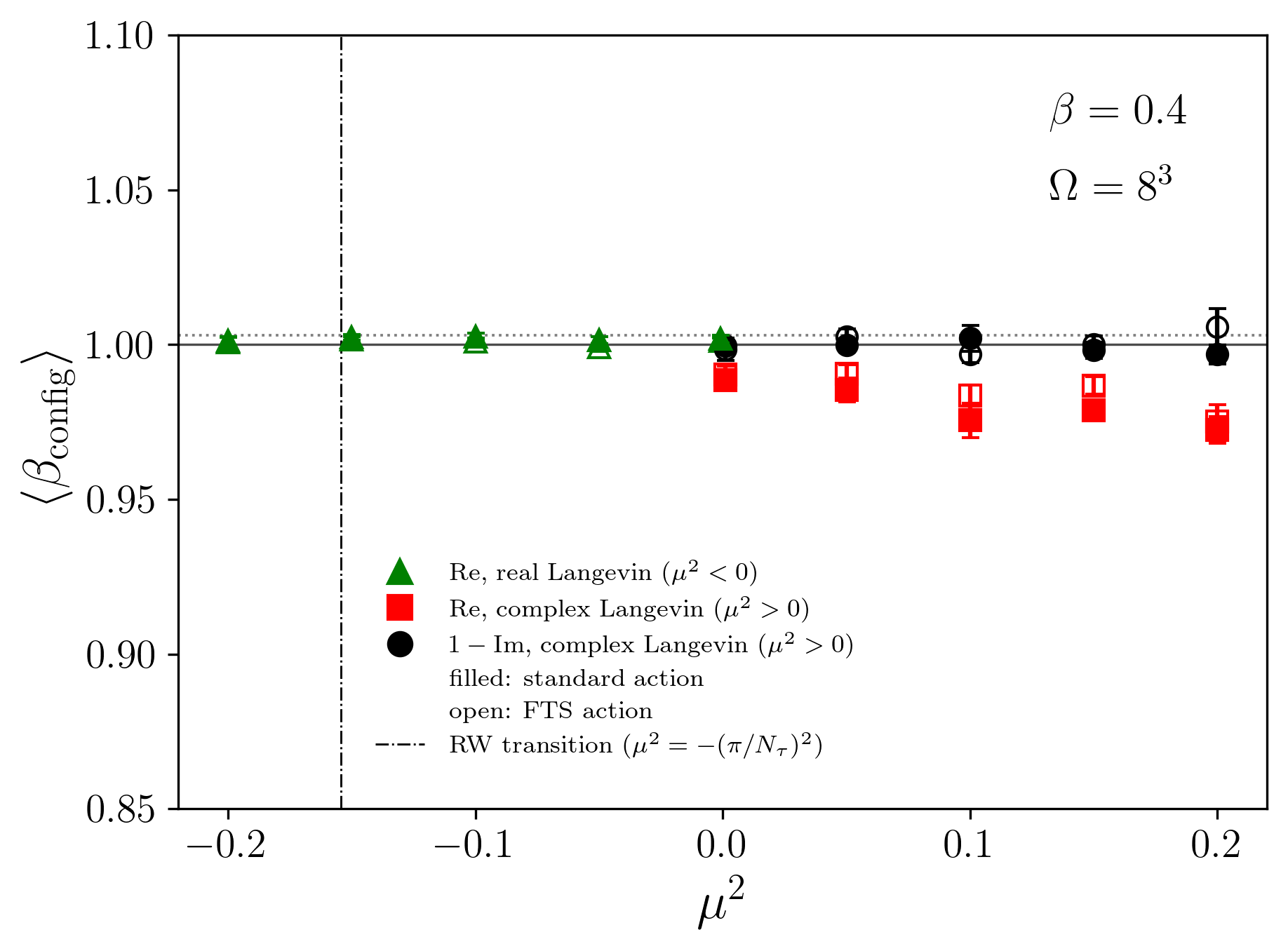} 

\includegraphics[width=0.49\columnwidth]{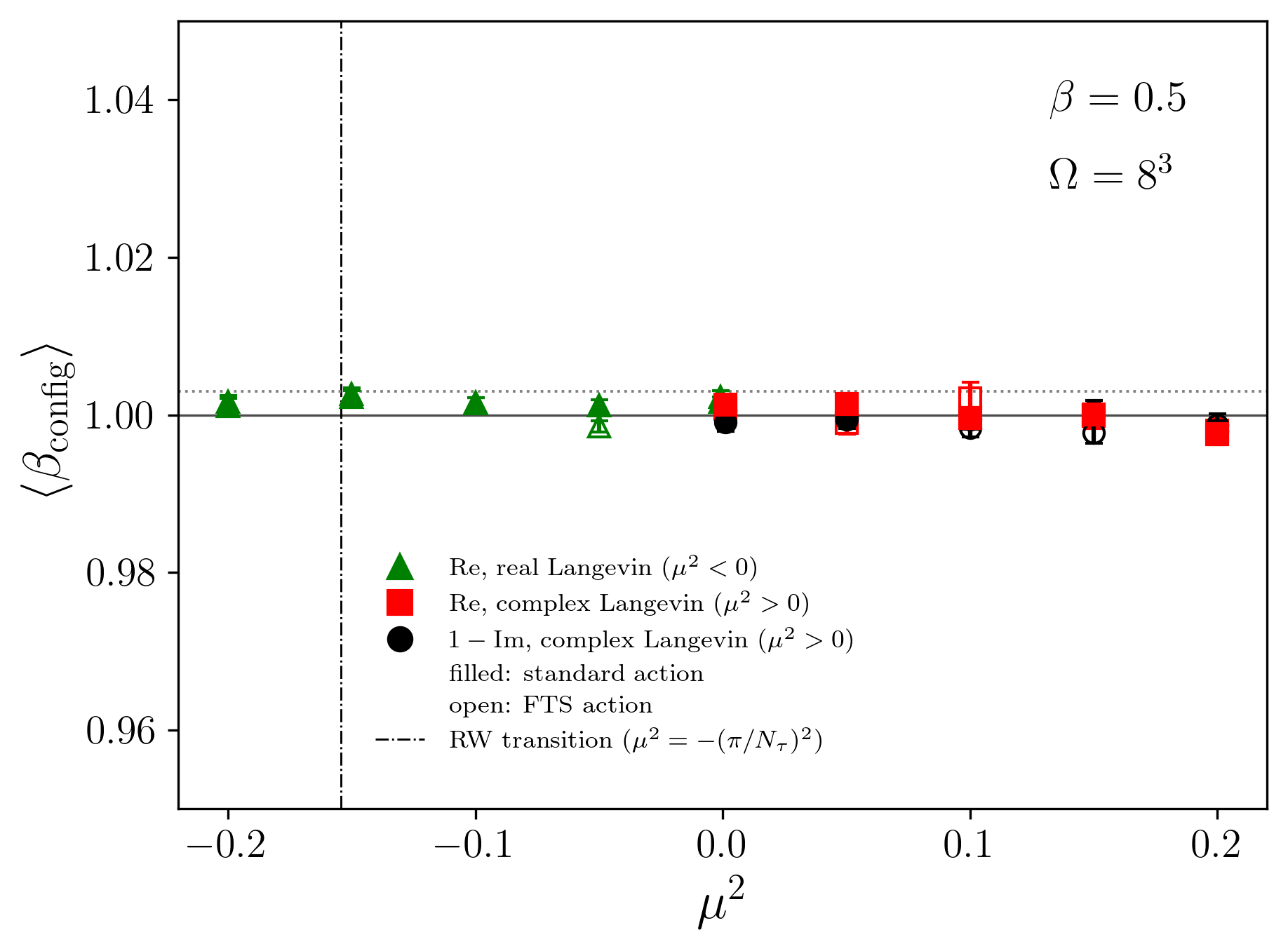}	\includegraphics[width=0.49\columnwidth]{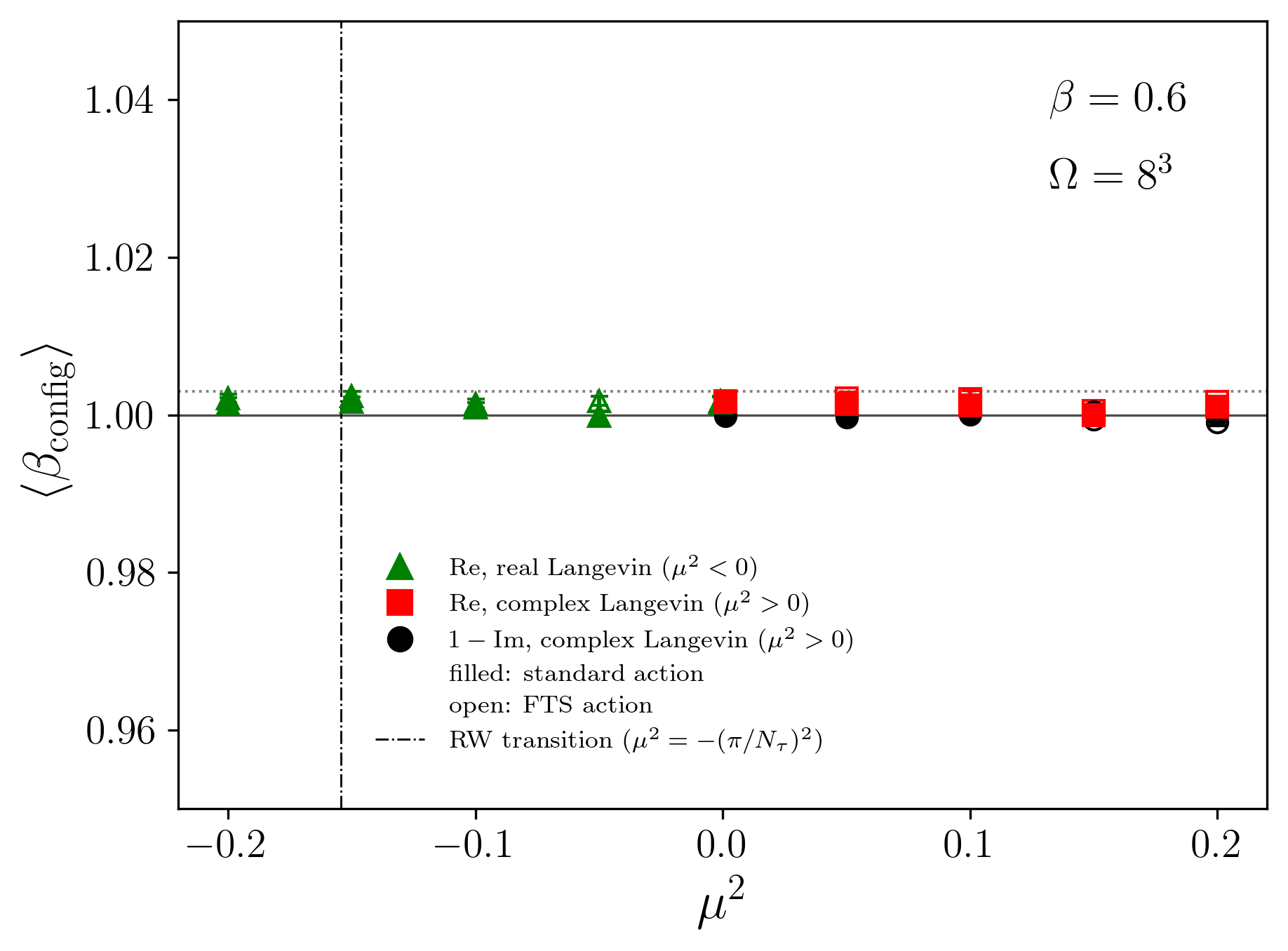} 
\caption {Measured configurational temperature $\langle \beta_{\rm config} \rangle$, Eq. \eqref{eq:beta_config_M_def}: real part and $1 - {\rm Im} \, \langle \beta_{\rm config} \rangle$ (black circles, complex Langevin only; for real Langevin the imaginary part vanishes), as a function of $\mu^2$, for $\beta = 0.1$ -- $0.6$ (disordered and near-transition regimes), on an $8^3$ lattice. 
Green triangles/red squares denote real/complex Langevin dynamics; filled/open symbols denote the standard/FTS action formulations. 
The solid horizontal line marks unity; the dotted line marks the $+ 0.3\%$ offset of the diagonal estimator on this lattice (Sec.~\ref{sec:lattice_estimator}). 
The dashed vertical line marks the Roberge--Weiss transition of the FTS action, $\mu^2 = - (\pi / N_\tau)^2$.} 
\label{fig:beta_beta_m_plots_a} 
\end{figure*} 

\begin{figure*}[htbp] 
\centering 

\includegraphics[width=0.49\columnwidth]{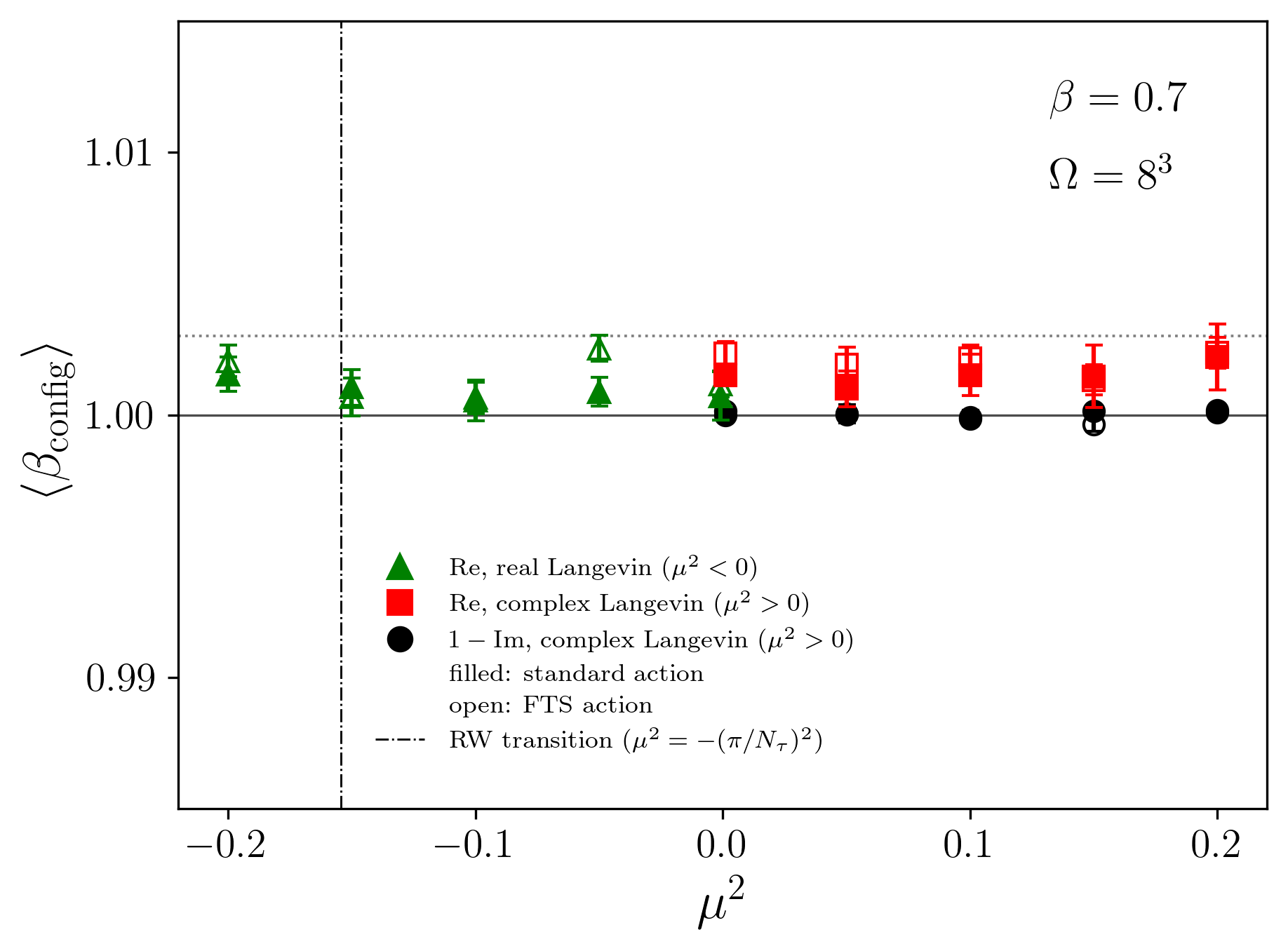} 
\includegraphics[width=0.49\columnwidth]{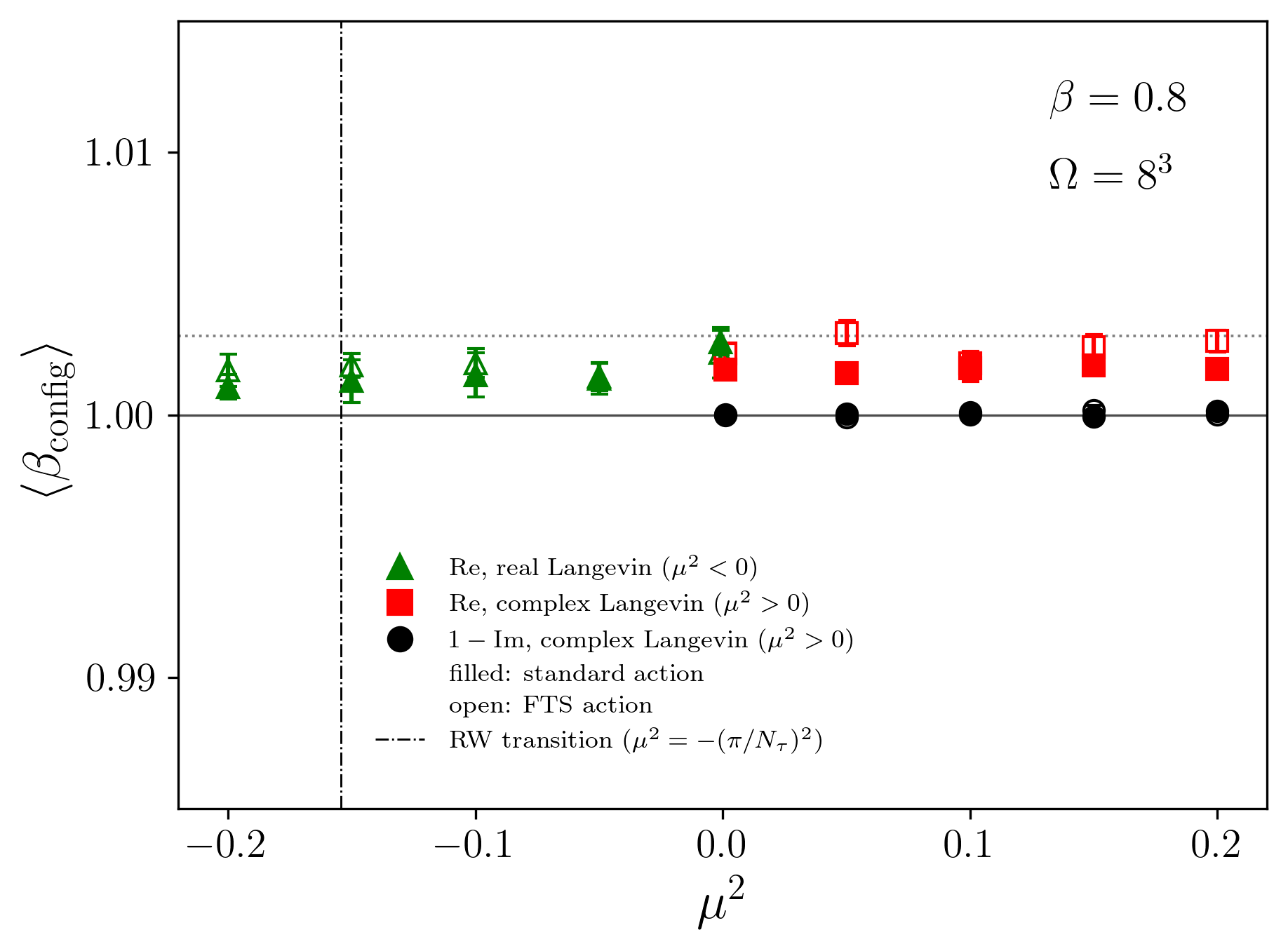} 

\includegraphics[width=0.49\columnwidth]{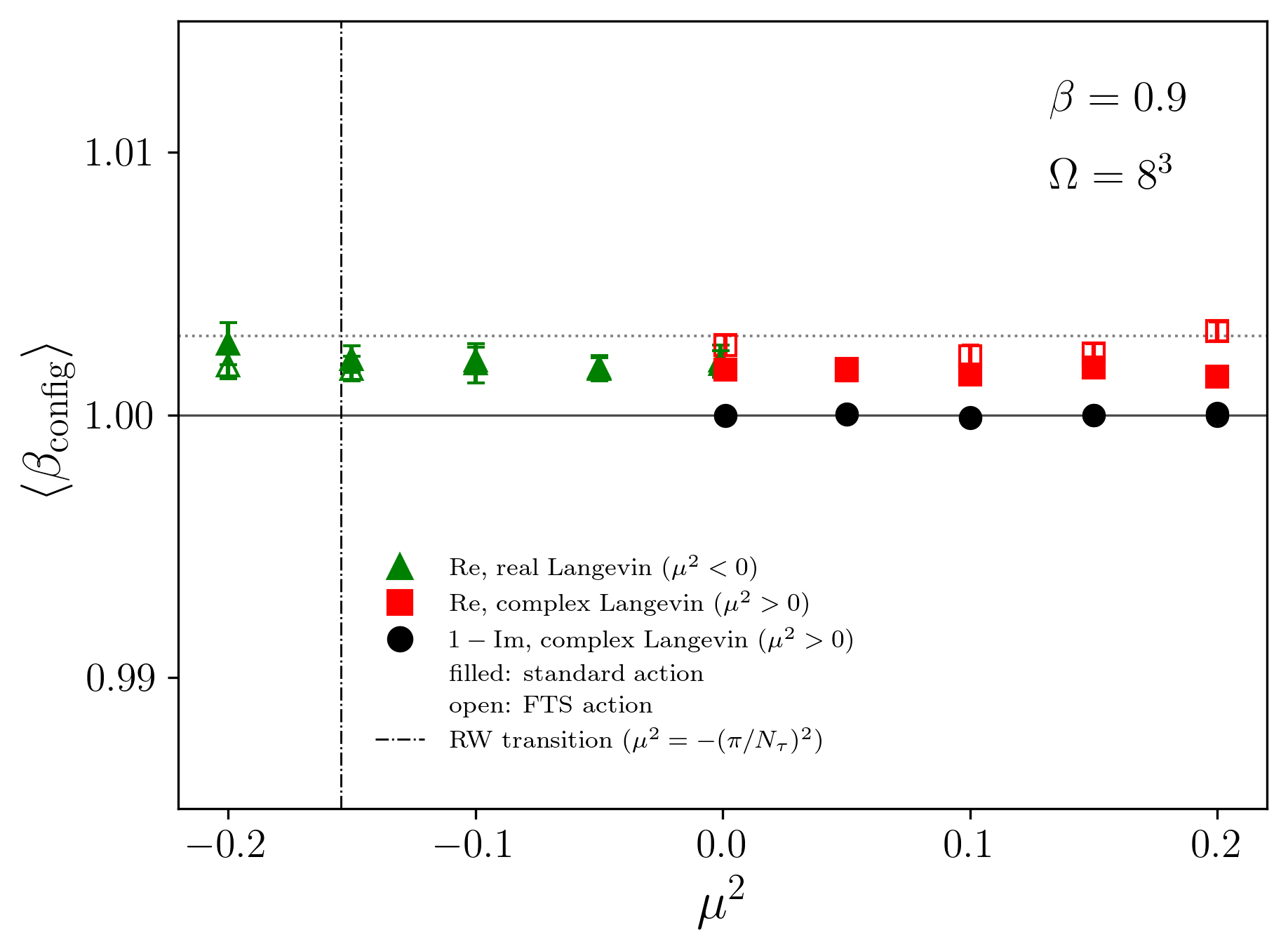} 
\includegraphics[width=0.49\columnwidth]{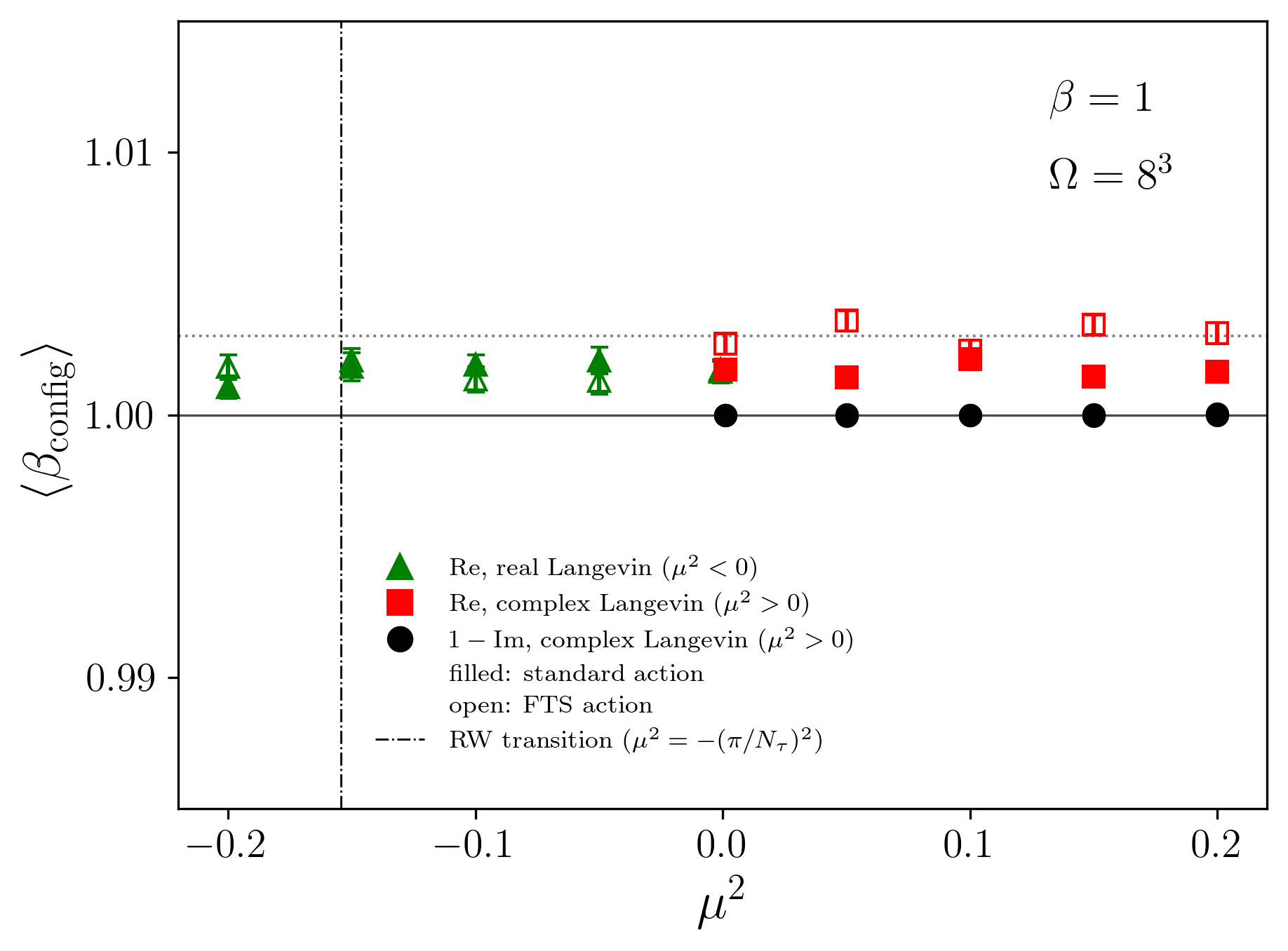} 

\caption {Same as Fig.~\ref{fig:beta_beta_m_plots_a}, but for $\beta = 0.7$ -- $1.0$ (ordered phase).} 
\label{fig:beta_beta_m_plots_b} 
\end{figure*} 

\begin{figure*}[htbp] 
\centering 
\includegraphics[width=0.49\columnwidth]{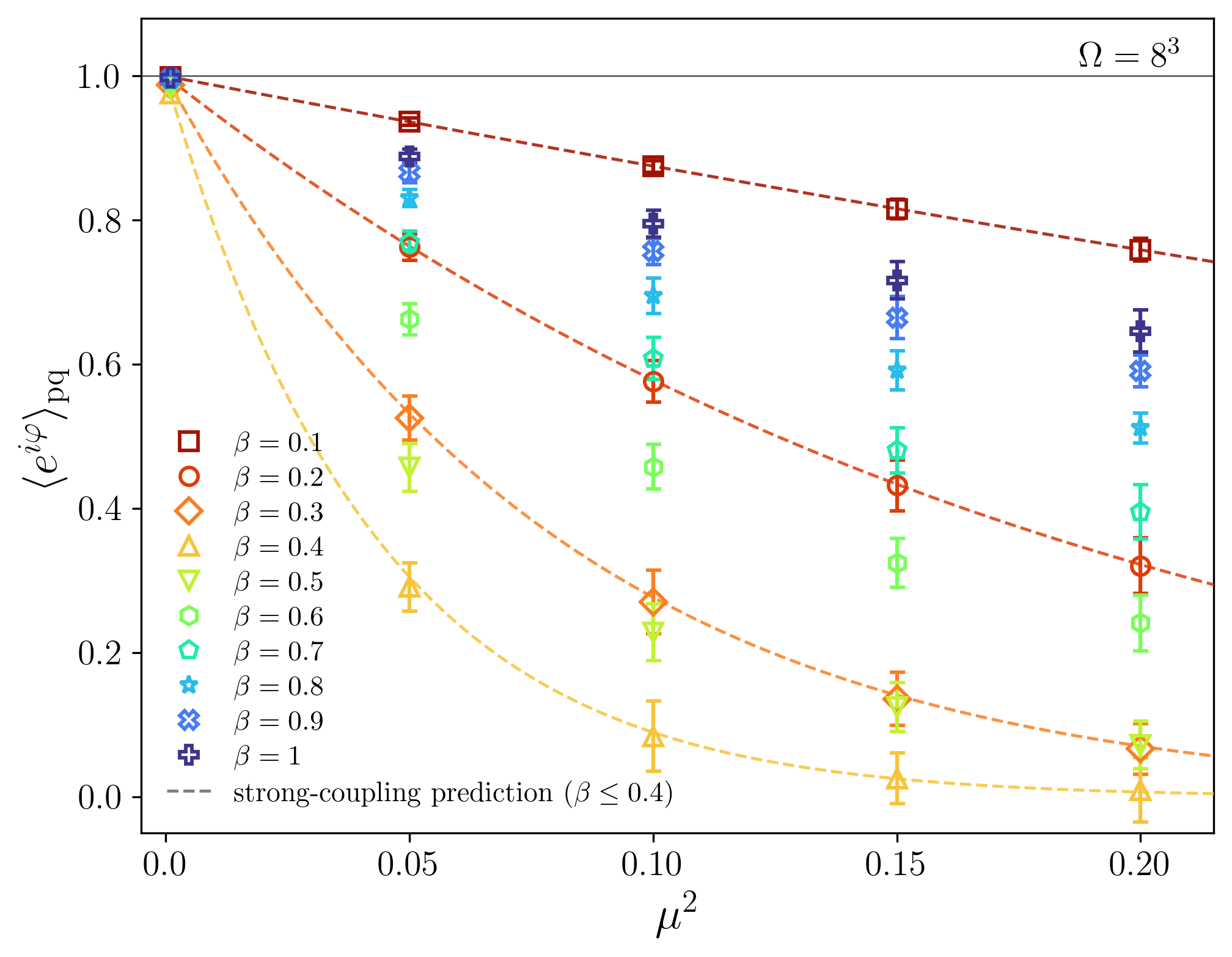} 
\includegraphics[width=0.49\columnwidth]{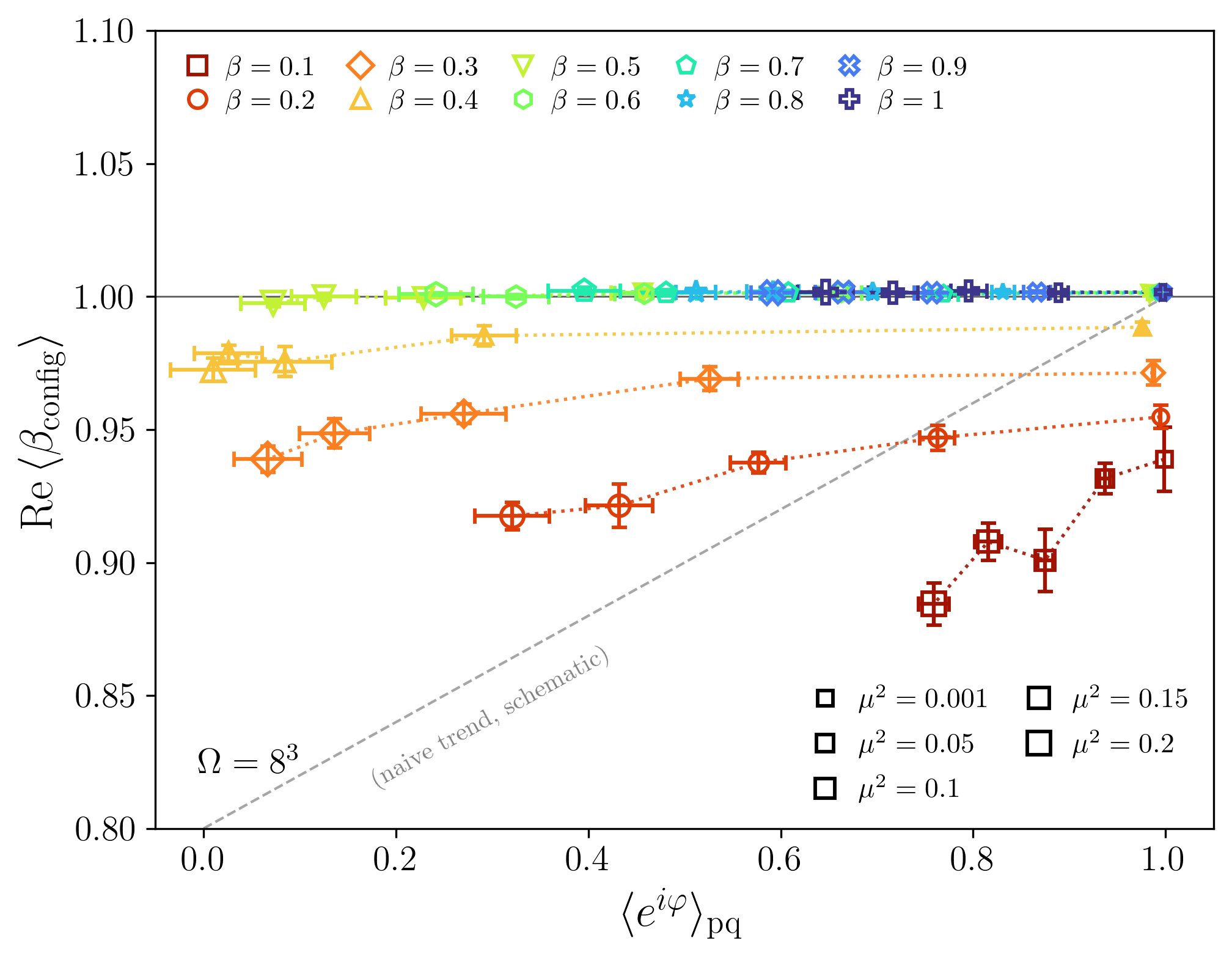} 
\caption{\textit{Left:} Average phase factor $\langle e^{i \varphi} \rangle_{\rm pq}$ as a function of $\mu^2$ for $\beta = 0.1$ -- $1.0$ on an $8^3$ lattice. 
Points with error bars are Metropolis measurements of the phase-quenched ensemble (anisotropic XY, $\beta_0 = \beta \cosh\mu$, $\beta_{1, 2} = \beta$). 
Dashed curves show the strong-coupling prediction $\exp(- \Omega \, \Delta f)$ for $\beta \leq 0.4$. 
\textit{Right:} ${\rm Re} \, \langle \beta_{\rm config} \rangle$ versus the sign-problem severity $\langle e^{i \varphi} \rangle_{\rm pq}$ for all $(\beta, \mu^2)$ grid points. 
The horizontal line at ${\rm Re} \, \langle \beta_{\rm config} \rangle = 1$ marks the target value (up to the $0.3\%$ offset of the diagonal estimator), and the gray dashed line is a schematic guide to the eye indicating a monotonic dependence on the phase factor (Sec.~\ref{sec:sign_problem_dependence}). 
Color encodes $\beta$ from small (red, $\beta = 0.1$) to large (blue, $\beta = 1.0$), and each $\beta$ has its own symbol. 
Dotted lines join the points of one $\beta$ in the order of $\mu^2$, and the marker size grows with $\mu^2$.} 
\label{fig:pq_vs_musq} 
\label{fig:scatter_pq_vs_clm} 
\end{figure*} 

Figs. \ref{fig:beta_beta_m_plots_a} and \ref{fig:beta_beta_m_plots_b} show the measured configurational temperature $\langle \beta_{\rm config} \rangle$ as a function of $\mu^2$, obtained on an $8^3$ lattice.
Fig.~\ref{fig:beta_beta_m_plots_a} covers $\beta = 0.1$ -- $0.6$, which spans the disordered and near-transition regimes.
Fig.~\ref{fig:beta_beta_m_plots_b} covers $\beta = 0.7$ -- $1.0$, the ordered phase.
We find that the standard (filled symbols) and final-time-slice (open symbols) action formulations give consistent values of the estimator.
This includes $\mu^2 = -0.2$ for $\beta \geq 0.5$, where the two formulations give different action densities (Sec.~\ref{sec:sim_details_results}): the estimator does not detect the metastability of the standard formulation.

For $\mu^2 < 0$ (imaginary chemical potential, green triangles), real Langevin dynamics gives $\langle \beta_{\rm config} \rangle$ consistent with its target within statistical errors for all values of $\beta$ explored.
The deviations from unity are below $0.6\%$ for $\beta \geq 0.5$ and below $1\%$ for $\beta \leq 0.4$, where the estimator fluctuates more strongly.
This confirms the implementation, and sets the resolution of the estimator for our statistics and step sizes.

For $\mu^2 > 0$ (complex Langevin, red squares), the behavior changes sharply.
In the disordered phase, $\beta \leq 0.4$, the estimator lies clearly below its target already at $\mu^2 = 0.001$, where the sign problem is negligible.
The deviation is $\sim 1.5\%$ at $\beta = 0.4$ and $\sim 6\%$ at $\beta = 0.1$, where $\langle \beta_{\rm config} \rangle$ = $0.9885(18)$ and $0.939(12)$, respectively.
At $\mu^2 = 0.2$ the deviation grows steadily as $\beta$ decreases from $0.4$ to $0.1$, from $\sim 3\%$ to $\sim 12\%$.
The values are $\langle \beta_{\rm config} \rangle$ = $0.972(4)$, $0.939(5)$, $0.918(5)$ and $0.884(8)$ at $\beta = 0.4$, $0.3$, $0.2$ and $0.1$.
The errors are larger in the disordered phase.
There, the Langevin distribution approaches the poles of the estimator, and its fluctuations become heavy-tailed (Sec.~\ref{sec:ds}).

In contrast, in the ordered phase, $\beta \geq 0.5$, the estimator agrees with its target within $\sim 0.6\%$ at all $\mu^2$ for the standard action.
At $\beta = 0.5$, the complex Langevin action density differs from the world-line result by $1$ -- $4\%$ (Fig.~\ref{fig:act_den_plots_a}).
Since the world-line representation is an exact rewriting of the same lattice partition function, this deviation indicates a small failure of complex Langevin, consistent with the extended drift tails at this coupling (Sec.~\ref{sec:Decay_Drift}), although a contribution from the finite step size cannot be excluded (Sec.~\ref{sec:sim_details_results}).
The estimator deviates from its target by only $\sim 0.3$ -- $0.5\%$ here, and does not resolve this failure beyond the uncertainty set by the directional term (Sec.~\ref{sec:lattice_estimator}).
Across the transition, the deviation of the estimator drops sharply: at $\mu^2 = 0.2$, from $\sim 3\%$ at $\beta = 0.4$ to $\sim 0.5\%$ at $\beta = 0.5$.
The imaginary part is consistent with zero.

\subsubsection{Dependence on the severity of the sign problem}
\label{sec:sign_problem_dependence}

The deviation of the estimator from unity grows as $\beta$ decreases.
We therefore ask whether the deviation scales with the severity of the sign problem, which we measure using the phase-quenched ensemble of Sec. \ref{sec:sign_problem_severity}.
Fig. \ref{fig:pq_vs_musq} (left panel) shows $\langle e^{i \varphi} \rangle_{\rm pq}$ as a function of $\mu^2$ for all ten values $\beta = 0.1$ -- $1.0$ on an $8^3$ lattice.
The $\mu^2$-dependence shows a non-trivial interplay with $\beta$, which goes beyond the leading-order strong-coupling suppression $e^{- \Omega \beta^2 \mu^2 / 4}$.

Consider first the disordered phase ($\beta \leq 0.4$).
The strong-coupling free-energy difference of Eq. \eqref{eq:delta_f_sc} (using $\cosh^2\mu - 1 = \sinh^2\mu$) grows with $\beta^2$.
A larger $\beta$ therefore produces a more severe sign problem at fixed $\mu^2$: at $\mu^2 = 0.2$ the phase factor falls from $\sim 0.76$ at $\beta = 0.1$ to $\sim 0.01$ at $\beta = 0.4$.
Dashed curves show this strong-coupling prediction of the phase factor for $\beta \leq 0.4$. 
They agree with our simulation data to within a few percent for $\beta \leq 0.2$. 
For $\beta = 0.3$ -- $0.4$, where higher-order terms in the $\beta$ expansion become important, the prediction deviates from the data by about $5\%$. 
The agreement at small $\beta$ validates both the formula and the simulation.

The behavior changes qualitatively above the phase transition.
For $\beta > 0.4$, the strong-coupling expansion is no longer reliable (it would predict near-zero phase factors, unlike the data), and direct Monte Carlo is required.
The measured values show that the sign problem is less severe in the ordered phase than near $\beta_c$, as anticipated by Eq.~\eqref{eq:pq_cumulant}.
At $\mu^2 = 0.2$, $\langle e^{i \varphi} \rangle_{\rm pq}$ rises from the value of $\sim 0.01$ at $\beta = 0.4$ back to $\sim 0.24$ at $\beta = 0.6$ and then to $\sim 0.65$ at $\beta = 1.0$.
The overall behavior is therefore non-monotonic.
The severity of the sign problem peaks near the phase boundary $\beta_c \approx 0.454$, and is moderate in the bulk of the ordered phase.
It is mild across the entire phase diagram at $\mu^2 = 0.001$, where the phase factor exceeds $0.97$ for all $\beta$.
At $\mu^2 = 0.2$, the couplings $\beta = 0.1$ and $0.4$ thus have the mildest and the most severe sign problem of the disordered phase.
We return to these two couplings in Sec.~\ref{sec:ds}.

We now know how the sign problem varies across the parameter space.
This allows us to pose a sharper question: does the severity of the sign problem quantitatively predict the deviation of $\langle \beta_{\rm config} \rangle$ from its target value?
If the severity of the sign problem controlled the failure, the deviation of the estimator from its target would vanish as $\langle e^{i \varphi} \rangle_{\rm pq} \to 1$ and grow monotonically as $\langle e^{i \varphi} \rangle_{\rm pq}$ decreases, at every $\beta$.
The gray dashed line in Fig.~\ref{fig:scatter_pq_vs_clm} (right) indicates such a monotonic trend schematically; it is a guide to the eye, not a prediction.

The data violate this expectation in both directions.
Consider the disordered phase ($\beta \leq 0.4 < \beta_c \approx 0.454$, warm-colored points).
There ${\rm Re} \, \langle \beta_{\rm config} \rangle$ lies $\sim 1$ -- $6\%$ below its target already at $\langle e^{i \varphi} \rangle_{\rm pq} \approx 1$ ($\mu^2 = 0.001$, where the sign problem is essentially absent), and up to $\sim 12\%$ below at larger $\mu^2$.
The points lie below the reference line $\langle \beta_{\rm config} \rangle = 1$ over the whole range of the phase factor, from $0.01$ to $1$.
Within the disordered phase and at fixed $\mu^2$, the deviation is largest where the sign problem is mildest.
At $\mu^2 = 0.2$ it is $\sim 12\%$ at $\beta = 0.1$, where $\langle e^{i \varphi} \rangle_{\rm pq} \sim 0.76$, and $\sim 3\%$ at $\beta = 0.4$, where $\langle e^{i \varphi} \rangle_{\rm pq} \sim 0.01$.

In the ordered phase ($\beta \geq 0.5 > \beta_c$, cool-colored points), the converse holds.
Here $\langle \beta_{\rm config} \rangle$ remains near unity even as $\langle e^{i \varphi} \rangle_{\rm pq}$ falls to $\sim 0.07$ for values of $\beta$ nearest to $\beta_c$ at $\mu^2 = 0.2$.
These points lie on the horizontal reference line at $\langle \beta_{\rm config} \rangle = 1$, far above the schematic monotonic trend.
The two clusters are separated not by the severity of the sign problem but by the phase boundary $\beta_c(\mu^2)$.
At fixed $\mu^2$, however, both the severity of the sign problem and the distance from the transition vary with $\beta$, so these data alone cannot separate the two.
The map of the estimator in Sec.~\ref{sec:phase_boundary_dependence} does so: as $\mu^2$ grows, $\beta_c(\mu^2)$ moves to smaller $\beta$, and the region of large deviations moves with it.
This agrees with the conclusion of Ref.~\cite{Aarts:2010aq} that the failure of complex Langevin in this model is not due to the sign problem, and with the success of complex Langevin in the relativistic Bose gas despite a severe sign problem \cite{Aarts:2008wh, Aarts:2009hn}. 

\subsubsection{Dependence on the phase boundary}
\label{sec:phase_boundary_dependence}

\begin{figure*}[htbp] 
\centering 
\includegraphics[width=1.0\columnwidth]{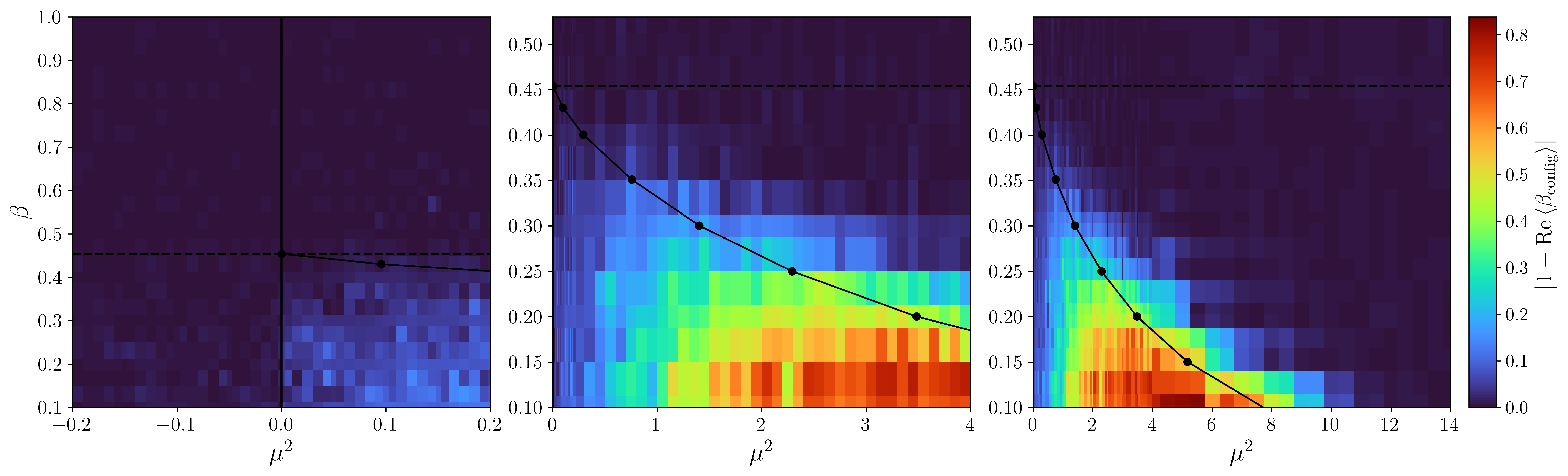} 
\caption{Heat map of $|1 - {\rm Re} \, \langle \beta_{\rm config} \rangle|$ in the $(\beta, \mu^2)$ plane. 
\textit{Left:} $\mu^2 \in [- 0.2, 0.2]$; \textit{center:} $\mu^2 \in [0, 4]$; \textit{right:} $\mu^2 \in [0, 14]$. 
All three panels share the same color scale, from dark blue (agreement with the target) through green and yellow to red (deviations up to $\sim 0.8$). 
The dashed horizontal line marks $\beta_c \approx 0.454$, the solid curve with circles marks the phase boundary $\beta_c(\mu^2)$ from the worm algorithm \cite{Banerjee:2010kc}, and the vertical line in the left panel marks $\mu^2 = 0$. 
In the right panel, cells without a simulation point are filled by inverse-distance interpolation for visual continuity only. Statistical errors are not shown.} 
\label{fig:beta_beta_m_heatmap} 
\end{figure*} 

The three panels of Fig. \ref{fig:beta_beta_m_heatmap} give a global view of the deviation of the estimator across the $(\beta, \mu^2)$ parameter space.
They display the deviation $|1 - {\rm Re} \, \langle \beta_{\rm config} \rangle|$ as a heat map at three successive resolutions.
The target value is unity up to the $0.3\%$ offset of the diagonal estimator, which is invisible on this color scale.
This quantity therefore measures the absolute departure from the target.
Blue pixels indicate agreement with the target, and colors from green to red indicate increasingly large deviations.
The three panels cover the resolutions for $\mu^2$ in $[- 0.2, 0.2]$ (left, narrow range), $[0, 4]$ (middle, wide range), and $[0, 14]$ (right, extended range).  
In all three panels, the dashed horizontal line marks $\beta_c \approx 0.454$, and the solid black curve with filled black circles shows the phase boundary $\beta_c(\mu^2)$ obtained from the worm algorithm \cite{Banerjee:2010kc}. 
This boundary is the infinite-volume onset $\mu_c = M$ of the transition from the normal to the superfluid phase, where $M$ is the particle mass at $\mu = 0$, as determined in Ref.~\cite{Banerjee:2010kc} for $0.04 \leq \beta \leq 0.43$.
Our simulations are on an $8^3$ lattice, on which the onset is shifted by finite-size effects, most strongly near $\beta_c$, where the correlation length is large \cite{Banerjee:2010kc}.
On the finite lattice, the transition is a crossover, and $\beta_c(\mu^2)$ serves as a reference line.
At small $\beta$, a simple hopping estimate of the onset, $\cosh \mu_c \approx 1/\beta - 2$, which follows from $I_1(\beta)/I_0(\beta) \approx \beta/2$ and neglects interactions between world lines, gives $\mu_c = 3.828$ at $\beta = 0.04$ and $1.763$ at $\beta = 0.2$, compared with $3.829(4)$ and $1.85801(3)$ in Ref.~\cite{Banerjee:2010kc}. 
At small values of $\mu^2$, the wide-range and extended-range panels are supplemented with data from the narrow-range and wide-range grids, respectively, to improve coverage in that region. 
The heat-map runs are single runs, with the same lattice and adaptive step size ($\bar\epsilon \leq 5 \times 10^{-4}$ for complex Langevin), and with run lengths equal to those of the first runs of Sec.~\ref{sec:sim_details_results}.
These maps are exploratory.
They use single runs, show no statistical errors, include interpolated cells in the right panel, and reach chemical potentials where we cannot exclude a localized pole contribution to the target value (Sec.~\ref{sec:lattice_estimator}).
They indicate where deviations occur on this lattice, but they are not a quantitative map of correctness, nor of the sign problem.
Our quantitative statements rest on the runs of Sec.~\ref{sec:estimator_results}.

In the left panel, the entire $\mu^2 < 0$ region appears uniformly dark for all $\beta$.
This reflects the agreement of real Langevin dynamics with the target.
For $\mu^2 > 0$, the color pattern separates the two phases, apart from a rounded edge near the boundary (see below).
The region above the phase boundary (the ordered phase) remains predominantly dark.
In contrast, the region below it (the disordered phase) shows deviations of up to $0.16$, which is light blue on the common color scale.
Warm colors appear only at larger $\mu^2$ in the other two panels.

The center panel extends the analysis into the real-chemical-potential quadrant, and follows the region of large deviations out to $\mu^2 = 4.0$.
The phase boundary $\beta_c(\mu^2)$ descends steadily from $\beta_c \approx 0.454$ near $\mu^2 = 0$ to $\beta_c \sim 0.20$ at $\mu^2 \sim 3.5$.
This reflects the tendency of the chemical potential to favor the ordered phase.
The deviation broadly follows this lowering trend across the entire range.
Well inside the ordered phase, the estimator stays close to its target in all three panels. Since these are single runs, and since the target itself is not controlled at large $\mu^2$ (Sec.~\ref{sec:lattice_estimator}), we do not quantify this agreement.
Close to the boundary, agreement with the target is only a weak indication of correct sampling.
Wherever it is available, the comparison with the world-line action density remains the more decisive test.
Within the disordered wedge, the deviations do not soften.
Instead, $|1 - {\rm Re} \, \langle \beta_{\rm config} \rangle|$ grows with $\mu^2$ and is largest in the deeply disordered, large-$\mu^2$ corner.
In this corner, the full theory is expected to show no dependence on $\mu$ below the onset of the density (Silver Blaze).
In this region, complex Langevin produces a spurious $\mu$ dependence of the action density, as seen at $\mu^2 \leq 0.2$ in Fig.~\ref{fig:act_den_plots_a} and found in Ref.~\cite{Aarts:2010aq}; the estimator map is consistent with this, although it does not measure the action density itself.

The right panel extends this picture further into the real-chemical-potential regime, reaching $\mu^2 = 14.0$ over $\beta \in [0.10, \sim0.5]$.
The phase boundary $\beta_c(\mu^2)$ continues its descent and reaches values $\beta_c \lesssim 0.15$ at large $\mu^2$.
The edge of the region of large deviations broadly follows this descent throughout.
The dark region expands, as more of the $(\beta, \mu^2)$ plane is swept into the ordered phase at large chemical potential.
At the same time, the disordered wedge narrows in its $\beta$ extent and intensifies in color.
In these single runs, the deviations thus grow with $\mu^2$ in the disordered region.
The edge of the region of large deviations is not sharp.
In the center panel, deviations of about $0.1$ -- $0.2$ extend into the ordered side of the boundary, by up to $\Delta\beta \approx 0.05$ -- $0.1$ for $\mu^2 \approx 1$ -- $4$.
In the right panel, deviations of about $0.2$ -- $0.4$ persist at $\beta \approx 0.10$ -- $0.15$ up to $\mu^2 \approx 10$, beyond the point where $\beta_c(\mu^2)$ crosses $\beta = 0.1$.
We attribute this to the rounding of the transition on the $8^3$ lattice, to the finite-size shift of the onset relative to the infinite-volume boundary, and to the proximity of the onset, but we have not verified this with other volumes.
This qualitative picture persists to the largest chemical potentials explored. It is consistent with a failure of complex Langevin tied to the phase structure of the theory; we discuss the limits of this conclusion at large $\mu^2$ below.

Taken together, the three panels show the following.
Throughout the parameter space explored, the region of large estimator deviations moves with the order--disorder phase boundary $\beta_c(\mu^2)$, which is determined independently by the worm algorithm \cite{Banerjee:2010kc}, and lies predominantly on its disordered side.
This is consistent with the conclusion drawn from Fig.~\ref{fig:scatter_pq_vs_clm} at $\mu^2 \leq 0.2$, that the deviations are governed by the phase of the system rather than by the magnitude of the sign problem.
At large $\mu^2$, however, the map alone does not separate the two effects, since we measured the sign problem only for $\mu^2 \leq 0.2$.
In the disordered region at large $\mu^2$, the sign problem is exponentially severe: the strong-coupling estimate, Eq.~\eqref{eq:delta_f_sc}, gives $\Omega \Delta f \approx 38$ at $\beta = 0.15$, $\mu^2 = 4$, i.e., $\langle e^{i \varphi} \rangle_{\rm pq} \approx e^{-38}$.
The decisive comparison would be with the ordered region at large $\mu^2$, where complex Langevin appears reliable, but where we have not quantified the sign problem; Eq.~\eqref{eq:pq_cumulant} would allow this through a measurement of $\langle J^2 \rangle_{\rm pq}$.
At large $\mu^2$ we therefore only conclude that, on this lattice, the failure region moves with $\beta_c(\mu^2)$.

Runaways to a non-finite drift occur in long runs at $\mu^2 > 0$ for $\beta = 0.2$ -- $0.5$ with $\bar\epsilon = 5 \times 10^{-4}$.
For these runs, we use only the measurements before the runaway; their averages agree within errors with those of the runs that reached the final Langevin time. 
The failure of complex Langevin in this regime is therefore not caused by the instabilities themselves, but reflects convergence to a wrong distribution.
The failure is localized predominantly within the disordered phase across all chemical potentials studied, regardless of whether the sign problem is mild or severe.
Together with the analysis of Sec.~\ref{sec:sign_problem_dependence}, this points to the phase structure of the underlying theory as the main factor controlling incorrect convergence on this lattice.

\subsection{Decay of the drift terms}
\label{sec:Decay_Drift}

A second, complementary criterion for the correctness of complex Langevin dynamics was formulated by Nagata \textit{et al.} \cite{Nagata:2016vkn, Nagata:2018net}.
It is based on the asymptotic behavior of the probability distribution of the Langevin drift terms.
It is argued to be necessary and sufficient for the validity of complex Langevin simulations, provided the Langevin process is stable and the observable converges.

\begin{figure*}[htbp] 
\centering 
\includegraphics[width=0.49\columnwidth]{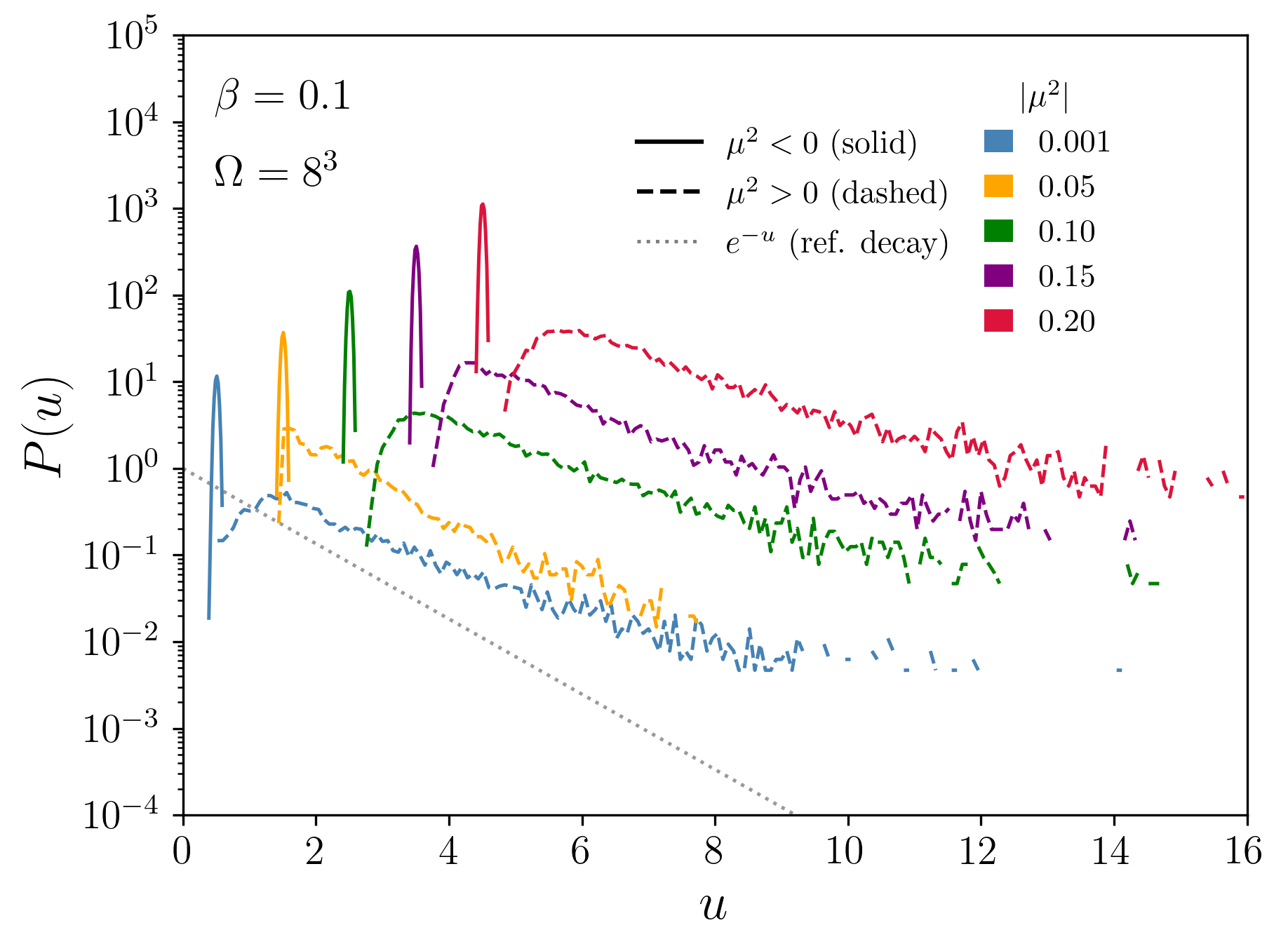} 
\includegraphics[width=0.49\columnwidth]{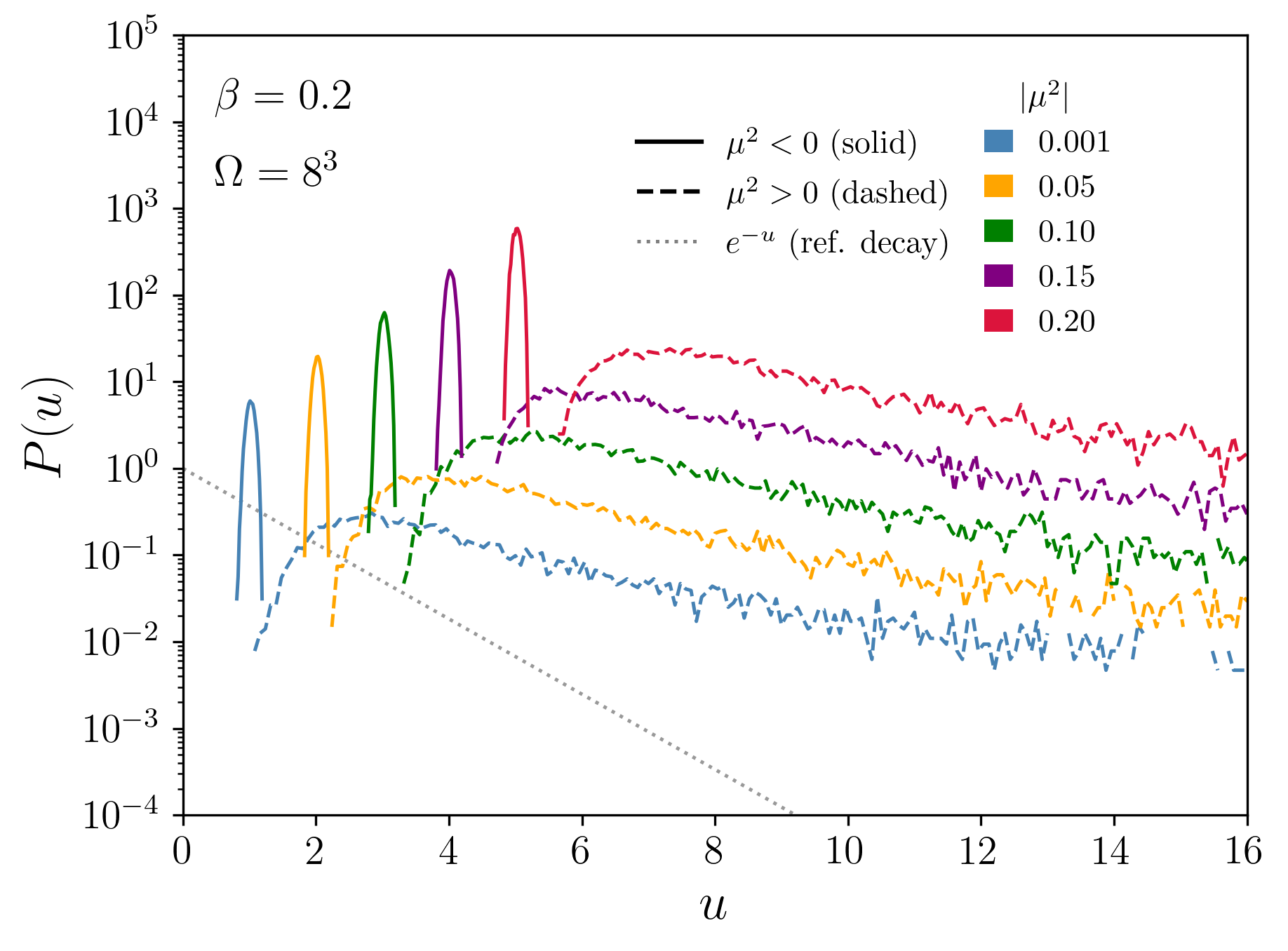} 

\includegraphics[width=0.49\columnwidth]{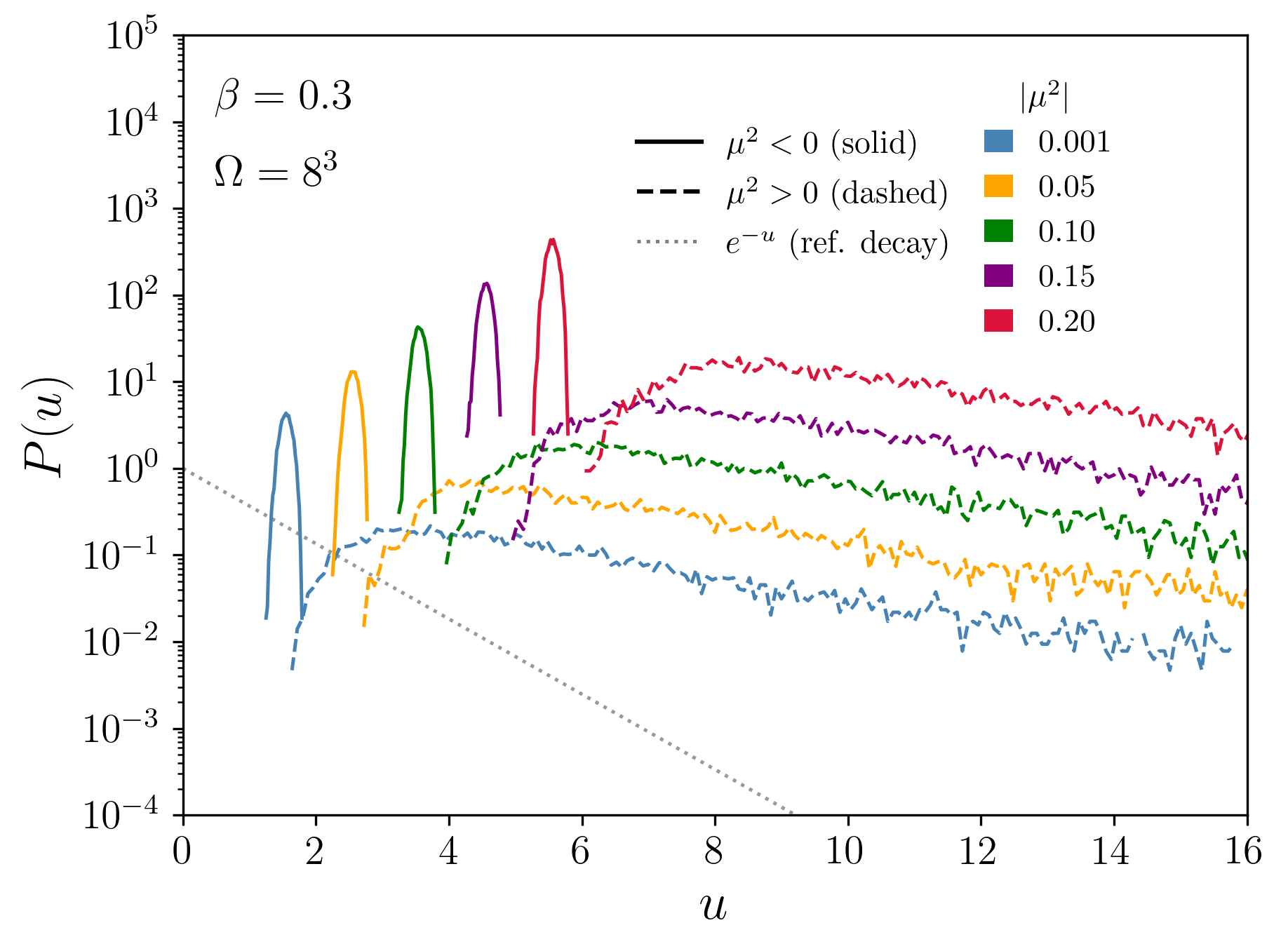} 
\includegraphics[width=0.49\columnwidth]{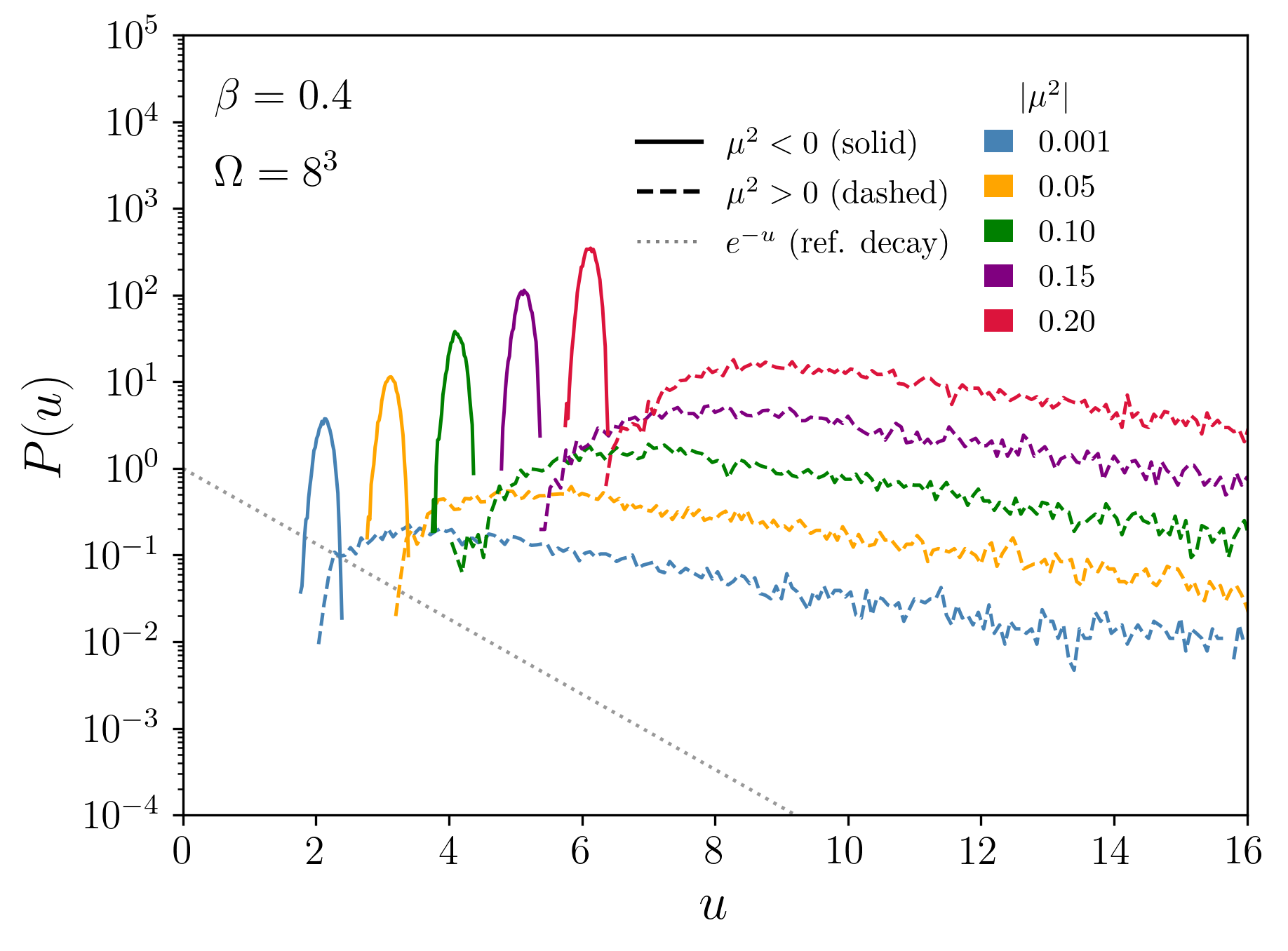} 

\includegraphics[width=0.49\columnwidth]{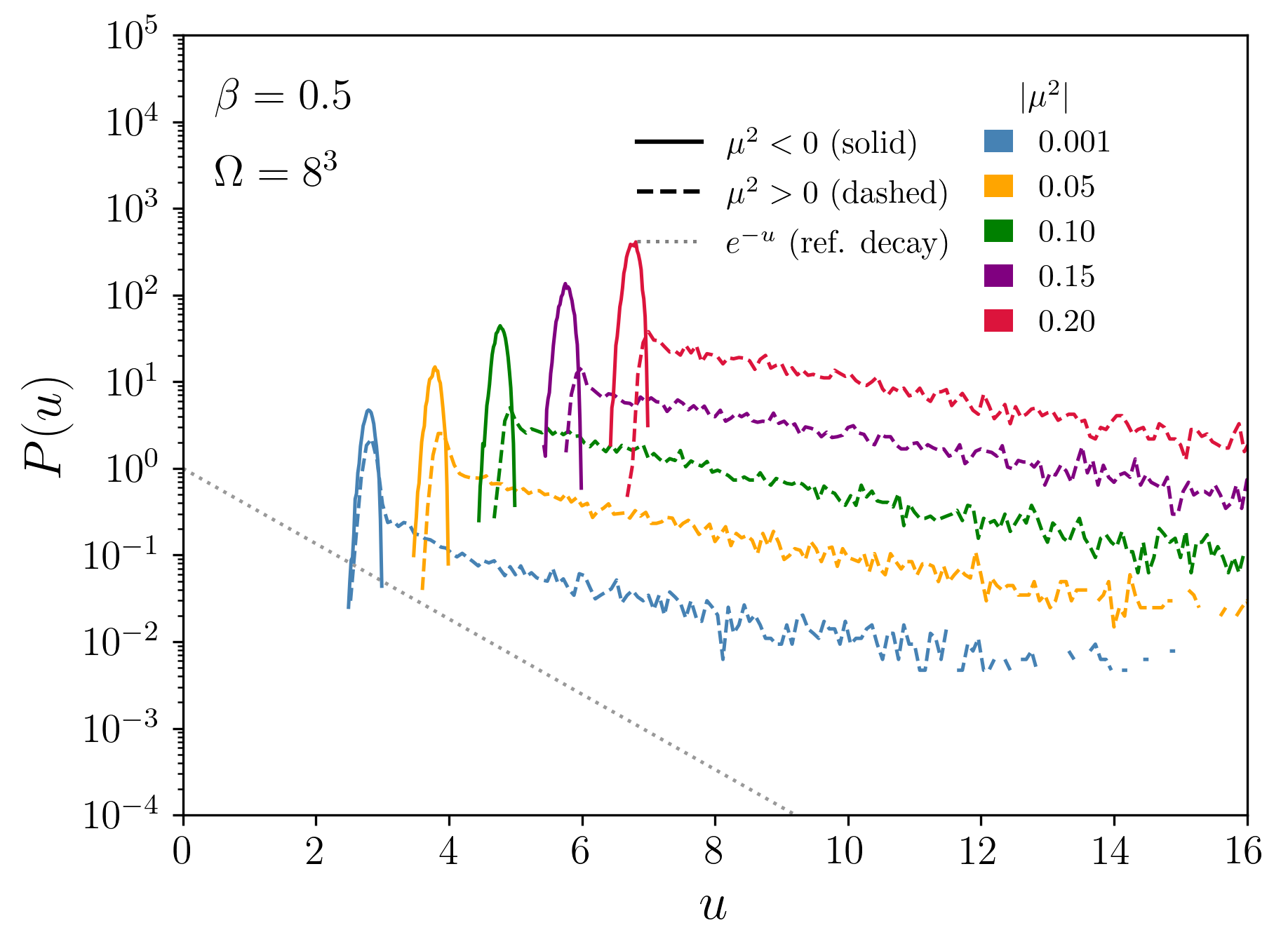} 
\includegraphics[width=0.49\columnwidth]{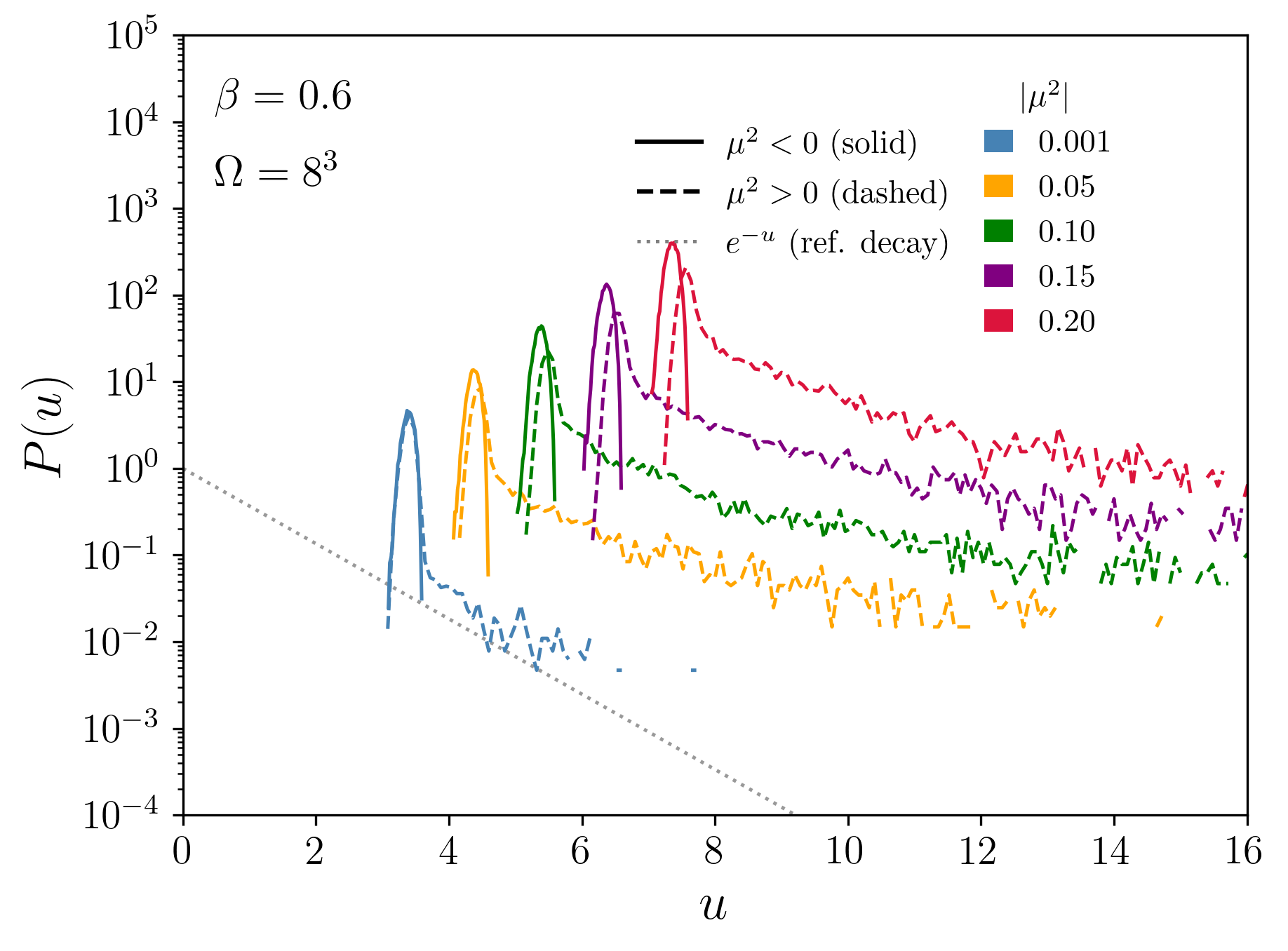} 

\caption{Probability distribution $P(u)$ of the drift magnitude $u = \max |K_x|$ for $\beta = 0.1$ -- $0.6$ on an $8^3$ lattice. 
Solid lines: imaginary chemical potential ($\mu^2 < 0$, real Langevin); dashed lines: real chemical potential ($\mu^2 > 0$, complex Langevin); dotted gray line: $e^{- u}$, a guide to the eye. 
Five values of $|\mu^2| \in \{0.001, 0.05, 0.10, 0.15, 0.20\}$ are shown. 
Curves for different $|\mu^2|$ are shifted for visibility (except $|\mu^2| = 0.001$).} 
\label{fig:decay_drifts_beta_0p4_0p2} 
\end{figure*} 

\begin{figure*}[htbp] 
\centering 
\includegraphics[width=0.49\columnwidth]{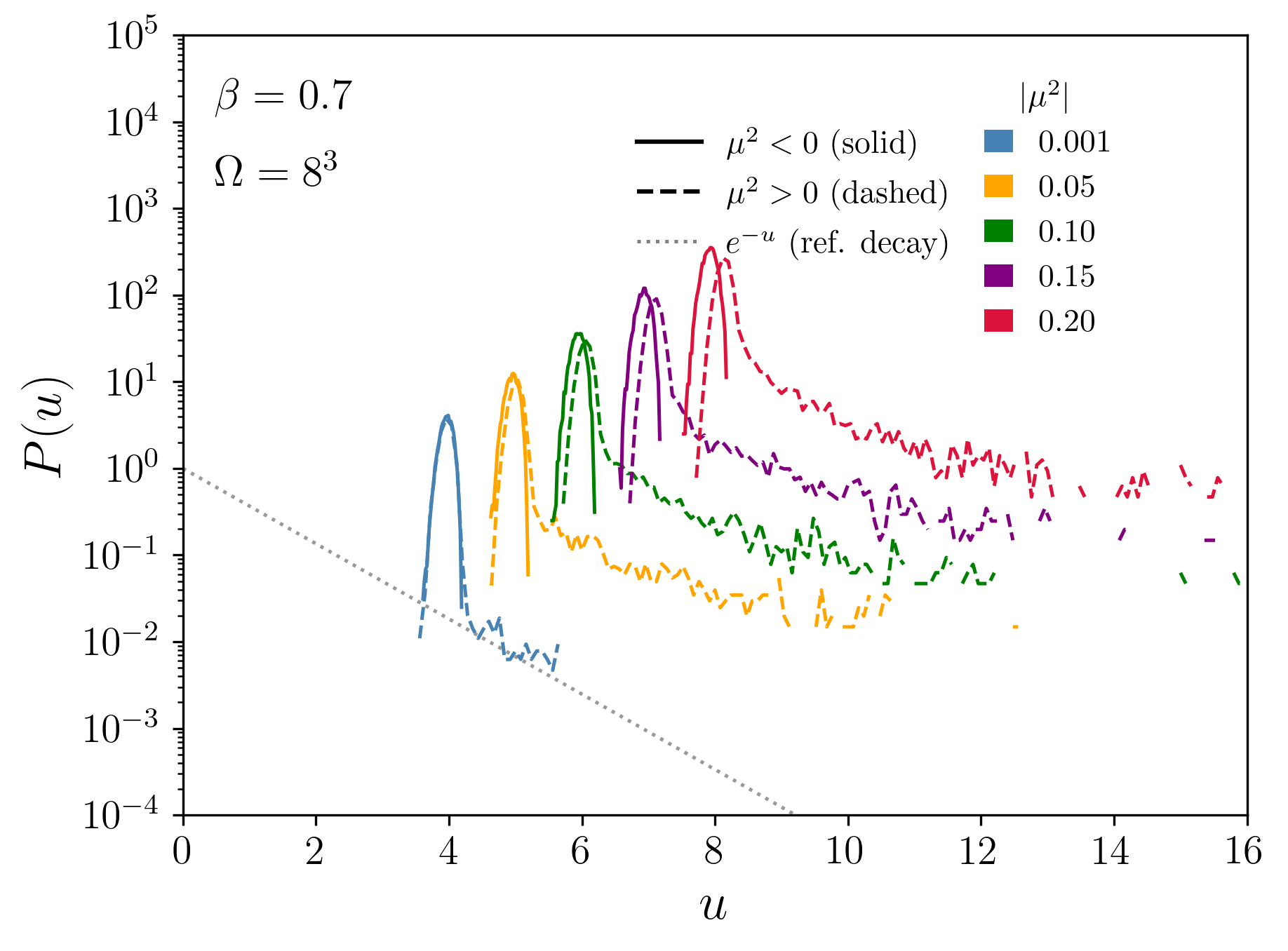} 
\includegraphics[width=0.49\columnwidth]{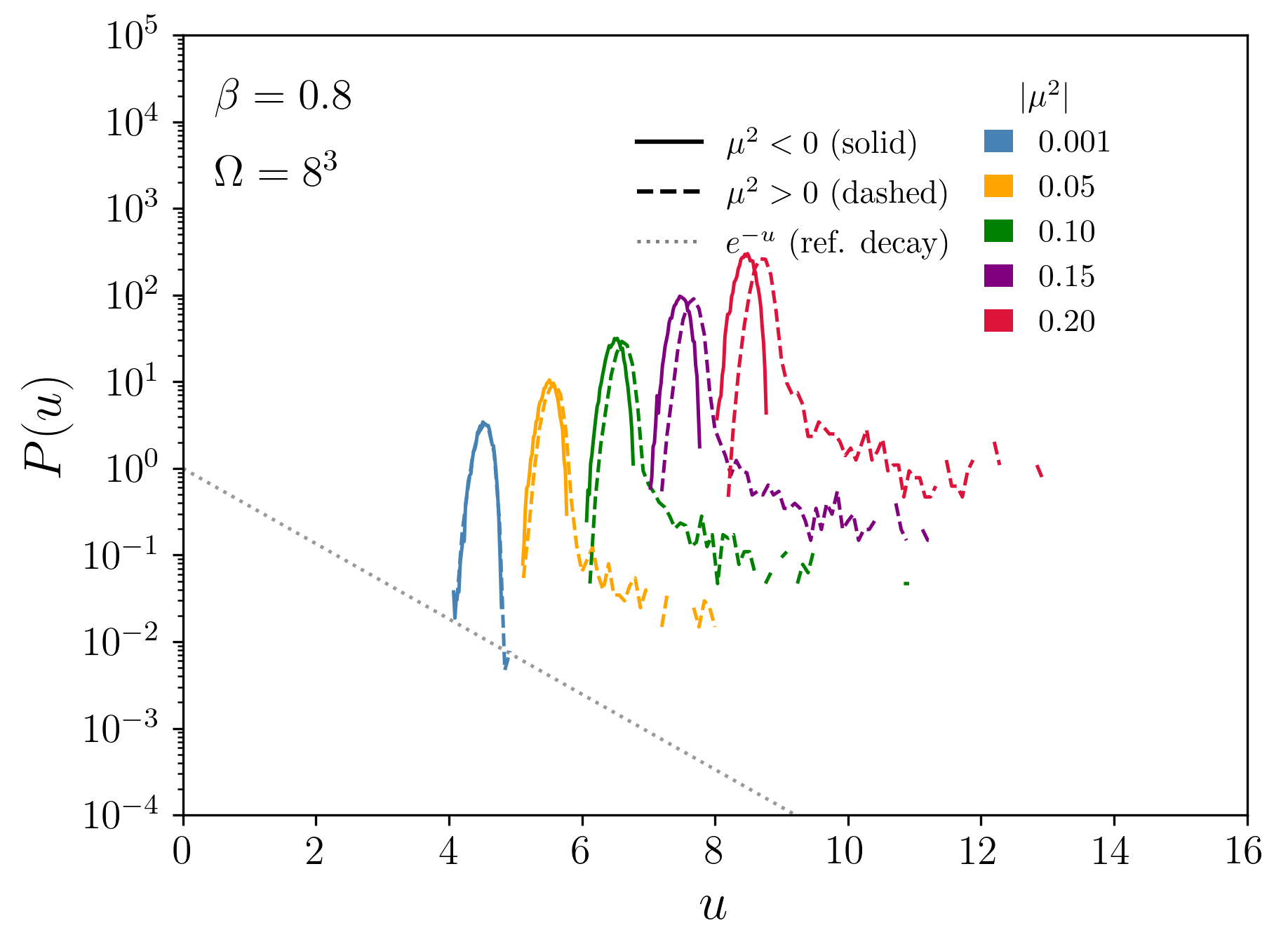} 

\includegraphics[width=0.49\columnwidth]{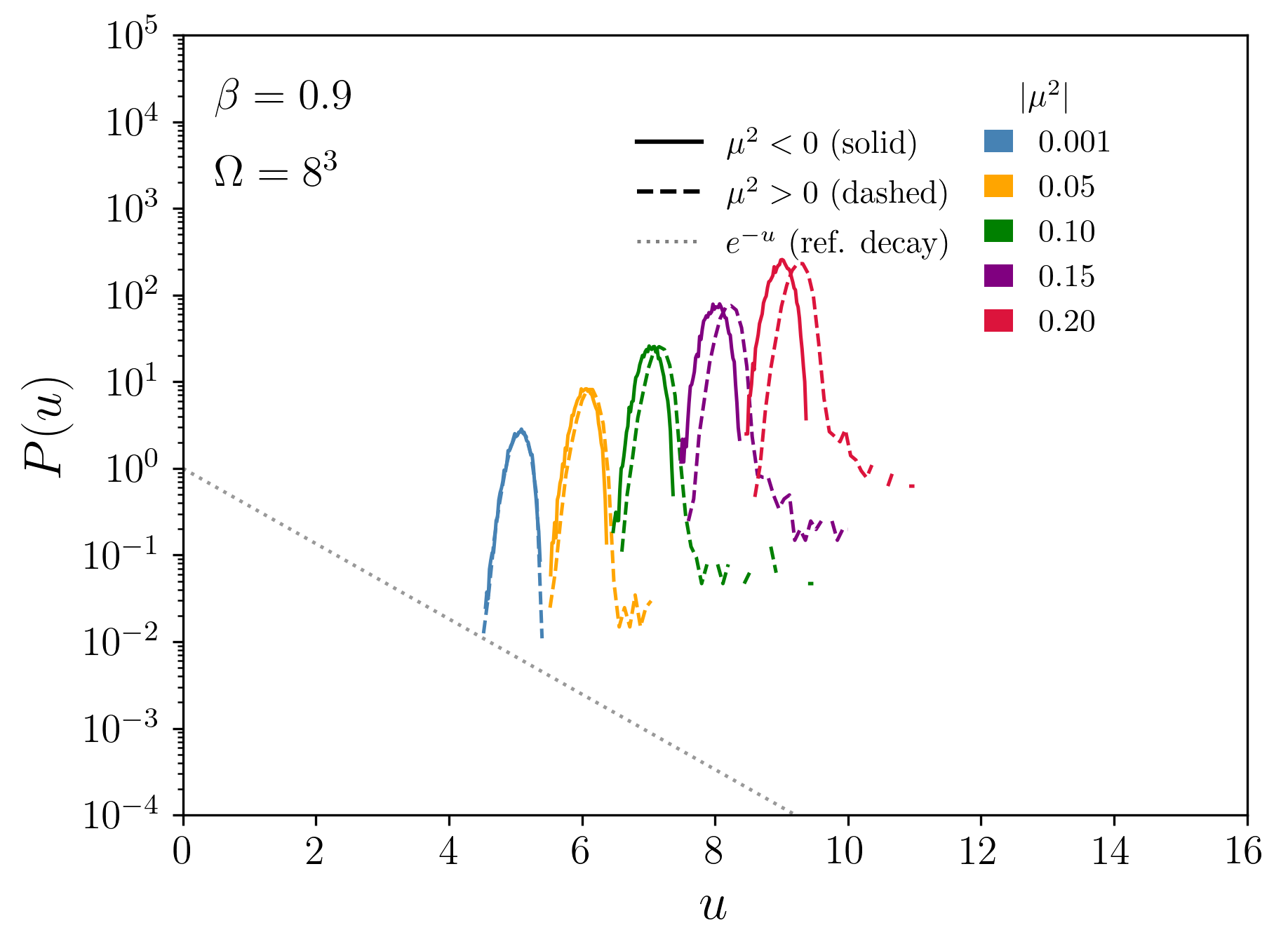} 
\includegraphics[width=0.49\columnwidth]{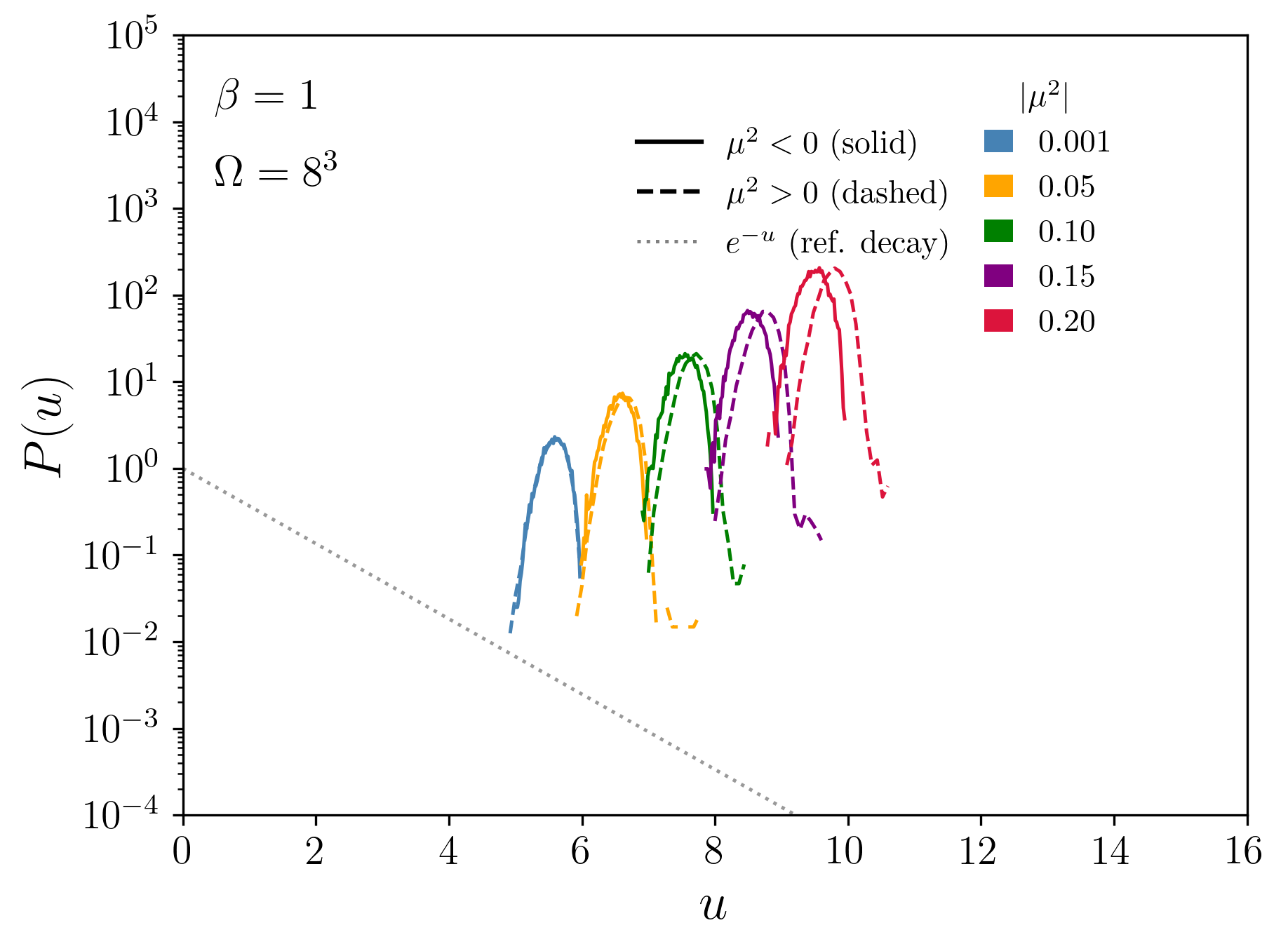} 

\caption{Probability distribution $P(u)$ of the drift magnitude $u = \max |K_x|$, same as Fig. \ref{fig:decay_drifts_beta_0p4_0p2} but for $\beta = 0.7$ -- $1.0$ (ordered phase).} 
\label{fig:decay_drifts_beta_0p7_0p5} 
\end{figure*} 

The central requirement is that the probability distribution of the drift magnitude must decay at least exponentially for large values.
This condition ensures that boundary terms vanish in the integration-by-parts steps that underlie the derivation of the complex Fokker--Planck equation.
If the decay is slower than exponential, these boundary terms can become non-negligible.
The formal justification of the stochastic process is then invalid, and the process may converge towards an incorrect distribution.

To implement this criterion, we define the drift magnitude as
\begin{equation} 
u \equiv \max_{x \in \Lambda} \sqrt{ \bigl( K_x^{\rm R} \bigr)^2 + \bigl( K_x^{\rm I} \bigr)^2}, 
\end{equation} 
where $K_x^{\rm R}$ and $K_x^{\rm I}$ denote the real and imaginary parts of the drift at lattice site $x$, and the maximum is taken over the entire lattice $\Lambda$ of volume $\Omega$.
The probability distribution $P(u)$ is then measured directly from the Langevin time history.

The correctness criterion can be stated as
\begin{equation} 
P(u) \lesssim e^{- c u} \qquad (u \to \infty; c > 0), 
\label{eq:drift_decay} 
\end{equation} 
i.e., the tail of the distribution must be suppressed at least exponentially. Histograms over a finite range of $u$ cannot establish this asymptotic behavior; we use them as indicators.
The criterion requires exponential decay with some $c > 0$; the value of $c$ depends on the normalization of $u$ and has no significance by itself.
On a semi-logarithmic plot over a limited range of $u$, an exponential with a small slope and a power law are hard to distinguish.
We therefore use the qualitative change of the tail, compared with real Langevin dynamics at the same coupling, as the indicator, and show $e^{-u}$ only as a guide to the eye.

Figs.~\ref{fig:decay_drifts_beta_0p4_0p2} and~\ref{fig:decay_drifts_beta_0p7_0p5} show the probability distributions $P(u)$.
Fig.~\ref{fig:decay_drifts_beta_0p4_0p2} covers $\beta = 0.1$ -- $0.6$, which spans the disordered and near-transition regimes.
Fig. \ref{fig:decay_drifts_beta_0p7_0p5} covers $\beta = 0.7$ -- $1.0$, the ordered phase.
Each panel displays five values of $|\mu^2|$, from $0.001$ (blue) to $0.20$ (red).
Solid lines correspond to imaginary chemical potential ($\mu^2 < 0$, real Langevin), dashed lines to real chemical potential ($\mu^2 > 0$, complex Langevin), and the dotted gray line shows $e^{-u}$ as a guide to the eye.
For visibility, the curves for $|\mu^2| = 0.05$, $0.10$, $0.15$ and $0.20$ are shifted to the right by $\Delta u = 1$, $2$, $3$ and $4$ and multiplied by $10^{1/2}$, $10$, $10^{3/2}$ and $10^2$.
This leaves the shape of each distribution, and hence its fall-off, unchanged.
However, positions and heights relative to the reference line carry no meaning.
We emphasize that the curves for $|\mu^2| = 0.001$ are not shifted and still show the actual values of $u$.
They can therefore be used to compare the behavior as $\beta$ changes.

For $\mu^2 < 0$ (solid lines), the distributions are sharply peaked and fall off steeply after the peak in all panels.
They remain consistent with exponential suppression across the full range of $\beta$ studied.
The peak position of the unshifted $|\mu^2| = 0.001$ curves grows with $\beta$, from $u \sim 0.5$ at $\beta = 0.1$ to $u \sim 5.6$ at $\beta = 1.0$.
This reflects the growth of the typical drift magnitude as the coupling increases.
At fixed $\beta$, the peak of the real-Langevin distributions depends only weakly on $\mu^2$.
Over the sampled range, this behavior is consistent with the drift criterion~\eqref{eq:drift_decay} for all values of $\beta$, as expected for a real action.

At $\mu^2 > 0$ (dashed lines), the picture differs between the two phases.
In the disordered phase ($\beta \leq 0.4 < \beta_c$, Fig.~\ref{fig:decay_drifts_beta_0p4_0p2}), the distributions develop broad tails that extend far beyond the range of the real-Langevin distributions at the same coupling.
At $\beta = 0.2$ -- $0.4$, they reach the end of the plotted range for all five values of $|\mu^2|$.
At $\beta = 0.1$, they extend to $u \approx 10$ for the smallest $|\mu^2|$ and to the end of the plotted range for $|\mu^2| = 0.2$.
It is in sharp contrast to the rapid fall-off of the real-Langevin distributions at the same couplings.
The tails become flatter with increasing $|\mu^2|$, in line with the deviations of the configurational temperature in the disordered phase (Sec.~\ref{sec:estimator_results}).
Over the sampled range, the drift distributions therefore do not support correct convergence in this regime.

At $\beta = 0.5$ and $0.6$ (Fig.~\ref{fig:decay_drifts_beta_0p4_0p2}), just above $\beta_c$, the dashed distributions still show long, slowly decaying tails that reach the end of the plotted range.
The drift distributions thus indicate failure there, in line with the $1$ -- $4\%$ deviation of the action density from the world-line result.
For $\beta \geq 0.7$ (Fig.~\ref{fig:decay_drifts_beta_0p7_0p5}), the dashed distributions are much narrower than in the disordered phase.
At $\beta = 0.7$ their tails still extend to $u \sim 10$ -- $14$ for $|\mu^2| \geq 0.05$; they shorten markedly for $\beta = 0.8$ and $0.9$.
At $\beta = 1.0$, the solid and dashed distributions become nearly indistinguishable for all five values of $|\mu^2|$.
The complex Langevin process then samples a drift distribution that is nearly the same as that of real Langevin dynamics.
This is consistent with ${\rm Re} \, \langle \beta_{\rm config} \rangle$ being at its target value at this coupling.
At $\beta = 0.7$ -- $0.9$, however, the dashed distributions retain tails beyond the real-Langevin range, although the complex Langevin action density agrees with the world-line result within $0.3\%$ (Sec.~\ref{sec:sim_details_results}).
Either these tails decay exponentially with a small slope, which the criterion allows, or the drift criterion is more conservative than necessary here; our data do not distinguish the two.

\subsection{Comparing the two correctness criteria}
\label{sec:compare_correctness}

\begin{table}[htbp]
\caption{Summary of the correctness tests at real chemical potential, $0 < \mu^2 \leq 0.2$, on the $8^3$ lattice. Phase factor: $\langle e^{i\varphi} \rangle_{\rm pq}$ at $\mu^2 = 0.2$ (Fig.~\ref{fig:pq_vs_musq}). World line: deviation of the complex Langevin action density from the world-line result (Figs.~\ref{fig:act_den_plots_a} and \ref{fig:act_den_plots_b}). Estimator: ${\rm Re}\,\langle \beta_{\rm config} \rangle$ at $\mu^2 = 0.2$ (standard action; Figs.~\ref{fig:beta_beta_m_plots_a} and \ref{fig:beta_beta_m_plots_b}). Estimator errors in the disordered phase are indicative (Sec.~\ref{sec:lattice_estimator}). 
Drift: tails of $P(u)$ over the sampled range, compared with real Langevin (Figs.~\ref{fig:decay_drifts_beta_0p4_0p2} and \ref{fig:decay_drifts_beta_0p7_0p5}).}
\label{tab:summary}
\begin{ruledtabular}
\begin{tabular}{llllll}
$\beta$ & Phase & Phase factor & World line & Estimator & Drift \\ \hline
0.1 & disordered & $0.76$ & fails ($25$--$37\%$) & $0.884(8)$ & broad tails \\
0.2 & disordered & $\approx 0.3$ & fails ($\approx 20$--$35\%$) & $0.918(5)$ & broad tails \\
0.3 & disordered & $\approx 0.07$ & fails ($\approx 30\%$) & $0.939(5)$ & broad tails \\
0.4 & disordered & $0.01$ & fails ($\approx 12$--$22\%$) & $0.972(4)$ & broad tails \\
0.5 & near transition & $0.07$ & off by $1$--$4\%$ & at target within $0.6\%$ & slowly decaying tails \\
0.6 & near transition & $0.24$ & off by up to $\sim 1\%$ & at target within $0.6\%$ & slowly decaying tails \\
0.7 & ordered & $\approx 0.4$ & agrees ($\leq 0.3\%$) & at target & tails to $u \sim 10$--$14$ \\
0.8, 0.9 & ordered & $0.5$, $\approx 0.6$ & agrees ($\leq 0.3\%$) & at target & shorter tails \\
1.0 & ordered & $0.65$ & agrees ($\leq 0.3\%$) & at target & as real Langevin \\
\end{tabular}
\end{ruledtabular}
\end{table}

Table~\ref{tab:summary} summarizes the verdicts of the two criteria and of the world-line comparison.
In the regime $\mu^2 < 0$, both criteria indicate correct sampling, as expected for a real action, which confirms our implementation.
For $\mu^2 > 0$, both indicate failure in the disordered phase, where the world-line comparison shows large deviations.
Close to the transition, at $\beta = 0.5$ and $0.6$, the action density deviates by up to a few percent; the drift distributions show slowly decaying tails, while the estimator does not resolve the failure.
At $\beta = 0.7$ -- $0.9$, the drift distributions still have tails, although the action density agrees with the world-line result; here the drift criterion is the more conservative of the two.
At $\beta = 1.0$, all three tests agree.

The two criteria are conceptually distinct.
The drift-decay criterion tests a condition of the formal justification of complex Langevin, namely the absence of boundary terms associated with large drifts \cite{Nagata:2016vkn}.
The estimator tests a Schwinger--Dyson identity with a known target value, so its deviation can be quoted with a statistical error, and its running average can be monitored during a run.
By construction, it is insensitive to large drifts, but it is affected by its poles at small drift (Sec.~\ref{sec:lattice_estimator}).
In the disordered phase, its deviation reaches several percent, for example ${\rm Re} \, \langle \beta_{\rm config} \rangle = 0.939(5)$ at $\beta = 0.3$, $\mu^2 = 0.2$.
The estimator needs the diagonal of the Hessian.
For the XY action, this costs about as much as the drift [Eq.~\eqref{eq:XY_hessian}], but it can be considerably more expensive in models such as QCD, which warrants further study.
The two criteria are therefore complementary, but not exhaustive.
Boundary-term observables \cite{Scherzer:2018hid, Scherzer:2019lrh} and further criteria compared in Ref.~\cite{Mandl:2026vdc} probe correctness in other ways.
Our results are consistent with the boundary-term analysis of this model \cite{Scherzer:2019lrh}, which finds large boundary terms in the disordered phase and small ones in the ordered phase.

\subsection{Towards correct convergence: dynamical stabilization and other directions}
\label{sec:ds}

\begin{figure*}[htbp] 
\centering 
\includegraphics[width=0.49\columnwidth]{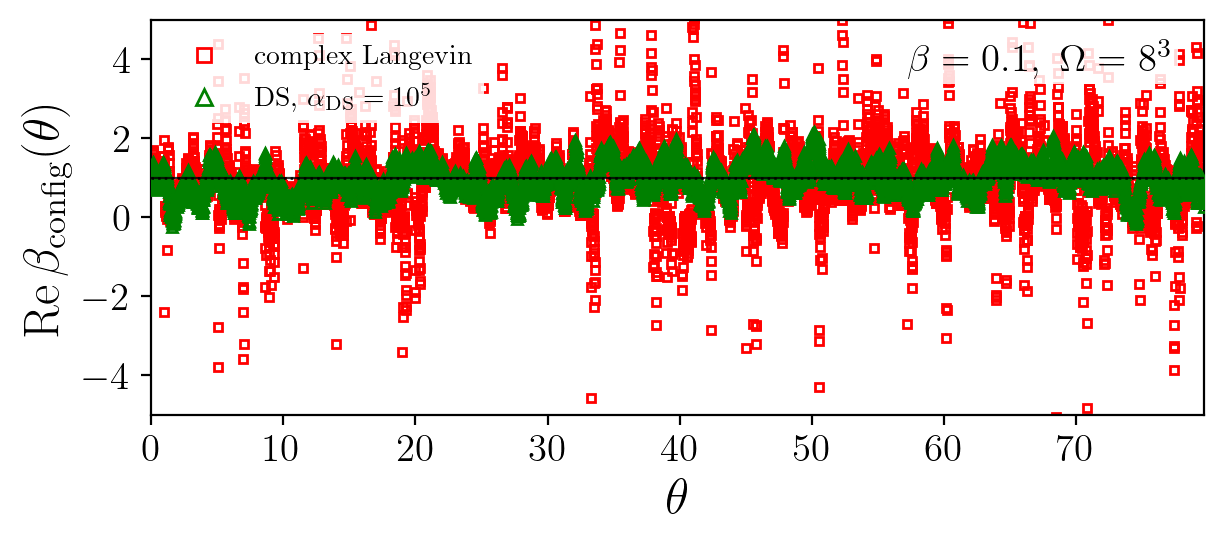} 
\includegraphics[width=0.49\columnwidth]{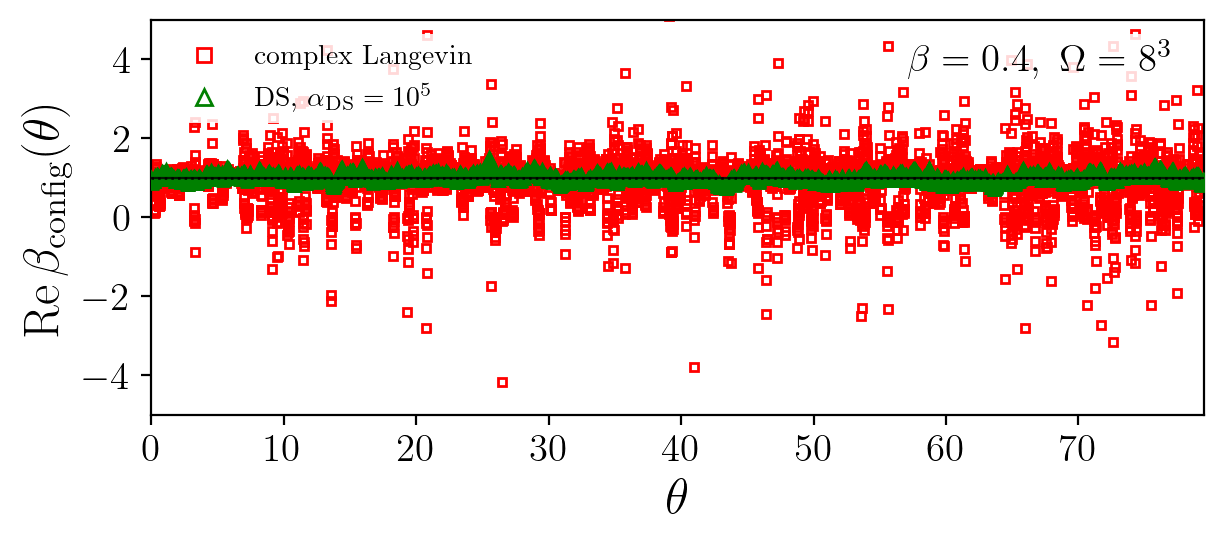} 

\includegraphics[width=0.49\columnwidth]{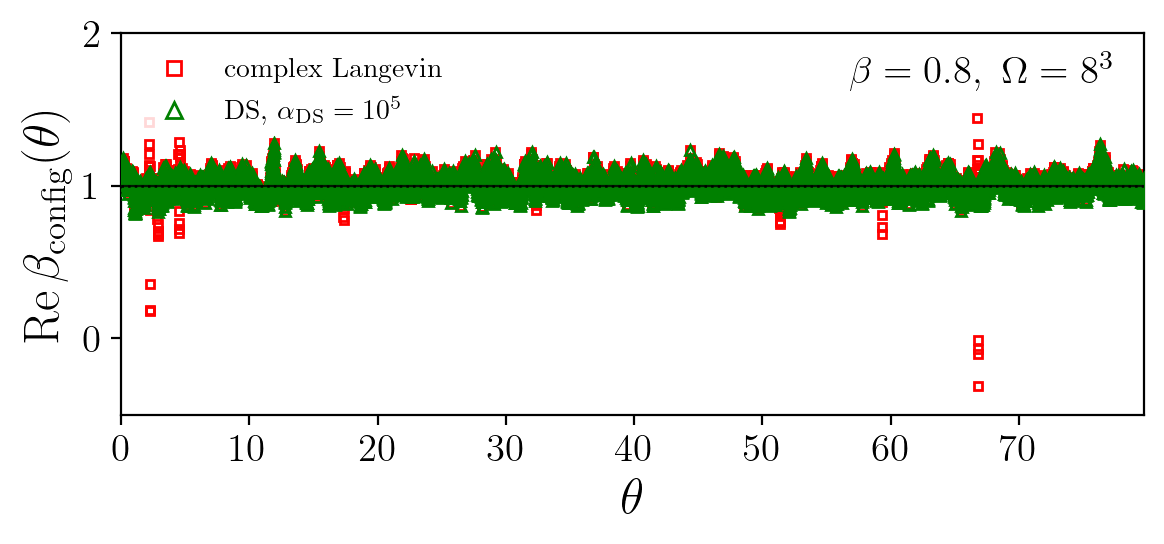} 
\caption{${\rm Re} \, \beta_{\rm config}$, Eq. \eqref{eq:beta_config_M_def}, against Langevin time $\theta$ at $\mu^2 = 0.2$ for $\beta = 0.1$ (disordered phase, mild sign problem), $0.4$ (disordered phase, severe sign problem) and, as a control, $0.8$ (ordered phase, moderate sign problem). 
Open red squares: complex Langevin without stabilization; open green triangles: with the force of Eq.~\eqref{eq:ds_force}, $p = 7$, $\alpha_{\rm DS} = 10^5$. 
The solid line marks $1$, the dotted line $1.003$. A window of the Langevin-time history is shown.} \label{fig:ds_history} 
\end{figure*} 

\begin{figure*}[htbp] 
\centering 
\includegraphics[width=0.485\columnwidth]{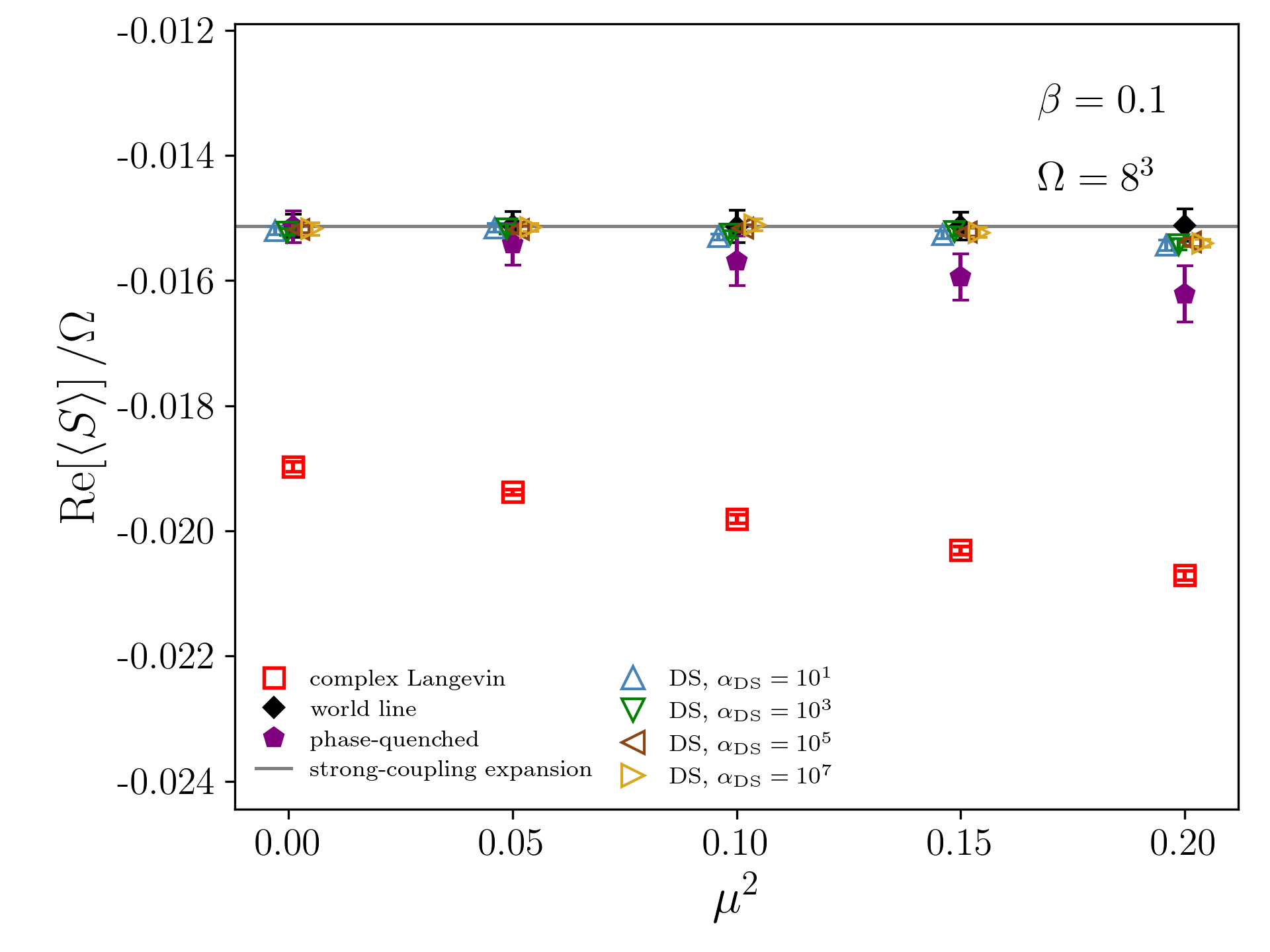} 
\includegraphics[width=0.485\columnwidth]{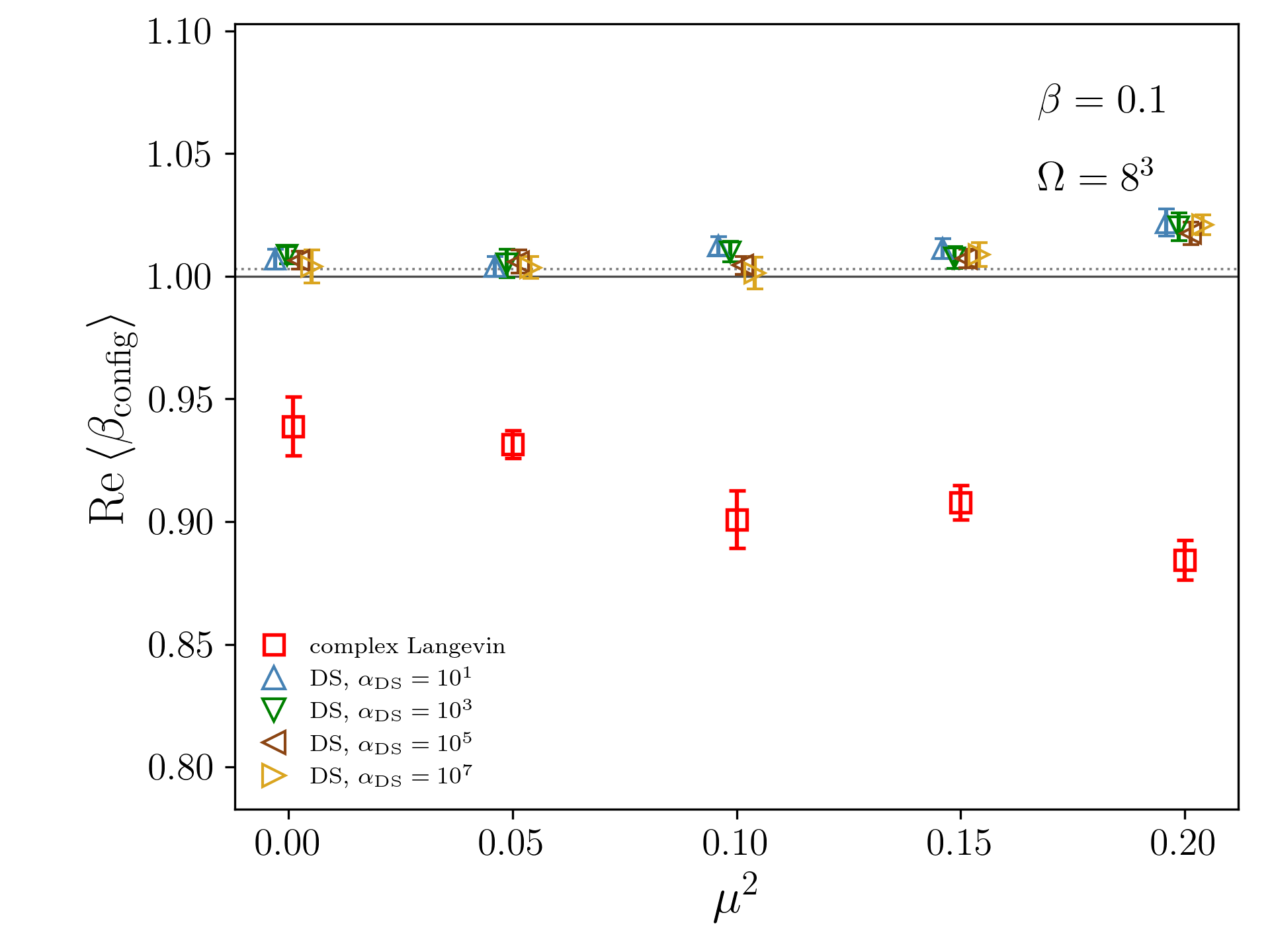} 

\includegraphics[width=0.485\columnwidth]{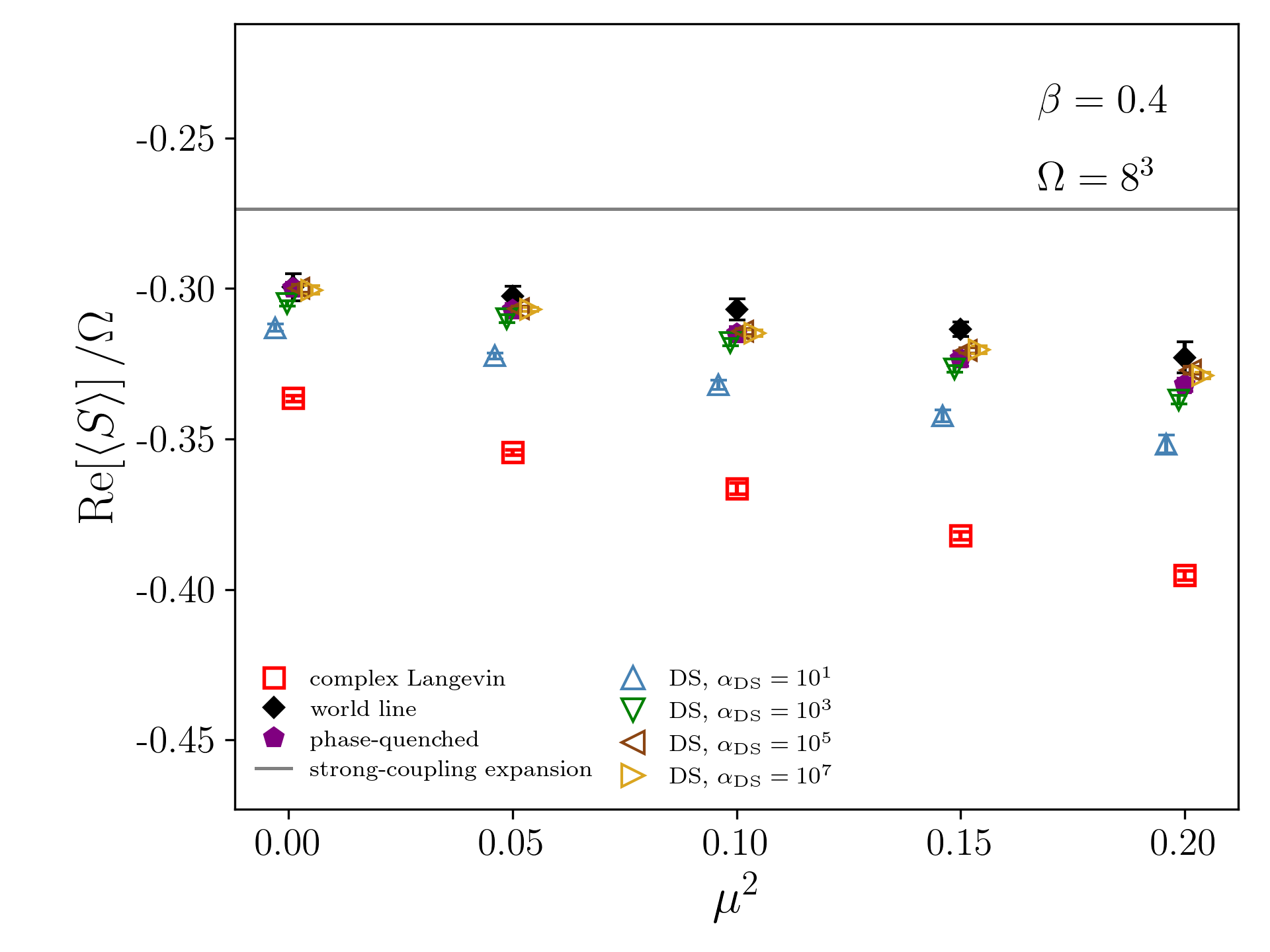} 
\includegraphics[width=0.485\columnwidth]{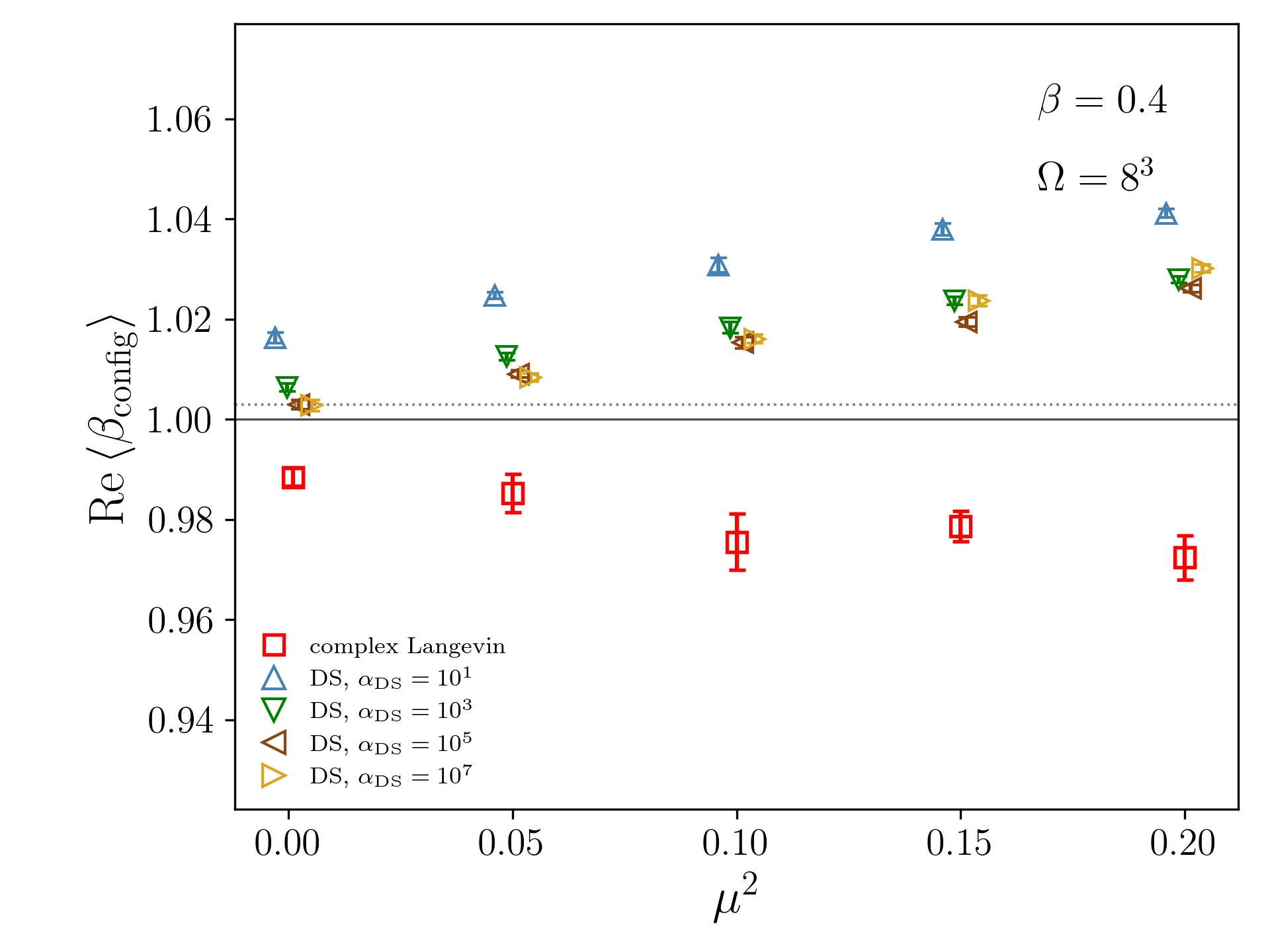} 

\includegraphics[width=0.485\columnwidth]{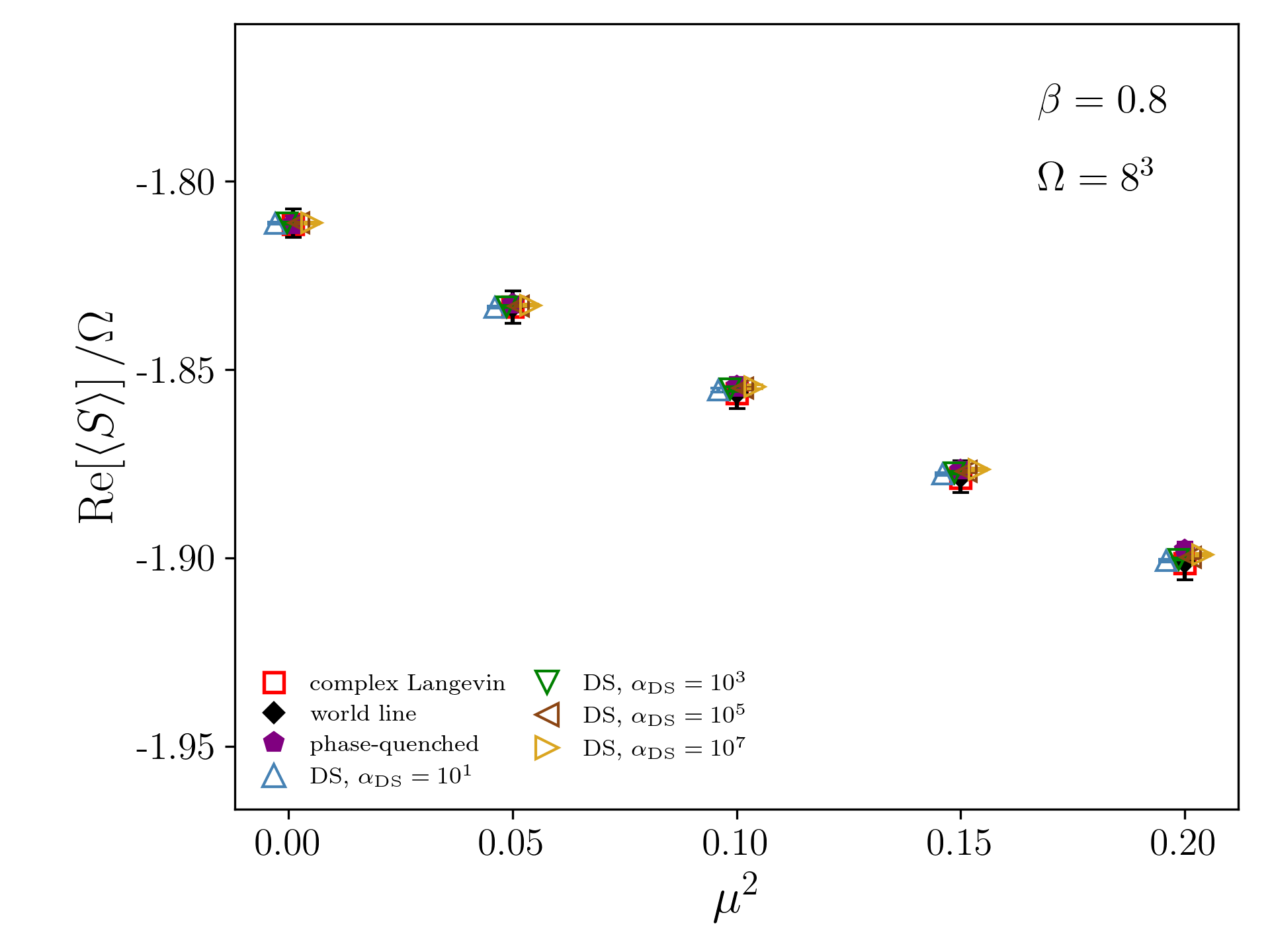} 
\includegraphics[width=0.485\columnwidth]{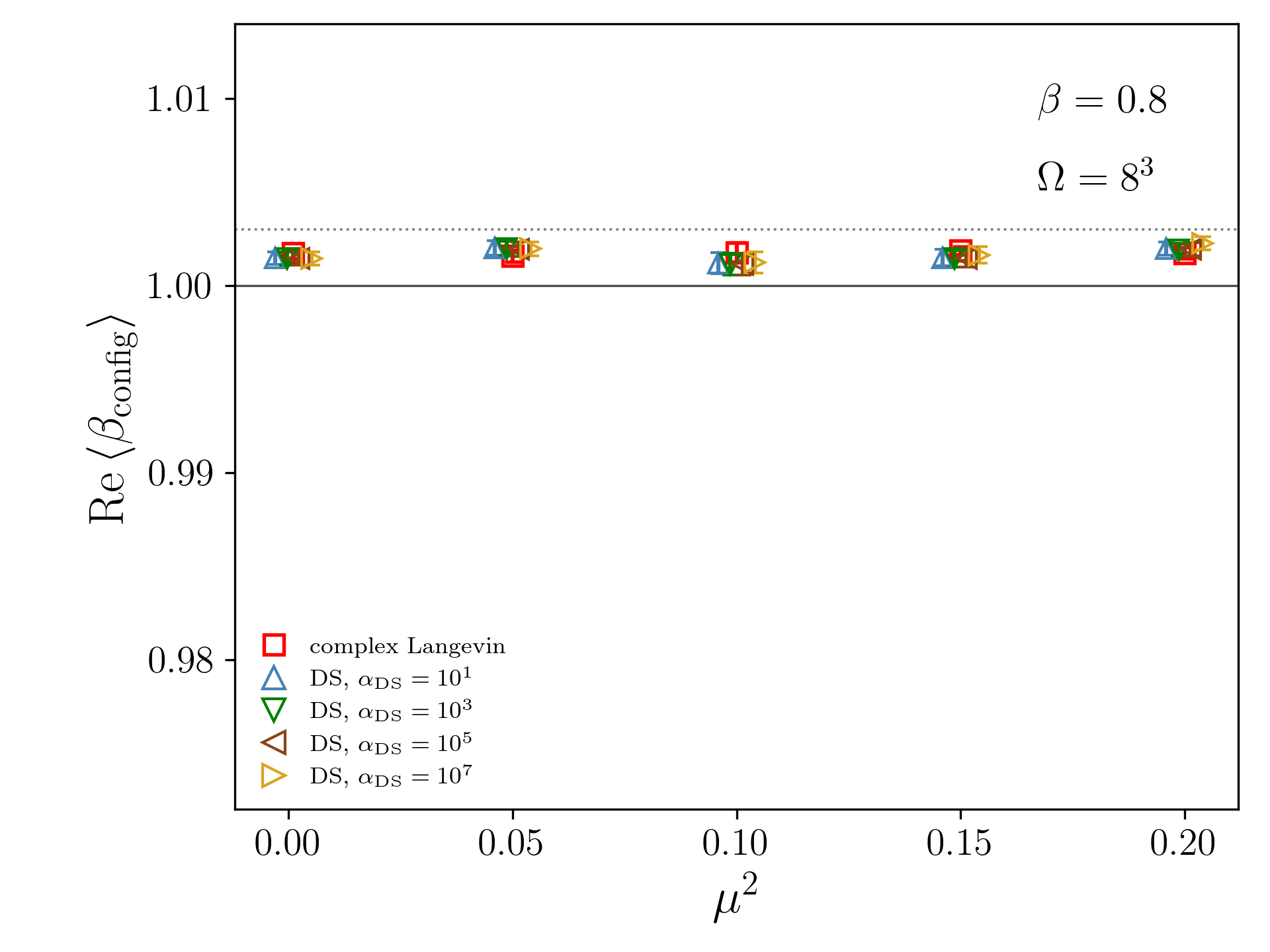} 
\caption{${\rm Re}[\langle S \rangle]/\Omega$ (left column) and ${\rm Re} \, \langle \beta_{\rm config} \rangle$ (right column) versus $\mu^2$ with dynamical stabilization, Eq. \eqref{eq:ds_force} with $p = 7$, for $\beta = 0.1$ (disordered phase, mild sign problem), $0.4$ (disordered phase, severe sign problem) and, as a control, $0.8$ (ordered phase, moderate sign problem). 
Open red squares: complex Langevin without stabilization (filled squares in Figs. \ref{fig:act_den_plots_a} and \ref{fig:beta_beta_m_plots_a}); open triangles up, down, left and right: $\alpha_{\rm DS} = 10$, $10^3$, $10^5$ and $10^7$, displaced horizontally for visibility; black diamonds: world-line result; purple pentagons: phase-quenched ensemble; gray line: strong-coupling expansion ($\beta \leq 0.4$). 
In the right column, the solid line marks $1$ and the dotted line $1.003$.} 
\label{fig:ds_vs_musq} 
\label{fig:ds_vs_musq_beta} 
\end{figure*} 

\begin{figure*}[htbp] 
\centering 
\includegraphics[width=0.49\columnwidth]{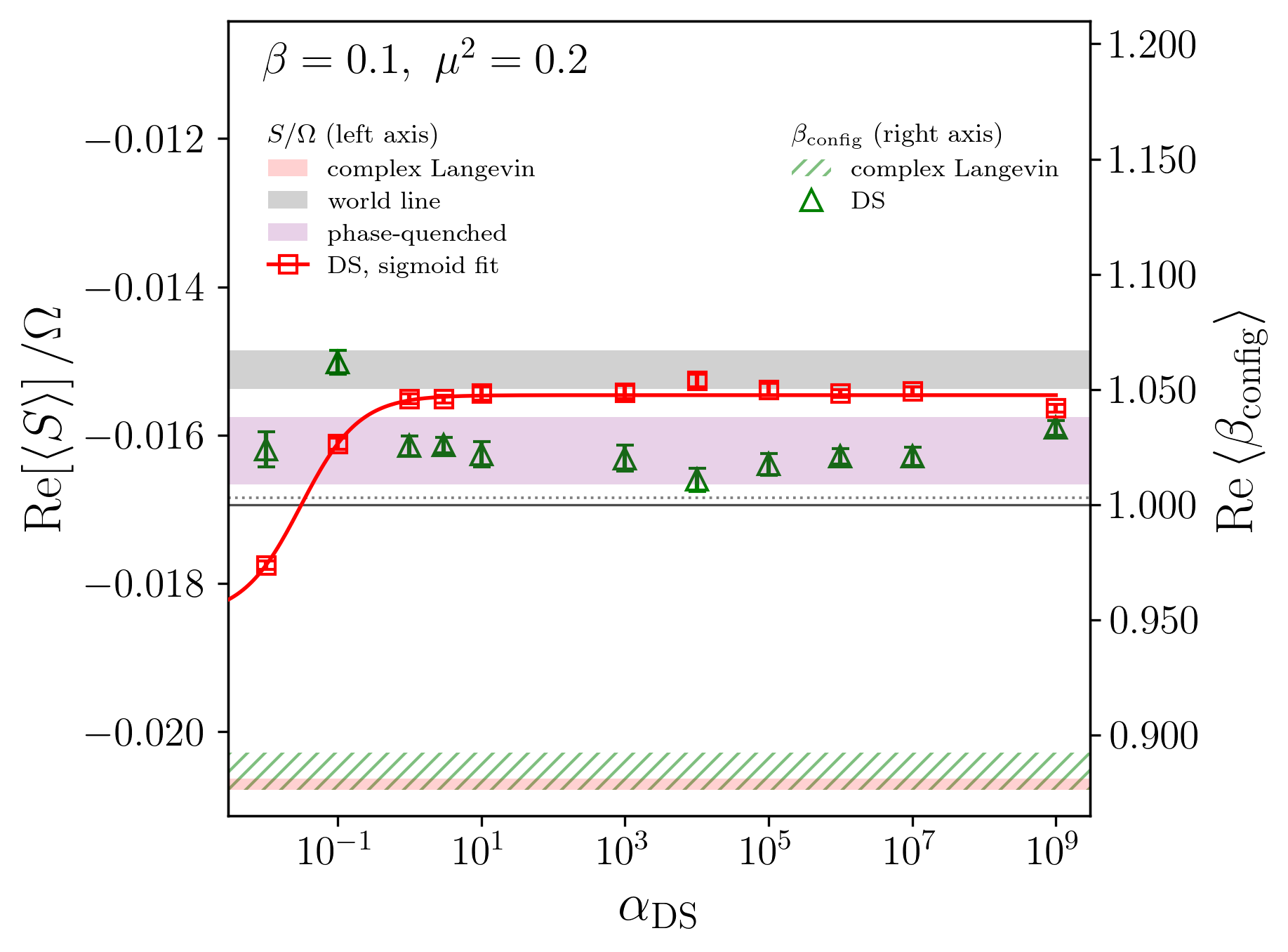} 
\includegraphics[width=0.49\columnwidth]{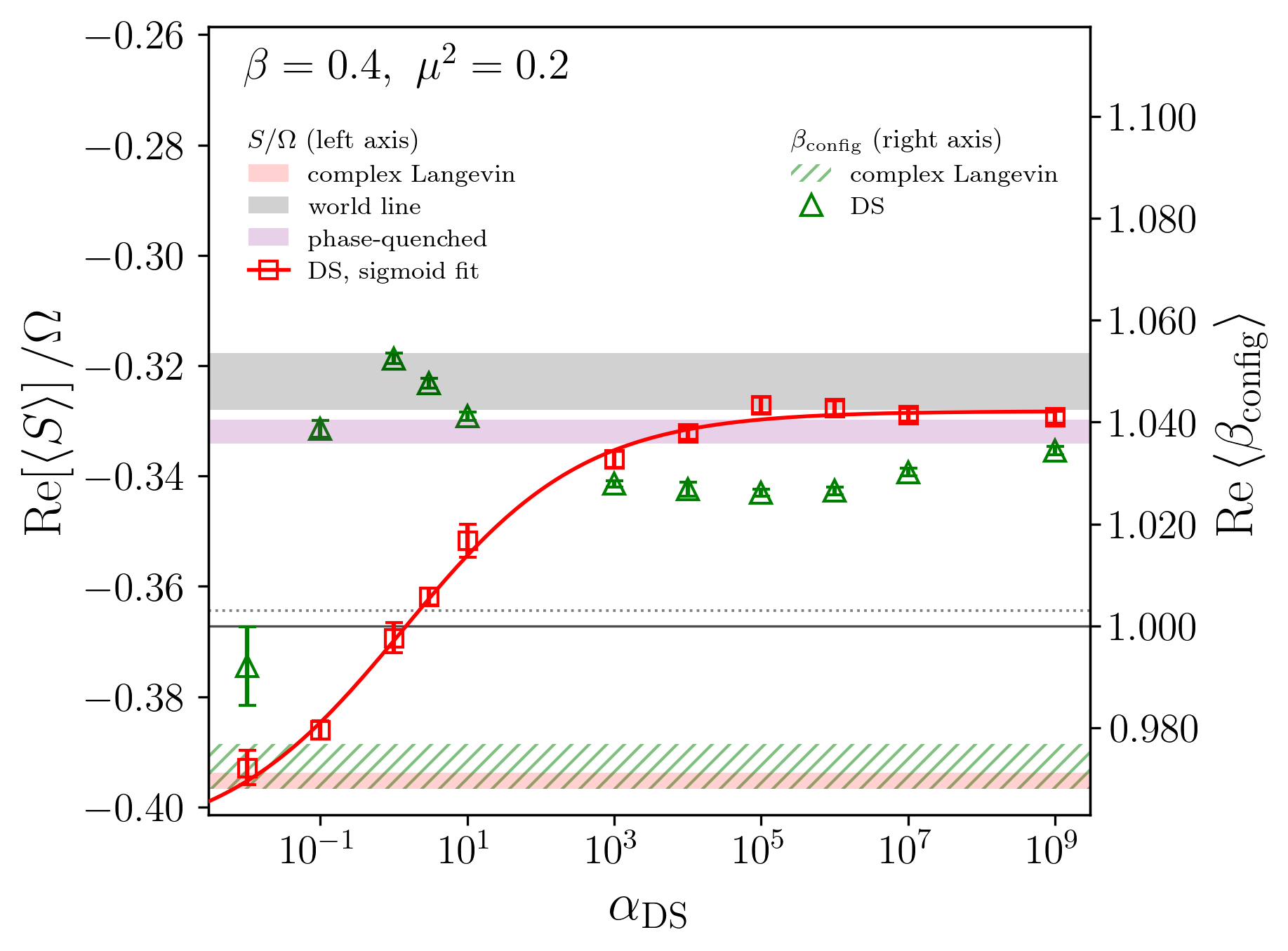} 

\includegraphics[width=0.49\columnwidth]{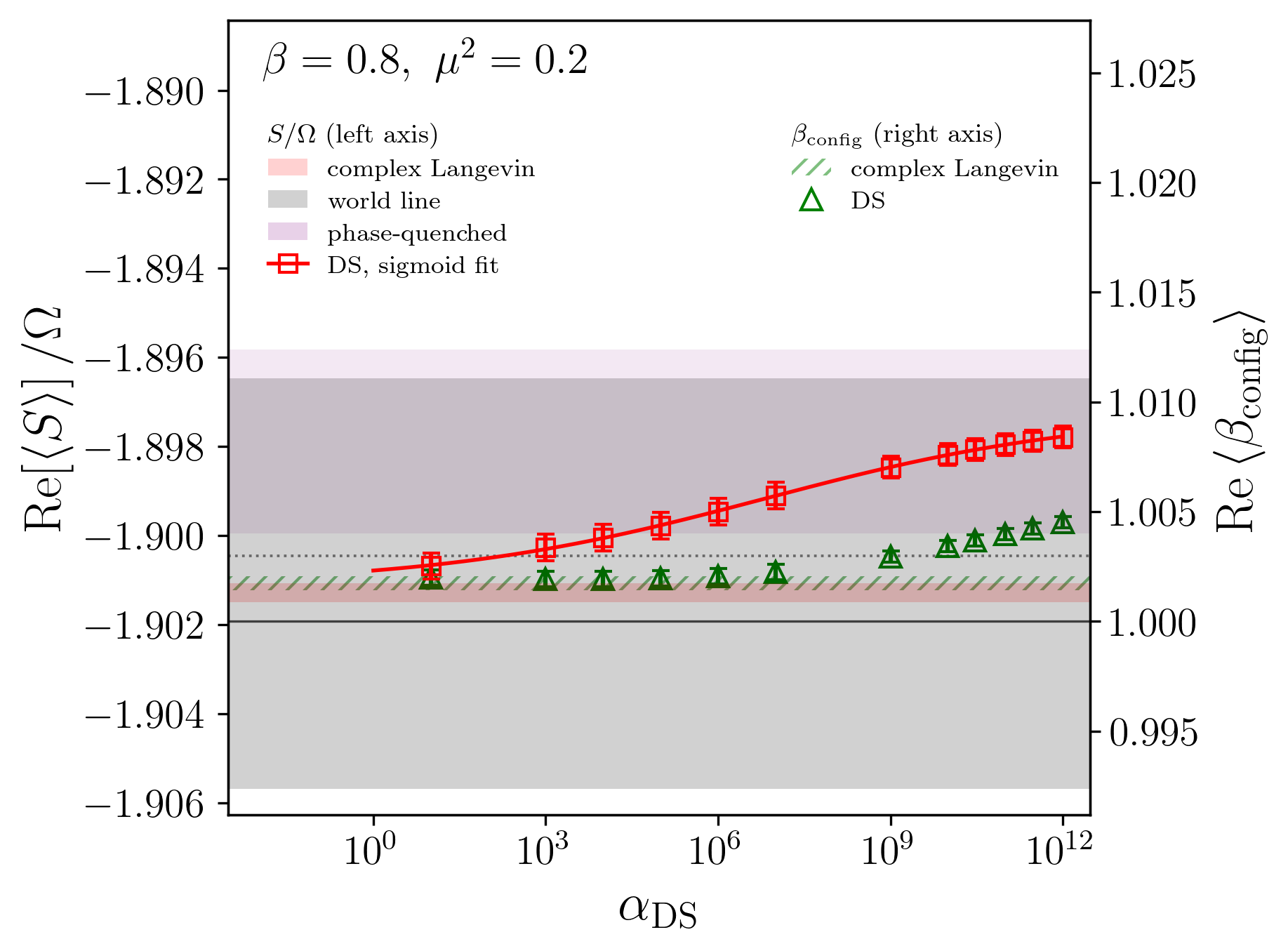} 
\caption{${\rm Re}[\langle S \rangle] / \Omega$ (left axis) and ${\rm Re} \, \langle \beta_{\rm config} \rangle$ (right axis) versus $\alpha_{\rm DS}$ at $\mu^2 = 0.2$, for $p = 7$ and $\beta = 0.1$ (disordered phase, mild sign problem), $0.4$ (disordered phase, severe sign problem) and, as a control, $0.8$ (ordered phase, moderate sign problem). 
Bands show the values and error of the world-line result (gray), of the phase-quenched ensemble (purple) and of complex Langevin without stabilization (red) for ${\rm Re}[\langle S \rangle]/\Omega$, and of complex Langevin without stabilization (hatched, green) for ${\rm Re} \, \langle \beta_{\rm config} \rangle$. 
Open red squares: stabilized ${\rm Re}[\langle S \rangle]/\Omega$, with sigmoid fit (red line); open green triangles: stabilized ${\rm Re} \, \langle \beta_{\rm config} \rangle$. 
On the right axis, the solid line marks $1$ and the dotted line $1.003$. 
At $\beta = 0.8$, the world-line and phase-quenched bands overlap.} 
\label{fig:ds_vs_alpha} 
\end{figure*} 

The failure of complex Langevin in the disordered phase raises two questions.
Does a modified Langevin process sample the correct distribution there?
Moreover, can the configurational temperature estimator distinguish this correct distribution?
Using the configurational estimator and the world-line results we can test the proposed modifications.  

Here we test one of them and list the others. 
We test the {\it dynamical stabilization} (DS) \cite{Attanasio:2018rtq}, which adds a force that pulls the fields towards the real manifold. 
The idea is to add a term $- \alpha_{\rm DS} (\phi_x^{\rm I})^p$ to the right-hand side of the Langevin equation for $\phi_x^{\rm I}$, Eqs. \eqref{eq:eqphic}.
The complex drift $-K_x$ becomes $-K_x - K^{\rm DS}_x$, where
\begin{equation} 
K^{\rm DS}_x = i \, \alpha_{\rm DS} \, (\phi_x^{\rm I})^p. 
\label{eq:ds_force} 
\end{equation} 
The power $p$ is odd, so that the force points towards $\phi_x^{\rm I} = 0$ for either sign of $\phi_x^{\rm I}$.
We use $p = 7$, as in Ref. \cite{Attanasio:2017rxk}.
Ref. \cite{Attanasio:2017rxk} tested this force for the XY model and found that it improved the agreement with the world-line result at $\beta = 0.2$, $\mu^2 = 0.2$.
A similar force is used in QCD; see also Ref. \cite{Hansen:2024lkn}.
The stabilizing force is not holomorphic.
The argument of Sec. \ref{sec:Complex_Langevin_Dynamics} therefore does not cover it, and correctness has to be tested a posteriori.
For a holomorphic observable $O$, stationarity of the stabilized process implies $\langle L_{\rm DS} O \rangle = 0$, where $L_{\rm DS} O = \sum_x \big[ \partial^2 O / \partial \phi_x^2 - (K_x + K^{\rm DS}_x) \, \partial O / \partial \phi_x \big]$, with $O$ regarded as a function of the complexified fields, provided there are no boundary terms \cite{Aarts:2011ax}.
For $O = S$, this gives, in place of Eq.~\eqref{eq:ds_trH},
\begin{equation} 
\left\langle \sum_x H_{xx} \right\rangle - \left\langle \sum_x K_x^2 \right\rangle = \left\langle \sum_x K_x K^{\rm DS}_x \right\rangle.  
\label{eq:ds_trH_ds} 
\end{equation} 
A nonzero right-hand side means that the stabilized distribution violates the identity $\langle \sum_x K_x^2 \rangle = -2 \langle S \rangle$ of the physical theory, so that $\langle S \rangle$, $\langle \sum_x K_x^2 \rangle$, or both are biased, unless boundary terms compensate.
The estimator then departs from its target by about $\langle \sum_x K_x K^{\rm DS}_x \rangle / \langle \sum_x K_x^2 \rangle$.
We find this ratio to be less than $3\%$ for $\alpha_{\rm DS} = 10^3$ -- $10^7$.
We quote the estimator without subtracting this term, so that its deviation from the target measures the violation of the physical Schwinger--Dyson identity by the stabilized process.
A direct, pole-free test of this violation would compare $\langle \sum_x K_x^2 \rangle$ in the stabilized ensemble with $-2 \langle S \rangle$ from the world-line results; we leave this for future work.

We simulated the standard action (that is, the action without the stabilization term) at three couplings, all at $\mu^2 = 0.2$: $\beta = 0.1$ (disordered phase, mild sign problem, $\langle e^{i \varphi} \rangle_{\rm pq} \sim 0.76$), $\beta = 0.4$ (disordered phase, severe sign problem, $\langle e^{i \varphi} \rangle_{\rm pq} \sim 0.01$) and, as a control, $\beta = 0.8$ (ordered phase, moderate sign problem, $\langle e^{i \varphi} \rangle_{\rm pq} \sim 0.5$).
For the complex Langevin simulations with the stabilized action, we used $\mu^2 = 0.001$ -- $0.2$, $p = 7$ and $\alpha_{\rm DS}$ between $0.01$ and $10^9$, and for $\beta = 0.8$, $\alpha_{\rm DS}$ up to $10^{12}$.
The run details are available in Ref. \cite{Joseph:2026data}.

Fig. \ref{fig:ds_history} shows the Langevin-time histories of ${\rm Re} \, \beta_{\rm config}$ at $\mu^2 = 0.2$, without and with stabilization.
The force (given by Eq.~\eqref{eq:ds_force})  removes large deviations (from unity) of the estimator values.
Without the force, a single measurement can reach about $\pm 10$ and beyond at $\beta = 0.1$ and $0.4$, whereas with the force, it stays close to the target value. 
At $\beta = 0.8$, both histories stay close to the target value. 
In these histories, the variance of ${\rm Re} \, \beta_{\rm config}$ over a run drops by one to two orders of magnitude at $\beta = 0.1$ and $0.4$.
We note that the large deviations of the estimator from unity originate in the poles of the estimator, where its denominator $\sum_x K_x^2$ vanishes. 
In the disordered phase, the complex Langevin process can come very close to the poles. 

Fig.~\ref{fig:ds_vs_musq} shows the action density and the estimator versus $\mu^2$, without stabilization, and for $\alpha_{\rm DS} = 10$ -- $10^7$.
Stabilization moves the action density towards the world-line value.
It reproduces this value where the sign problem is mild, and stays close to the phase-quenched value where it is severe.
This is expected: for $\alpha_{\rm DS} \to \infty$, the fields are pinned to $\phi_x^{\rm I} = 0$, where the real part of the drift reduces to the drift of $S_{\rm pq}$ and ${\rm Re}\, S$ to $S_{\rm pq}$.
The strong-stabilization limit is therefore the phase-quenched theory.

At $\beta = 0.1$ (disordered phase, mild sign problem), the action density agrees with the world-line result for all $\alpha_{\rm DS} \geq 10^3$, within $\sim 2$ standard deviations. 
At the same time, it is $\sim 25$ -- $37\%$ below it without stabilization.
Without stabilization, the estimator lies $\sim 6$ -- $12\%$ below its target.
With stabilization, it lies close to its target at small $\mu^2$ and moves above it with increasing $\mu^2$, by up to $\sim 2\%$ at $\mu^2 = 0.2$, where the action density also deviates slightly more from the world-line value.

At $\beta = 0.4$ (disordered phase, severe sign problem), the action density stays $\sim 1$ -- $4\%$ below the world-line value for $\mu^2 \geq 0.1$, by up to $\sim 5$ standard deviations.
It agrees within $\sim 2$ standard deviations with the phase-quenched value, which lies only $\sim 3\%$ below the world-line one there.
With stabilization, the estimator lies $\sim 1$ -- $3\%$ above its target.

At $\beta = 0.8$ (ordered phase, moderate sign problem), the action density agrees with the world-line value within $\sim 0.2\%$ for all $\alpha_{\rm DS}$, as without stabilization, and the estimator lies within $\sim 0.2\%$ of its target.

At $\mu^2 = 0.001$, where $\langle e^{i \varphi} \rangle_{\rm pq} > 0.97$, complex Langevin without stabilization already fails in the disordered phase.
It lies $\sim 25\%$ and $\sim 12\%$ below the world-line values at $\beta = 0.1$ and $0.4$, whereas with stabilization it agrees with them within $\sim 1$ standard deviation.
The deviation that remains with stabilization appears at $\beta = 0.4$, $\mu^2 \geq 0.1$, where the sign problem is severe, $\langle e^{i \varphi} \rangle_{\rm pq} \lesssim 0.1$.
The data thus suggest that stabilization removes the portion of the failure unrelated to the sign problem.
We have not established that the sign problem causes the remaining part, since this rests on two couplings of the disordered phase on a single lattice.

Fig.~\ref{fig:ds_vs_alpha} shows the dependence on $\alpha_{\rm DS}$ at $\mu^2 = 0.2$.
The action density moves from the incorrectly converged value without stabilization to a plateau with stabilization.
A sigmoid fit (similar to what was used in Ref. \cite{Hansen:2024lkn}) places the plateau between the world-line and the phase-quenched values at $\beta = 0.1$ and $0.4$.
The estimator responds to the stabilization immediately.
Already for the weakest forces, $\alpha_{\rm DS} = 0.01$ -- $0.1$, it moves from $\sim 12\%$ and $\sim 3\%$ below its target at $\beta = 0.1$ and $0.4$ to within $\sim 6\%$ above it, and to within $\sim 3\%$ for $\alpha_{\rm DS} \geq 10^3$, by the stabilizing term, Eq. \eqref{eq:ds_trH_ds}.
It also follows small changes in the action density at large $\alpha_{\rm DS}$.
At $\beta = 0.1$, the action density comes closest to the world-line value around $\alpha_{\rm DS} = 10^4$, where the estimator comes closest to its target, and both move away again towards $10^9$.
At $\beta = 0.8$, both agree with the world-line value and the target.
Only at very large $\alpha_{\rm DS}$, up to $10^{12}$, does the action density deviate slightly from the world-line central value, and the estimator responds with a rise of $\sim 0.2\%$.
There is, however, a drawback.
For weak forces, the estimator crosses its target on its way from below to above, so that agreement with the target does not imply correct results.
This is most apparent at $\beta = 0.4$ and $\alpha_{\rm DS} = 0.01$: the estimator is $0.992(8)$, at its target within errors, while the action density is still $\sim 22\%$ below the world-line value.
A plausible explanation is that even a weak force suppresses the large excursions of the estimator, which originate near its poles (Fig.~\ref{fig:ds_history}), while the bulk of the distribution, which determines the action density, is corrected only at larger $\alpha_{\rm DS}$.
The estimator alone is therefore not a reliable criterion for tuning the stabilizing force.

Dynamical stabilization is one of the several proposed methods for obtaining correct results where complex Langevin fails.
Other directions we have not tested are as follows.
\begin{itemize}
\item[--] \emph{Re-weighting.} Complex Langevin configurations are generated for an auxiliary ensemble in which convergence is correct.
They can then be re-weighted to the target ensemble, provided there is sufficient overlap between the two parameter regimes \cite{Bloch:2017rwcl}.

\item[--] \emph{Kernels.} A kernel in the Langevin equation can give correct convergence \cite{Okamoto:1988ru, Boguslavski:2022dee, Mandl:2024zvz}.
The deviation of the estimator from its target could serve as part of the loss for optimizing kernels~\cite{Alvestad:2022abf}, alongside other criteria.

\item[--] \emph{Boundary-term corrections.} Boundary terms measure the error of complex Langevin for a given observable and can be used to correct it.
For this model, this reduced the deviation of the action density at $\beta = 0.2$, $\mu^2 = 0.1$ from $27\%$ to $8\%$ \cite{Scherzer:2019lrh, Cai:2021hjq}.

\item[--] \emph{Contour deformations.} Instead of complexifying the fields stochastically, one can deform the integration contour to reduce the sign problem.
For the XY model in $2 + 1$ dimensions at finite density, this reduces the severity of the sign problem exponentially in $\mu^2$ and in the spatial volume, as measured by the average phase factor \cite{Tulipant:2022vtk}.
\end{itemize}

In each case, the estimator, together with a sign-problem-free benchmark where one exists, provides a test of the result.

\section{Discussion and Outlook}
\label{sec:discussion}

We have benchmarked complex Langevin dynamics for the three-dimensional XY model at finite chemical potential against the world-line formulation, and used this benchmark to assess the configurational temperature estimator and the drift-decay criterion.

On the methodological side, we derived the target value of the estimator from a Schwinger--Dyson identity.
The identity holds unchanged for the complex weight, provided the estimator is built from the holomorphic drift square $\sum_x K_x^2$, and the term localized on its singular set is negligible for the chemical potentials of our main scan.
We related the simpler identity $\langle \sum_x K_x^2 \rangle = -2 \langle S \rangle$ to the consistency conditions and boundary terms of Refs.~\cite{Aarts:2011ax, Scherzer:2018hid}.
Complex Langevin reproduces the identity when it converges correctly, so agreement with the target is a necessary, but not a sufficient, condition for correctness.

Our results confirm and quantify the phase dependence of complex Langevin in this model found in Refs.~\cite{Aarts:2010aq, Scherzer:2019lrh}.
At imaginary chemical potential, where real Langevin is correct, the estimator agrees with its target to within $\sim 1\%$ for all $\beta$.
At real chemical potential, complex Langevin reproduces the world-line action density within $0.3\%$ for $\beta \geq 0.7$, and the estimator agrees with its target.
In the disordered phase, the estimator deviates from its target by $\sim 1$ -- $12\%$, and the world-line comparison confirms that complex Langevin fails.
This happens already at $\mu^2 = 0.001$, where $\langle e^{i\varphi} \rangle_{\rm pq} > 0.97$ for every $\beta$, while for $\beta \geq 0.7$ complex Langevin is correct even where the phase factor is substantially reduced.
At $\mu^2 \leq 0.2$, where we measured the severity of the sign problem, the failure is therefore controlled by the phase of the system rather than by that severity.
The exploratory map of the estimator up to $\mu^2 = 14$ indicates that, on our $8^3$ lattice, the region of large deviations moves with the order--disorder boundary $\beta_c(\mu^2)$ as $\mu^2$ increases, lies predominantly on its disordered side, and extends somewhat into the ordered side.
At large $\mu^2$ we did not measure the sign problem, so the map alone does not separate the two effects there.

The comparison also shows the limitations of the estimator.
Near the transition, at $\beta = 0.5$, the action density is off by up to a few percent, but the estimator does not resolve this, while the drift distributions show slowly decaying tails over the sampled range.
The estimator tests a local identity and cannot detect ergodicity problems, such as the metastability of the standard discretization beyond the Roberge--Weiss point.
In the disordered phase, the complex Langevin process comes close to the poles of the estimator, which makes its fluctuations heavy-tailed, renders its jackknife errors only indicative, and can contribute to its deviation.
The drift criterion, in turn, appears more conservative: at $\beta = 0.7$ -- $0.9$ the drift distributions retain tails while the results are correct.
All our results are for a single lattice volume and a single target step size, without extrapolation to $\bar\epsilon \to 0$.
Ref.~\cite{Aarts:2010aq} found the region of failure to be independent of the volume, but the volume and step-size dependence of the estimator remain to be studied.

Dynamical stabilization brings the action density into agreement with the world-line result at $\beta = 0.1$ for all $\mu^2$ studied, within about two standard deviations, and leaves the correct results at $\beta = 0.8$ unchanged.
At $\beta = 0.4$, where the sign problem is severe, a deviation of $1$ -- $4\%$ remains for $\mu^2 \geq 0.1$, and the results approach those of the phase-quenched theory, which is the limit of strong stabilization.
With stabilization, the estimator lies above its target by an amount set by the stabilizing term, which signals that the stabilized process violates the physical Schwinger--Dyson identity at the percent level.
A weak force can bring the estimator to its target while the action density is still far off, so the estimator alone cannot be used to tune the stabilization.

These results suggest that the configurational temperature estimator is most useful as one of several complementary checks.
It has a known target value, it can be monitored during a run, and in this model it detected the large failures in the disordered phase; where the process approaches its poles, however, its statistical error must be interpreted with care.
It should be combined with drift-based diagnostics, boundary-term analyses and, where available, sign-problem-free benchmarks.
Natural next steps are to extend it to abelian and non-abelian gauge theories, where the cost of the Hessian needs to be assessed, to study its volume and step-size dependence, to characterize the distribution of $\sum_x K_x^2$ near its zeros, and to use it together with other criteria to delineate the regions of parameter space where complex Langevin can be trusted, for example in the optimization of kernels \cite{Okamoto:1988ru, Boguslavski:2022dee, Alvestad:2022abf, Mandl:2024zvz}.
Given its quantitative character, the estimator might also serve as a consistency check for sampling in generative models, such as diffusion models for lattice field theories~\cite{Habibi:2024fbn, Aarts:2025lpi}.

\begin{acknowledgments} 

We extend our gratitude to Navdeep Singh Dhindsa, Diaa E. Habibi, Vamika Longia, Michael Mandl, and Piyush Kumar for their invaluable discussions.
The work of A.J. was supported in part by a Start-up Research Grant from the University of the Witwatersrand.
The work of A.K. was partly supported by the Deutsche Forschungsgemeinschaft (DFG, German Research Foundation) through the CRC-TR 211 ``Strong-interaction matter under extreme conditions'' -- project number 315477589 -- TRR 211, by the National Natural Science Foundation of China under Grants No. 12293064, No. 12293060, and No. 12325508, and by the National Key R$\&$D Program of China under Contract No. 2022YFA1604900. 
Parts of the computations in this work were performed on the CPU resources of the GPU cluster at Bielefeld University. 
\end{acknowledgments} 

\section*{Data availability} 

The data that support the findings of this article are openly available \cite{Joseph:2026data}. 

\raggedright 
\bibliographystyle{apsrev4-2} 
\bibliography{bibfile} 

\end{document}